\documentclass{aa} 
\usepackage{graphicx}
\usepackage{color}
\usepackage{txfonts}
\usepackage{amsmath}
\usepackage{natbib}\usepackage{hyperref}
\hypersetup{
    colorlinks=true,
    linkcolor=blue,
    filecolor=magenta,      
    urlcolor=cyan,
    citecolor=blue
}

\newcommand{\msun}{{\rm M_\odot}}
\newcommand{\feh}{\rm{[Fe/H]}}
\newcommand{\teff}{T_{\rm eff}}
\newcommand{\logg}{\log{g}}
\newcommand{\vmic}{v_{\rm mic}}

\newcommand{\vsini}{v_{\rm sin~i}}
\newcommand{\rsun }{{\rm R_\odot}}
\newcommand{\vbrd}{v_{\rm brd}}
\newcommand{\numax}{\nu_{\rm max}}
\newcommand{\dnu}{\Delta\nu}
\begin{document}
\title{\vspace{-0.25cm}The SAPP pipeline for the determination of stellar abundances and atmospheric parameters of stars in the core program of the PLATO mission}

\author{
Matthew Raymond Gent\inst{1} \and
Maria Bergemann\inst{1, 2}\and
Aldo Serenelli\inst{3,4,1}\and
Luca Casagrande \inst{8}\and
Jeffrey M. Gerber\inst{1}\and
Ulrike Heiter\inst{6}\and
Mikhail Kovalev \inst{1,23} \and
Thierry Morel\inst{5}\and
Nicolas Nardetto\inst{7}\and 
Vardan Adibekyan \inst{11,13}\and
Víctor Silva Aguirre \inst{24}\and
Martin Asplund \inst{19}\and
Kevin Belkacem \inst{14}\and
Carlos del Burgo \inst{15,16}\and 
Lionel Bigot \inst{7} \and
Andrea Chiavassa \inst{7}\and
Luisa Fernanda Rodríguez Díaz \inst{24}\and
Marie-Jo Goupil \inst{14}\and 
Jonay I. Gonz\'alez Hern\'andez \inst{16,17}\and
Denis Mourard \inst{7}\and
Thibault Merle\inst{9}\and
Szabolcs Mészáros\inst{10,11,12}\and
Douglas J. Marshall \inst{20,21} \and
Rhita-Maria Ouazzani \inst{14}\and 
Bertrand Plez \inst{22}\and
Daniel Reese \inst{26}\and
Regner Trampedach \inst{25}\and
Maria Tsantaki \inst{18}
}
\institute{
Max Planck Institute for Astronomy, 69117, Heidelberg, Germany \email{gent@mpia-hd.mpg.de}
\and
Niels Bohr International Academy, NBI, University of Copenhagen, Blegdamsvej 17, DK-2100 Copenhagen, Denmark
\and
Institute of Space Sciences (ICE, CSIC), Carrer de Can Magrans S/N, E-08193, Cerdanyola del Valles, Spain
\and
Institut d'Estudis Espacials de Catalunya (IEEC), Carrer Gran Capita 2, E-08034, Barcelona, Spain
\and
Space sciences, Technologies and Astrophysics Research (STAR) Institute, Universit\'e de Li\`ege, Quartier Agora, All\'ee du 6 Ao\^ut 19c, B\^at. B5C, B4000-Li\`ege, Belgium
\and
Observational Astrophysics, Department of Physics and Astronomy, Uppsala University, Box 516, 751 20 Uppsala, Sweden
\and
Université C\^ote d'Azur, Observatoire de la C\^ote d'Azur, CNRS, Lagrange, CS 34229, Nice, France
\and
Research School of Astronomy $\&$ Astrophysics, Mount Stromlo Observatory, Australian National University, ACT 2611, Australia
\and
Institut d'Astronomie et d'Astrophysique, Universit\'e Libre de Bruxelles, CP. 226, Boulevard du Triomphe, 1050 Brussels, Belgium
\and
ELTE E\"otv\"os Lor\'and University, Gothard Astrophysical Observatory, 9700 Szombathely, Szent Imre H. st. 112, Hungary
\and
MTA-ELTE Exoplanet Research Group, Hungary
\and
MTA-ELTE Lend{\"u}let Milky Way Research Group, Hungary
\and
Instituto de Astrof\'isica e Ci\^encias do Espa\c{c}o, Universidade do Porto, CAUP, Rua das Estrelas, 4150-762 Porto, Portugal
\and
Observatoire Paris-Site de Meudon, LESIA, 5 Pl Jules Janssen, 92195 Meudon Cedex, France
\and
Instituto Nacional de Astrofísica, \'Optica y Electr\'onica, Luis Enrique Erro 1, Sta. Ma. Tonantzintla, Puebla, CP 72840, Mexico
\and
Instituto de Astrof{\'\i}sica de Canarias, 38205 La Laguna, Tenerife, Spain
\and
Departamento de Astrof{\'\i}sica, Universidad de La Laguna, 38206 La Laguna, Tenerife, Spain
\and
Istituto Nazionale di Astrofisica - INAF,Osservatorio Astrofisico di Arcetri, Largo E. Fermi 5, 50125 Firenze, Italia
\and
Max Planck Institute for Astrophysics, Karl-Schwarzschild-Str. 1, D-85741 Garching, Germany
\and
Universit\'e de Toulouse, UPS-OMP, IRAP, F-31028 Toulouse cedex 4, France
\and
CNRS, IRAP, 9 Av. colonel Roche, BP 44346, F-31028 Toulouse cedex 4, France
\and
LUPM, Univ Montpellier, CNRS, Montpellier, France
\and
Yunnan Observatories, Chinese Academy of Sciences, P.O. Box 110, Kunming 650216, China
\and
Department of Physics and Astronomy, Aarhus University, Ny Munkegade 120/1520, 8000, Aarhus, Denmark
\and 
Space Science Institute,4765 Walnut Street, Boulder, CO 80301, U.S.A.
\and
LESIA, Observatoire de Paris, B\^at. 14 - Soleil-Plan\`etes, 5 place Jules Janssen, 92195 Meudon, France
}
\date{Received March 23, 2021; accepted November 8, 2021}
\abstract{We introduce the SAPP (Stellar Abundances and atmospheric Parameters Pipeline), the prototype of the code that will be used to determine parameters of stars observed within the core program of the PLATO space mission. The pipeline is based on the Bayesian inference and provides effective temperature, surface gravity, metallicity, chemical abundances, and luminosity. The code in its more general version can have a much wider range of applications. It can also provide masses, ages, and radii of stars and can be used for stars of stellar types not targeted by the PLATO core program, such as red giants. We validate the code on a set of $27$ benchmark stars that includes $19$ FGK-type dwarfs, $6$ GK-type sub-giants, and $2$ red giants. Our results suggest that combining various observables is the optimal approach, as it allows to break degeneracies between different parameters and yields more accurate values of stellar parameters and more realistic uncertainties. For the PLATO core sample, we obtain a typical uncertainty of $27 (\rm{syst.}) \pm 37$ ($\rm{stat.}$) K for $\teff$, $0.00 \pm 0.01 $ dex for $\logg$,  $0.02 \pm 0.02$ dex for metallicity [Fe/H], $-0.01 \pm 0.03$ $\rsun$ for radii, $-0.01 \pm 0.05$ $\msun$ for stellar masses, and $-0.14 \pm 0.63$ Gyrs for ages. We also show that the best results are obtained by combining the $\nu_{max}$ scaling relation and stellar spectra. This resolves the notorious problem of degeneracies, which is particularly important for F-type stars.}
\titlerunning{SAPP pipeline}
\authorrunning{Gent et al.}
\maketitle 



\section{Introduction}
\label{sec:intro}
The past decade has seen revolutionary developments in astronomical surveys and large-scale observational programs aimed at assembling high-quality data for millions of stars in our Galaxy. Many of the ongoing efforts are motivated by using stars as tracers of Galaxy structure and evolution (e.g., \textit{Hipparcos}: \citealt{Perryman1997}, 2MASS: \citealt{Skrutskie2006}, SDSS/SEGUE: \citealt{Yanny2009}, GCS: \citealt{Nordstrom2004a}, Gaia-ESO: \citealt{2012Msngr.147...25G,Randich2013}, LAMOST: \citealt{2012RAA....12.1197C}, RAVE: \citealt{2006AJ....132.1645S}, Gaia: \citealt{Gaia-Collaboration2016}, APOGEE: \citealt{2017AJ....154...94M}, GALAH: \citealt{DeSilva2015}). With the advent of exoplanet science, the role of stars as exoplanet hosts is becoming increasingly important. Space-based missions - CoRoT, Kepler, and TESS - are discovering new exoplanets at an astonishing rate \citep[e.g.][]{Borucki2010, Buchhave2014, Ricker2015}, but also yield precise data for the studies of stellar interiors with asteroseismology techniques 
\citep[e.g.][]{Christensen-Dalsgaard2002, Huber2013, Serenelli2017}. Future facilities, such as PLATO \citep{Rauer2014}, are opening entirely new  perspectives for studies of exoplanets in different environments and for studies of stars at the level of details that have so far been only available for our Sun. 
 
In this work, we introduce the SAPP pipeline\footnote{We stress that the detailed approach to the analysis of PLATO targets will evolve.} that will be used for the determination of atmospheric parameters of stars observed within the core program of the PLATO space mission \citep{Montalto2021}. The code will provide stellar effective temperature ($\teff$), metallicity ([Fe/H]), surface gravity ($\logg$), radial velocities, and detailed chemical composition, among other parameters. Our ambitious goal is to reach uncertainties as low as 1$\%$ on $\teff$ and [Fe/H], for instance, in order to match the high spectral data quality provided by upcoming ground-based facilities, such as WEAVE and 4MOST. The SAPP is written in Python and the source code is public and available for use\footnote{\url{https://github.com/mg477/SAPP}}.
 
Our approach to the analysis of our targets is somewhat different from other available pipelines and codes (e.g., SME: \citealt{Valenti1996,Piskunov2017}, The Cannon: \citealt{Ness2015}, The Payne: \citealt{Ting2016}, \citealt{Rix2016}, \citealt{Ting2019}, MATISSE: \citealt{Recio-Blanco2006}, CNN StarNet: \citealt{Bialek2020}), although our analysis shares many of the elements with these software units. It is beyond the scope of this paper to provide a detailed comparison of our code with the other programs, as ultimately the design of a computer program follows the needs and objectives of a given research project, and this sets the conceptual and numerical basis of a code. In our case, motivated by the need to provide accurate and precise astrophysical parameters for several $10^5$ (and more) stars to enable their exploitation in studies of exoplanets, we have chosen a Bayesian inference method as the basis of the code \citep[see, e.g.][for a review of  methods and models]{2019ARA&A..57..571J}. The idea of using Bayesian techniques is not new, and it has been already presented in a number of studies \citep[e.g.][]{Pont2004, Jorgensen2005, Serenelli2013, Schoenrich2014, Bailer-Jones2018,McMillan2018,delBurgo2018,Steinmetz2020}. In this work, we closely follow the methodology outlined in \citet{Schoenrich2014} and combine various sources of observational information, including electromagnetic spectra, parallaxes, photometry, and seismic constrains to determine astrophysical parameters of stars. It has been already shown \citep[e.g.][]{Pont2004, Jorgensen2005,2012Gruberbauer,2012Bazot} that the analysis of fundamental parameters of stars, masses and metallicities, especially benefits from the Bayesian approach. Therefore, in order to take the full advantage of the Bayesian formalism, we also include stellar evolution models that give us the ability to determine masses, luminosities, radii, and ages of stars in a single consistent framework taking into account the correlations between all relevant parameters. 

The paper is organised as follows. Section \ref{sec:observations} gives an overview of the observed sample of stars that are used to validate the approach developed in this work. In Sect. \ref{sec:theory} we outline the basic concepts behind the numerical part of the code and review the input parameters and input models, including stellar structure models, stellar atmospheres, and the grids of stellar spectra. In Sect. \ref{sec:results} we present the results of the analysis of the benchmark stars, as well as clusters and in Sect. \ref{sec:BO-SAPP} we describe the operation of SAPP briefly in the context of the PLATO space mission \citep{Rauer2014}. We close the paper with the discussion of forthcoming improvements to the pipeline in Sect. \ref{sec:future} and draw conclusions in Sect. \ref{sec:Conclusion}. 
\begin{figure}[ht!]
\centering
\includegraphics[scale=0.45]{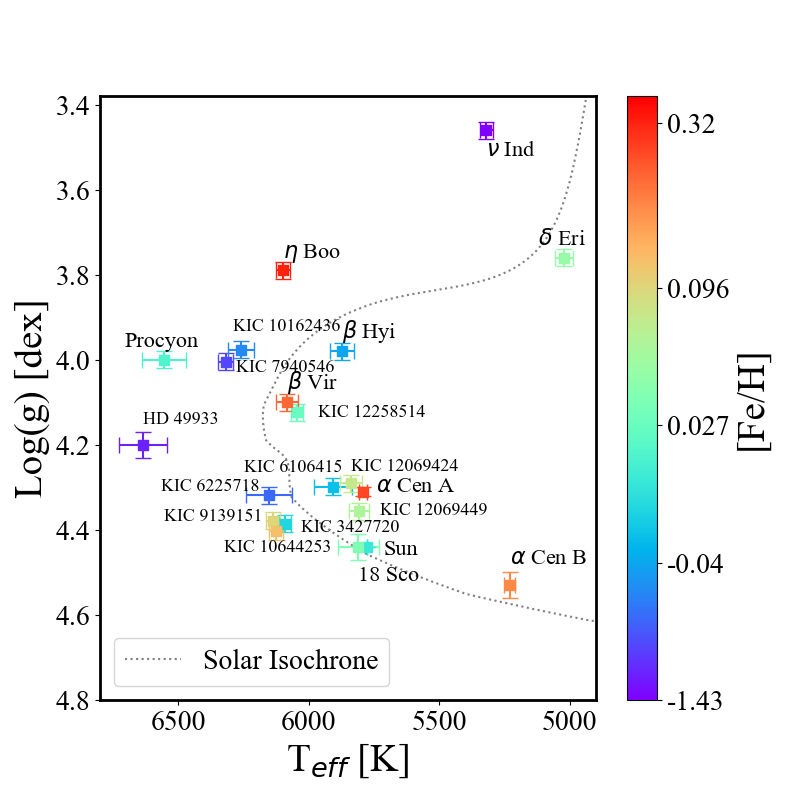}
\caption{$\teff$-$\logg$ diagram of the benchmark targets. Colour scale represents metallicity (blue end is metal poor and red end is just above solar metallicity). Each star is plotted with its reference values (Table \ref{tab:ref_primary}) on top of a solar isochrone.}
\label{fig:fig1}
\end{figure}
\begin{table*}
\small
\begin{center}
\setlength{\tabcolsep}{1pt}
\caption{Reference parameter sample of benchmark stars}
\label{tab:ref_primary}
\begin{tabular}{l l c c c c c c c c c c c}
\hline
\hline
\noalign{\smallskip} Star ID  & HD  & V &  Ks & Parallax & $\teff$  & $\log g$ &  [Fe/H] & Mass & Age & Radius & Luminosity  \\ 
        &      &  mag & mag & mas & K  & dex & dex & M$_\odot$ & Gyrs & R$_\odot$ & L$_\odot$ \\
\noalign{\smallskip}
\hline
\noalign{\smallskip}
 Gaia-ESO & &  & & & & & & & & & \\
18 Sco        &  \object{HD 146233} & 5.50  &  ~~4.19 $\pm$  0.29 & 70.74 $\pm$ 0.06  & 5810 $\pm$ 80 & 4.44 $\pm$ 0.03 &  ~~0.03 $\pm$ 0.03 & 1.02 $\pm$ 0.06 & 2.90 $\pm$ 0.50 & 1.01 $\pm$ 0.01 & 1.05 $\pm$ 0.06 \\
$\alpha$ Cen A & \object{HD 128620} & 0.01  &  -1.52  $\pm$  0.05\tablefootmark{$a$}& 743.00 $\pm$ 1.30 \tablefootmark{$(3)$}  & 5792 $\pm$ 16 & 4.31 $\pm$ 0.01 & ~~0.26 $\pm$ 0.08 & 1.11 $\pm$ 0.00 & 5.26 $\pm$ 0.95 & 1.22 $\pm$ 0.01 & 1.52 $\pm$ 0.01 \\
$\alpha$ Cen B & \object{HD 128621} & 1.33  &  -0.64  $\pm$  0.05\tablefootmark{$a$}& 743.00 $\pm$ 1.30  \tablefootmark{$(3)$} & 5231 $\pm$ 20 & 4.53 $\pm$ 0.03 & ~~0.22 $\pm$ 0.10 & 0.94 $\pm$ 0.00 & 5.26 $\pm$ 0.95 & 0.86 $\pm$ 0.00 & 0.50 $\pm$ 0.01 \\
$\beta$ Hyi    & \object{HD 2151}   & 2.70  &  ~~1.3 $\pm$ 0.04\tablefootmark{$b$}& 133.72 $\pm$ 0.27 \tablefootmark{$(1)$}  & 5873 $\pm$ 45 & 3.98 $\pm$ 0.02 & $-0.04$ $\pm$ 0.06 & 1.14 $\pm$ 0.05 & 6.40 $\pm$ 1.40 & 1.89 $\pm$ 0.03 & 3.52 $\pm$ 0.09 \\
$\beta$ Vir    & \object{HD 102870} & 3.60  &  ~~2.28 $\pm$ 0.01\tablefootmark{$c$}& 90.89 $\pm$ 0.19 \tablefootmark{$(1)$} & 6083 $\pm$ 41 & 4.10 $\pm$ 0.02 & ~~0.24 $\pm$ 0.07 & 1.34 $\pm$ 0.04 & 4.00 $\pm$ 1.00 & 1.68 $\pm$ 0.01 & 3.58 $\pm$ 0.04 \\
$\delta$ Eri   & \object{HD 23249}  & 3.54  &  ~~1.62 $\pm$ 0.29 & 110.03 $\pm$ 0.19 \tablefootmark{$(1)$} & 5022 $\pm$ 34 & 3.76 $\pm$ 0.02 & ~~0.06 $\pm$ 0.05 & 1.33 $\pm$ 0.07 & 6.19   & 2.35 $\pm$ 0.01 & 2.94 $\pm$ 0.00\\
$\eta$ Boo     & \object{HD 121370} & 2.68  &  ~~1.31 $\pm$ 0.02\tablefootmark{$d$} & 87.75 $\pm$ 1.24 \tablefootmark{$(2)$} & 6099 $\pm$ 28 & 3.79 $\pm$ 0.02 & ~~0.32 $\pm$ 0.08 & 1.64 $\pm$ 0.07 & 2.67 $\pm$ 0.10 & 2.67 $\pm$ 0.02 & 8.97 $\pm$ 0.12\\
CoRoT 20       & \object{HD 49933} & 5.76   &  ~~4.72  $\pm$ 0.02 & 33.53 $\pm$ 0.04  &  6635 $\pm$ 91 & 4.20 $\pm$ 0.03 & $-0.41$ $\pm$ 0.08 & 1.28 $\pm$ 0.01 & 1.83 $\pm$ 0.10 & 1.46 $\pm$ 0.01 & 3.52 $\pm$ 0.04\\
Procyon        &  \object{HD 61421} & 0.37  &  -0.70 $\pm$ 0.01\tablefootmark{$e$} & 284.56 $\pm$ 1.26 \tablefootmark{$(2)$} & 6554 $\pm$ 84 & 4.00 $\pm$ 0.02 & ~~0.01 $\pm$ 0.08 & 1.50 $\pm$ 0.07 & 1.87 $\pm$ 0.13 & 2.05 $\pm$ 0.03 & 6.90 $\pm$ 0.35\\
Sun     & \ \ \ \ \ \ \ ...   & ...   & ...   &  ... & 5777 $\pm$ 1  & 4.44 $\pm$ 0.01 & ~~0.03 $\pm$ 0.05 & 1.00 $\pm$ 0.00 & 4.56 $\pm$ 0.00 & 1.00 $\pm$ 0.00 & 1.00 $\pm$ 0.00\\
   & & & & & & & & & & & \\
 Kepler Legacy & & & & & & & & & & & \\
KIC 10162436   & \object{HD 188819} & 8.66  & ~~7.36 $\pm$ 0.02 & 7.30 $\pm$ 0.01  & 6259 $\pm$ 49 & 3.98 $\pm$ 0.02 & $-0.07$ $\pm$ 0.02 & 1.41 $\pm$ 0.05 & 2.57 $\pm$ 0.43 & 2.01 $\pm$ 0.03 & ...  \\
KIC 10644253   & \object{BD+47 2683} & 9.26  & ~~7.87 $\pm$ 0.03 & 10.35 $\pm$ 0.01  & 6126 $\pm$ 27 & 4.40 $\pm$ 0.02 & ~~0.13 $\pm$ 0.02 & 1.16 $\pm$ 0.02 & 1.37 $\pm$ 0.72 & 1.12 $\pm$ 0.00 & 1.45 $\pm$ 0.09\\
16 Cyg A   & \object{HD 186408} & 5.95  & ~~4.43 $\pm$  0.02 & 47.32 $\pm$ 0.02  & 5839 $\pm$ 42 & 4.29 $\pm$ 0.02 & ~~0.09 $\pm$ 0.01 & 1.07 $\pm$ 0.01 & 7.36 $\pm$ 0.31 & 1.22 $\pm$ 0.01 & 1.52 $\pm$ 0.05\\
16 Cyg B   & \object{HD 186427} & 6.20  & ~~4.65 $\pm$  0.02 & 47.33 $\pm$ 0.02  & 5809 $\pm$ 39 & 4.36 $\pm$ 0.02 & ~~0.06 $\pm$ 0.01 & 1.04 $\pm$ 0.05 & 7.05 $\pm$ 0.63 & 1.11 $\pm$ 0.02 & 1.21 $\pm$ 0.11\\
KIC 12258514   & \object{HD 183298} & 8.16  & ~~6.76 $\pm$ 0.02 & 12.25 $\pm$ 0.01  & 6046 $\pm$ 24 & 4.12 $\pm$ 0.02 & ~~0.03 $\pm$ 0.02 & 1.25 $\pm$ 0.02 & 5.50 $\pm$ 0.40 & 1.59 $\pm$ 0.02 & 2.63 $\pm$ 0.12\\
KIC 3427720  & \object{BD+38 3428} & 9.22  & ~~7.83 $\pm$ 0.02 & 10.74 $\pm$ 0.01  & 6092 $\pm$ 24 & 4.39 $\pm$ 0.02 & $-0.02$ $\pm$ 0.02 & 1.10 $\pm$ 0.02 & 2.97 $\pm$ 0.78 & 1.12 $\pm$ 0.01 & 1.37 $\pm$ 0.08\\
KIC 6106415  & \object{HD 177153} & 7.21   & ~~5.83 $\pm$ 0.02 & 24.16 $\pm$ 0.01  & 6090 $\pm$ 17 & 4.30 $\pm$ 0.02 & $-0.04$ $\pm$ 0.01 & 1.04 $\pm$ 0.02 & 4.55 $\pm$ 0.28 & 1.21 $\pm$ 0.01 & 1.61 $\pm$ 0.09\\
KIC 6225718  & \object{HD 187637} & 7.53   & ~~6.28 $\pm$ 0.02 & 19.03 $\pm$ 0.02  & 6308 $\pm$ 33 & 4.32 $\pm$ 0.02 & $-0.11$ $\pm$ 0.01 & 1.17 $\pm$ 0.04 & 2.23 $\pm$ 0.20 & 1.23 $\pm$ 0.02 & 2.08 $\pm$ 0.11\\
KIC 7940546  & \object{HD 175226} & 7.42   & ~~6.17 $\pm$ 0.02 &  12.96 $\pm$ 0.02  & 6319 $\pm$ 28 & 4.00 $\pm$ 0.02 & $-0.10$ $\pm$ 0.01 & 1.51 $\pm$ 0.09 & 2.42 $\pm$ 0.17 & 1.97 $\pm$ 0.04 & 5.69 $\pm$ 0.35\\
KIC 9139151  & \object{BD+45 2796} & 9.29  & ~~7.95 $\pm$ 0.02 & 9.76 $\pm$ 0.01  & 6136 $\pm$ 27 & 4.38 $\pm$ 0.02 & ~~0.10 $\pm$ 0.02 & 1.18 $\pm$ 0.03 & 1.49 $\pm$ 0.68 & 1.16 $\pm$ 0.01 & 1.81 $\pm$ 0.11\\
   &  & & & & & & & & & & \\
 Other & & & & & & & & & & & \\
$\nu$ Ind    & \object{HD 211998} & 5.29   & ~~3.54 $\pm$ 0.26 & 35.13 $\pm$ 0.06  & 5320 $\pm$ 24 & 3.46 $\pm$ 0.02 & $-1.43$ $\pm$ 0.06 & 0.85 $\pm$ 0.05 & 11.00 $\pm$ 1.06 &  ... & ... \\
\noalign{\smallskip}
\hline
\end{tabular}
\tablefoot{The reference parameter sample of well-studied FGK stars used for the tests reported in this work. The errors of V-mag are 0.01 mag. KIC 12069424 is 16 Cyg A and KIC 12069449 is 16 Cyg B. See Sect. \ref{sec:benchmark_stars_parameters} for a more detailed description. vsini values are not tabulated here however they range from 1.1 kms$^{-1}$ ($\alpha$ Cen B) to 12.7 km$s^{-1}$ ($\eta$ Boo).
The Ks magnitudes are taken from \citet{cutri03}, except for the stars marked in the table, for which they were calculated from K magnitudes taken from the following references, transformed to Ks using Eq.~(A1) from \citet{carpenter01}: 
\tablefoottext{$a$}{\citet{1981A&AS...45....5E},}
\tablefoottext{$b$}{\citet{1974MNSSA..33...53G,1981A&AS...45....5E,1990MNRAS.242....1C},}
\tablefoottext{$c$}{\citet{1966CoLPL...4...99J, 1968ApJ...152..465J, 1991ApJ...368..264A},}
\tablefoottext{$d$}{\citet{1966CoLPL...4...99J, 1967ApJ...149..345J, 1979MNRAS.188..847B, 1984MNRAS.206..611G, 1988A&AS...74..127S, 1989A&A...215..305A},}
\tablefoottext{$e$}{\citet{1964ApJ...139.1130L, 1966CoLPL...4...99J, 1974MNSSA..33...53G, 1978AJ.....83..651V, 1981A&AS...45....5E, 1984RMxAA...9...65T, 1984A&A...137L...9R, 1987ApJ...320..364E, 1994A&A...282..684A}.}
The photo-geometric distances of the stars are taken from \citet{Bailer-Jones2021}, these use Gaia eDR3 parallaxes which are not marked in the table \citep{Gaia-Collaboration2020}. The stars that are marked, the distances were calculated using the formula d = 1/$\pi$ and thus parallaxes shown are from, 
\tablefoottext{$1$}{\citet{Gaia-Collaboration2020},}
\tablefoottext{$2$}{\citet{2007Van},}
\tablefoottext{$3$}{\citet{2016Pourbaix}.}
Mean and standard deviation are given in case of more than one reference. \\
All parameters that have an uncertainty of $0.00$ are accurate to less than $0.00$ of the given unit. \\ The current solar effective temperature is 5772 $\pm$ 1 K \citep{Prsa2016}, the use of the older value of 5777 K within this study is justified by the SAPP's systematic uncertainties which are more than 10 K.
}
\end{center}
\end{table*}

\section{Observations}
\label{sec:observations}
\subsection{Observed data}
The core PLATO sample focuses on main-sequence stars and sub-giants \citep{Montalto2021}. Therefore, we first focus on these types of stars, but owing to the versatility of the code and input physics, we also present in Sect. \ref{subsec:med_spectra} the performance of the code on spectra of red giants in the Galactic field and in stellar clusters.

Our core validation stellar sample includes Gaia benchmark stars \citep{Heiter2015}, Kepler Legacy stars \citep{Silva-Aguirre2017, Nissen2017}, and a very metal-poor star $\nu$ ind \citep{Chaplin2020}. These stars are shown in the $\teff - \logg$ plane in Fig \ref{fig:fig1}, and they provide a representative coverage of the current PIC \citep{Montalto2021}. The stars cover a broad range in effective temperatures, from 5022 K ($\delta$ Eri) to 6635 K (HD 49933), and surface gravities from $\logg = 3.46$ ($\nu$ Ind) to $\logg = 4.53$ dex ($\alpha$ Cen B). One of the stars in the sample ($\eta$ Boo) is a relatively fast rotator (for FGK-type stars) with a projected equatorial rotational velocity $\vsini = 12.7$ kms$^{-1}$. Some of the stars have a super-solar metallicity, in excess of $+0.2$ dex ( $\eta$ Boo, $\alpha$ Cen A,B system), whereas the most metal-poor star in the sample is $\nu$ Ind with [Fe/H] $=-1.43$ dex \citep{Chaplin2020}.

The spectra of the benchmark stars are taken from the ESO public archive. For the majority of stars, spectra obtained with the high-resolution UVES spectrograph mounted on VLT are available. For the Kepler stars, reduced combined spectra taken with HARPS-N facility were kindly provided by P. E. Nissen (priv.comm.). The UVES spectra (4800 to 6800 $\AA$) have a resolving power of $\lambda/\delta \lambda \approx 47\,000$  \citep{Dekker2000} and the HARPS data of $115\,000$ \citep{Mayor2003}.The signal-to-noise ratio (S/N) of the UVES and HARPS spectra is in the range from $200$ to over $1\,000$. The spectra are all pre-processed (described in Appendix \ref{subsec:cont_normalisation}) by the SAPP using different procedures developed, this includes continuum normalisation, radial velocity correction and contaminant treatment. Then we degrade the resolving power of the spectra to $R = 20\,000$ and reduce the wavelength coverage to 5300 - 5600 \AA\ in the optical spectral window. This is done to ensure that our tests are as predictive as possible: spectra with such characteristics represent a plausible scenario, in which for the vast majority of PLATO targets only spectra observed with medium-resolution (R $\approx 20\,000$) facilities, such as 4MOST, WEAVE, and APOGEE, will be available. 

\begin{figure}
    \centering
    \includegraphics[scale=0.35]{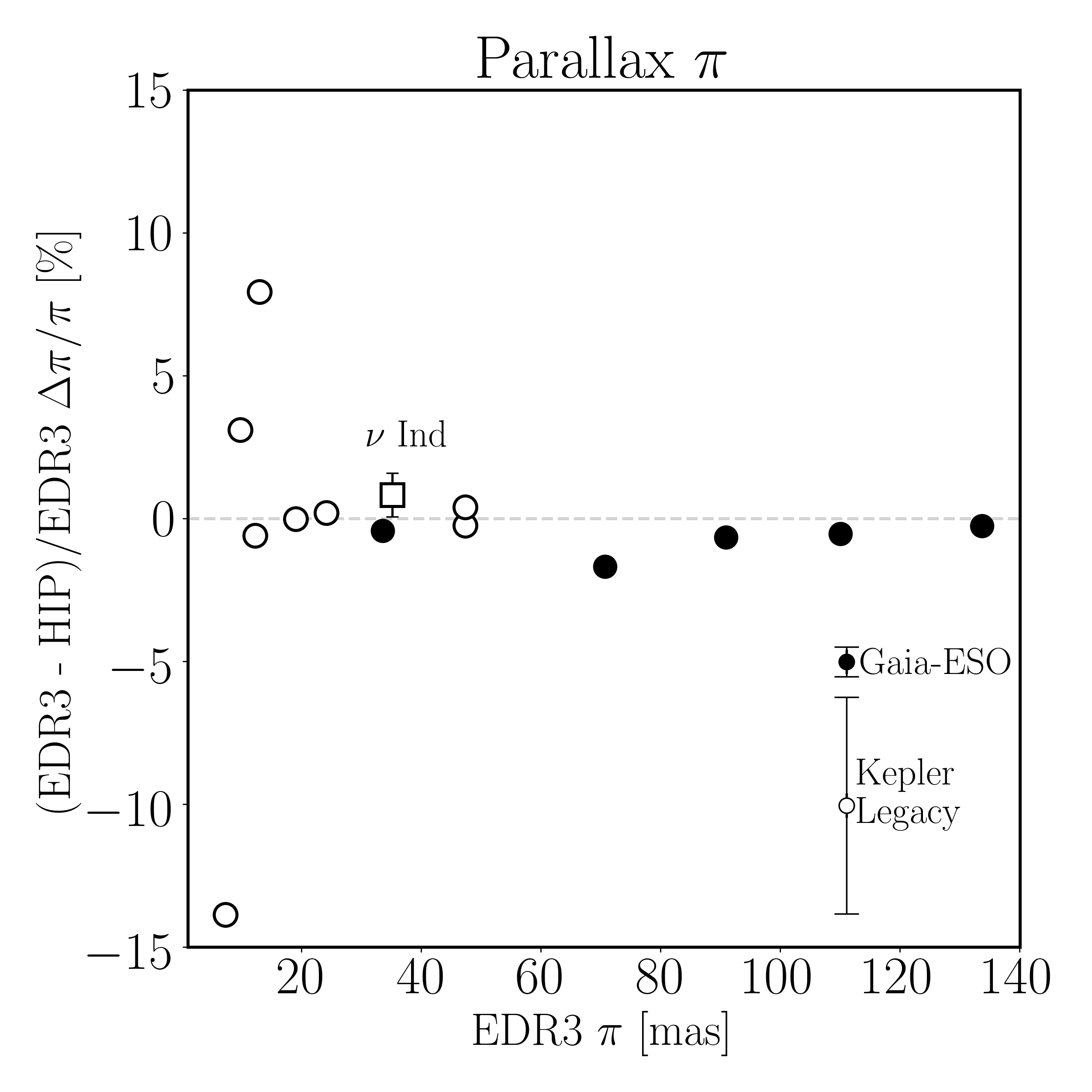}
    \caption{Percentage difference between Gaia EDR3 parallax and Hipparcos parallax \citep{2007Van} with respect to Gaia EDR3 in mas. The error represents the uncertainty of EDR3 parallaxes and Hipparcos parallaxes propagated through the percentage difference.}The open circles represent the Kepler legacy stars, the solid black circles represent Gaia-ESO stars, and the single open square represents our metal poor star $\nu$ Ind. The average uncertainty for the Kepler Legacy and Gaia ESO stars are in the bottom right corner.
    \label{fig:parallax}
\end{figure}

The observed photometry for these stars was extracted from the Gaia EDR3 catalogue \citet{Gaia-Collaboration2016,Gaia-Collaboration2018,Gaia-Collaboration2020}, which we supplemented with 2MASS magnitudes (JHK$_s$) and Johnson-Cousins photometric data, where available. For most stars in the sample, we used the photo-geometric distances from \cite{Bailer-Jones2021} catalogue. If they were not available (see Table \ref{tab:ref_primary}), then Gaia parallaxes $\pi$ are converted into distances via $d\ [\text{pc}] = 1/\pi$. This is a suitable approximation in this work, because the benchmark stars are all nearby targets, with the most distant system being KIC 10162436 at around $138$ pc. If Gaia parallaxes were not available, then values from the Hipparcos catalogue were supplemented \citep{1997HIPPARCOS,2007Van}. The largest percentage difference between distances from \cite{Bailer-Jones2021} and using the inverse parallax values are 0.02$\%$ for Gaia EDR3 and 0.39 $\%$ for Hipparcos, respectively. The comparison between EDR3 and Hipparcos parallaxes can be seen in Fig. \ref{fig:parallax}, where it can be seen that the maximum propagated percentage difference for Kepler Legacy stars are $\sim5\%$ and for Gaia-ESO less than 0.5$\%$.  

The observed magnitudes were corrected for interstellar extinction using the \texttt{Stilism} tool \citep{Capitanio2017}. The tool provides line-of-sight reddening interpolated on a given distance with associated uncertainties. For the benchmark stars, the reddening is small (E(B-V) $\lesssim 0.015$ mag) and is implemented via the extinction correction (see equation \ref{eqn:phot_to_norm_mag}).

\subsection{Stellar parameters of the benchmark stars}\label{sec:benchmark_stars_parameters}
Stellar parameters for our benchmark stars were obtained from several literature sources, the priority being given to \citet{Jofre2018}, because they provide a comprehensive homogeneous analysis of all classical stellar parameters, including metallicity and detailed chemical abundances, derived with constraints from independent data. This study is based on the analysis presented by \citet{Jofre2014} and \citet{Heiter2015}. All parameters for $\nu$ Ind were adopted from \citet{Chaplin2020}.

In short, the determination of parameters for the Gaia benchmark stars is as follows. The $\teff$ values rely on interferometric angular diameters measured with CHARA and VLTI facilities and bolometric fluxes obtained either via integrating the observed Spectral Energy Distributions (SEDs) or indirectly by converting the photometry into $F_{\rm bol}$ using synthetic calibrations. $\delta$ Eri is the only star, for which we give a preference to the recent accurate $\teff$ and radius values based on the interferometric angular diameter measured with VLTI/PIONEER \citep{Rains2020}. The surface gravities were determined using masses estimated from evolutionary tracks, angular diameters, and Hipparcos parallaxes, which are mostly in excellent agreement with Gaia DR2 and Gaia EDR3 values for the reference sample (see Fig. \ref{fig:parallax}). The surface gravities were verified using asteroseismic scaling relations, where possible. Metallicities are model-dependent quantities: here we adopt the Non-Local Thermodynamic Equilibrium (NLTE) estimates. Luminosity measurements are taken from \cite{Heiter2015}, whereas masses are adopted from different sources. For $\delta$ Eri's mass is taken from \cite{2010Bruntt} with age adopted from \cite{2005Thevenin}. $\alpha$ Cen A $\&$ B have masses and radii from \cite{2017Kervella} with age from \cite{2018Joyce}. The mass and age of $\nu$ Ind are taken from \cite{Chaplin2020}, but no luminosity or radius are provided in that study. HD 49933 has mass, age, and radius estimated by \cite{2013Liu}. The age and radius sources vary star-by-star, for 18 Sco age \cite{2013Monroe} and radius \cite{2018Bazot}, for $\beta$ Hyi age and radius \cite{2011Brandao},  $\beta$ Vir age \cite{2006Eggenberger} and radius \cite{2012Boyajian}, $\eta$ Boo age \cite{2005Carrier} and radius \cite{2007vanBelle}, Procyon age \cite{2013Liebert} and radius from \cite{2004Kervella}.

For the Kepler Legacy sample, stellar parameters were adopted from \citet{Nissen2017}, except the 16 Cyg A \& B binary, for which the interferometric $\teff$ estimate from \citet{White2013} is given a preference. We do not use $\nu$ Ind or the Kepler stars in the analysis of $\teff$ (except 16 Cyg A \& B), because their $\teff$ were not determined in the same fundamental way as those of Gaia benchmark stars and large scatter exists between different estimates (e.g.  \citealt{Wu2017} for KIC 6225718). The interferometric angular diameter (AD) measurements for KIC 6225718 and KIC 6106415 exist, however, the errors of the $\teff$ estimates are unfortunately too large ($70$ to $90$ K) to provide a meaningful constraint on the methods. The estimates of mass (M), radius (R), luminosity (L), and age ($\tau$) for the majority of these stars come from \cite{Serenelli2017} (KIC 10162436, KIC 10644253, KIC 3427720, KIC 9139151) and \cite{2017Creevey} (16 Cyg A $\&$ B, KIC 12258514, KIC 6106415, KIC 6225718, KIC 7940546). 
\section{Methods}
\label{sec:theory}
\subsection{Bayesian probability approach} \label{sec:Bayes_approach}

The conditional probability of a set of model parameters \textbf{X} = X$_1$,...,X$_n$ given a set of observations \textbf{O} = O$_1$,...,O$_m$ is given by Bayes theorem and results from from the combined probability P(\textbf{X},\textbf{O}) = P(\textbf{X$|$O})P(\textbf{O}). Hence,
\begin{equation}
    \label{eqn:BayesTheorem}
    P(\textbf{X$|$O}) = \frac{P(\textbf{X})}{P(\textbf{O})}{P(\textbf{O$|$X})},
\end{equation}
where P(\textbf{X$|$O}) is the posterior probability, that is the conditional probability of the parameter set \textbf{X} given the set of observables \textbf{O}. P(\textbf{O$|$X}) is the likelihood, i.e. the probability of the observations occurring given the set of parameters \textbf{X}. P(\textbf{X}) is the prior probability ascribed to a set of predefined parameters. The denominator P(\textbf{O}) is function only of the observations, and it is not relevant for determination of stellar model parameters, as it acts only as a normalisation factor. Therefore, the expression above can be simplified to
\begin{equation}
    \label{eqn:BayesPont}
    P(\textbf{X$|$O}) \propto P(\textbf{X})P(O_1,...,O_m|\textbf{X}),
\end{equation}
where the posterior P(\textbf{X$|$O}) is a probability distribution function (PDF) on the chosen parameter space. For the remainder of this paper the observational likelihoods are abbreviated by a prime given the set of parameters, i.e.  $P'(\textbf{X})$. The current implementation of SAPP assumes observables are statistically independent, which allows decomposing $P(O_1,...,O_m|\textbf{X})$ as, 
\begin{equation}
    \label{eqn:IndividualLhood}
    P'(\textbf{X})=P(O_1,...,O_m|\textbf{X}) = \prod^m_{j=1}P(O_j|\textbf{X}).
\end{equation}

More specifically, based on the observables used in this work, 
\begin{equation}
    \label{eqn:PDF}
    \begin{aligned}
    P(\textbf{X$|$O}) \propto P(O_{spec}|\textbf{X})\cdot P(O_{mag,dist}|\textbf{X}),\\
    \cdot P(O_{seism}|\textbf{X})\cdot P_{prior}(\textbf{X}),
    \end{aligned}
\end{equation}
where the subscripts refer to a specific observable, $j$: "spec" for stellar spectra (fluxes against wavelength), "mag" for photometric magnitudes and their products, "seism" for asteroseismic quantities ($\dnu$, $\numax$), and "prior" for any priors. Here P$_{prior}$($\textbf{X}$) is the prior probability distribution, this can represent the initial mass function (IMF), selection function, or any other source of information that further constrains the parameter space. All observables are assumed to be normally distributed, i.e. 
\begin{equation}
    \label{eqn:red_gaussian}
    P(O_j|\textbf{X}) = \prod_k^{N_j} G_j(x-\bar{x}_k,\sigma_k),
\end{equation}
where N$_j$ is the number of parameter points (index running over k) for the given observable $j$. Therefore,

\begin{equation}
    \label{eqn:norm_gaussian}
    G_j(x-\bar{x}_k,\sigma_k) =\frac{1}{(\sigma_k^{N_j}2\pi)^{\frac{1}{N_j}}} \text{exp}\left[-\frac{(x-\bar{x}_k)^2}{2\sigma_k^2}\right]
\end{equation}
It is easy to modify the expression for the likelihood to include statistically dependent observables by introducing the use of their correlation matrix. This is usually not necessary, but in this work we explore the influence of covariance in the analysis (see Sect. \ref{subsec:bayes_covariance}) and therefore, we also perform additional calculations with equation \ref{eqn:norm_gaussian} modified to:

\begin{equation}
    \label{eqn:cov_gaussian}
    G_j(x-\bar{x}_k,\sigma_k) = \frac{\text{exp}(-\frac{1}{2}(\mathbf{x}-\mathbf{\bar{x}_k})^\text{T}\Sigma^{-1}(\mathbf{x}-\mathbf{\bar{x}_k}))}{\sqrt{(2\pi)^{\mathbf{N_j}}|\Sigma|}},
\end{equation}

where $\Sigma$ is the covariance matrix. The individual likelihoods are described in the following sections. Similar to  \citet{Schoenrich2014}, we include spectra, photometry, parallax, stellar evolution models, and asteroseismic constraints, when available.

The core parameter space is defined as follows. The quantities $\teff$, $\logg$, and $\feh$ represent the \textit{key parameter space}, which all grids (photometric, asteroseismic, spectroscopic) of the SAPP share. On the other hand, mass M, age $\tau$, and radius R, Luminosity L, and abundances are secondary parameters. They can only be inferred from some of the grids and so have dependencies on each of the three parameters in the \textit{key parameter space}.
\subsection{Synthetic photometry}\label{sec:photometry}

The photometry PDF P(O$_{mag,dist}|$\textbf{X}) relates magnitudes and parallaxes to the model predictions. We denote the stellar model magnitudes at model point i and magnitude k$_1$ by m$_{i,k_1}$  and the photometric observation in the filter k$_1$ (such as G, G$_{\text{Bp}}$ or G$_{\text{Rp}}$) with O$_{k_1}$:

\begin{equation}
    \label{eqn:phot_gaussian_mag}
    P_i(O_{\text{mag,dist}}|\textbf{m},d,r) = \prod_{k_1}^{N_{k_1}}P(O_{k_1}|m_{i,k_1},\mu(d),A_{k_1}(r)),
\end{equation}
in accordance with equation (\ref{eqn:norm_gaussian}),
\begin{equation}
    \label{eqn:phot_to_norm_mag}
    P(O_{k_1}|m_{i,k_1},\mu(d),r) = G_j(m_{i,k_1}-(O_{k_1}-\mu(d)-A_{k_1}),\sigma_{k_1}),
\end{equation}
where the extinction value is A$_{k_1}$(r) is a function of reddening,r, and $\mu$(d) = 5~log(d)-5 is the distance modulus and $\sigma_{k_1}$ is the photometric error combined for the $k_1$ band.

The extinction in an individual photometric band is calculated using: 
\begin{equation}
\label{eqn:extinction_conversion}
R_{k_1} = \frac{A(k_1)}{E(B-V)},
\end{equation}
where the values of R were adopted from \cite{Casagrande2011b}:
\begin{equation}
\label{eqn:extinction}
\begin{aligned}
R_{k_1} = 
\begin{cases}
4.23,\ k_1 = \text{B}_{\text{T}} \\
3.24,\ k_1 = \text{V}_{\text{T}} \\
0.86,\ k_1 = \text{J} \\
0.5,\ k_1 = \text{H} \\
0.3,\ k_1 = \text{Ks}
\end{cases}
\end{aligned}
\end{equation}

For the Gaia EDR3 dataset, the A$_G$ value can be given, however it is not always available. In this case, (G$_{\text{Bp}}$-G$_{\text{Rp}}$) color-dependent extinction coefficients presented in \cite{2021Casagrande} are used for G, G$_{\text{Bp}}$ and G$_{\text{Rp}}$. For the magnitudes, we separate the PDFs entirely (G, G$_{\text{Bp}}$, G$_{\text{Rp}}$) and (H, J, Ks, B, V).

\subsection{Distance and extinction uncertainties with respect to photometry}
To include the uncertainty of distance modulus and extinction in computing the photometric PDF, $\mu$(d) and A$_{k_1}$ must be considered as separate quantities. In equation \ref{eqn:phot_to_norm_mag}, $\mu$(d) and A$_{k_1}$ enter as parameters in each band with no error, to account for the measured value and error, an additional factor in the likelihood is introduced,

\begin{equation}
\begin{aligned}
\label{eqn:DM_error}
    P_{i,\alpha,\beta} = P_i(O_{\text{mag,dist}}|\textbf{m},d,r) \times \text{exp}\left[-\frac{(\mu_\alpha - \mu(d))^2}{2\delta\mu(d)^2}\right]\\
    \times\ \text{exp}\left[-\frac{(A_{k_1,\beta} - A_{k_1}(r))^2}{2\delta A_{k_1}(r)^2}\right],
\end{aligned}
\end{equation}

where $P_{i,\alpha,\beta}$ is marginalised over parameters $\mu_\alpha$ and A$_{k_1,\beta}$ which vary over a large range in values centering around the observed quantities $\mu$(d) and A$_{k_1}$.

%
\subsection{Asteroseismic constraints}\label{sec:asteroseismology}
Cool stars, typically FGK, show ubiquitous oscillations which are excited by convective motions in their envelopes, the same mechanism responsible for solar oscillations. These so-called solar-like oscillations are characterised by a regular pattern showing modes of consecutive radial order to be almost equally spaced in frequency. The difference in frequency values of such consecutive modes determines the large frequency separation $\dnu$. Moreover, the distribution of power as a function of frequency shows a well defined peak at the so called frequency of maximum power, $\nu_{\rm max}$. These two quantities are related through standard asteroseismic scaling relationships (e.g. \citealt{Chaplin2013, Serenelli2017}) to fundamental stellar parameters as follows:
\begin{equation}
\label{eqn:delta_nu}
\Delta \nu \simeq \Delta \nu_\odot \sqrt{\frac{\rho}{\rho_\odot}} = \Delta \nu_\odot \sqrt{\frac{M}{R^3}}
\end{equation} 
where $\Delta\nu_\odot=135.1$ $\mu$Hz \citep{Huber2011}, $\rho$ is the mean density of the star, and $\rho_\odot$ is the solar mean density, and the last equality assumes the stellar mass $M$ and radius $R$ are expressed in solar units. 

For $\nu_{\rm max}$, the relation is:
\begin{equation}
\label{eqn:nu_max}
\nu_{\text{max}} \simeq \nu_{\odot}\frac{g}{g_\odot}\sqrt{\frac{T_{\text{eff},\odot}}{T_{\text{eff}}}} = \nu_{\odot} \frac{M}{R^2} \sqrt{\frac{T_{\text{eff},\odot}}{T_{\text{eff}}}},
\end{equation}
where the solar values are $\nu_\odot$ = 3090 $\mu$Hz \citep{Huber2011}, $\log{g_\odot} = 4.44$, and T$_{\text{eff},\odot}$ = 5777~K \citep{Heiter2015} and, again, the last expression assumes solar units for $M$ and $R$.

The $\numax$ taken from the stellar evolution model is calculated using the aforementioned relationship, given the model values of $\teff$, $M$, and $R$. However, $\dnu$ is computed from the radial orders (l=0) according to the prescription in \cite{White2011}.  We denote the stellar model asteroseismic quantities at model point $i$ and quantity k$_2$ by $\nu_{\text{seism},i,k_2}$. It should be stressed that $\numax$ and $\dnu$ are, strictly speaking, not direct observables. The analysis of light curves and the oscillation power-spectrum is a complex and non-trivial procedure that involves a number of parameters, and it is far beyond the scope of this work to incorporate it into the present scheme\footnote{It shall be kept in mind that dedicated PLATO WPs are in charge of the light curve analysis.}. We therefore refer to these quantities as observables \citep{Belkacem2011}. The asteroseismology likelihood is constructed as:
\begin{equation}
	\label{eqn:seism_gaussian_prod}
	P_i(O_{\text{seism}}|\boldsymbol{\nu}) = \prod_{k_2}^{N_{k_2}}P(O_{k_2}|\nu_{i,k_2})
\end{equation}

\begin{equation}
\label{eqn:seism_gaussian}
P(O_{k_2}|\nu_{i,k_2}) = G_j(\nu_{i,k_2}-O_{k_2},\sigma_{k_2})
\end{equation}
\subsection{Spectroscopy}\label{sec:spectroscopy}

\subsubsection{Spectroscopic models}\label{subsec:spectroscopic_models}
The main difference with the approach outlined in \citet{Schoenrich2014} is our code does not use synthetic grids directly, but relies on the Payne algorithm. This is a fast model-reconstruction technique, which is based on artificial neural networks (ANN) \citep{Ting2016, Ting2019}. The conceptual idea of the method is to use an ANN (or a set of ANNs) to represent how the flux at each point across the stellar spectrum varies with each individual stellar parameter, e.g. the $\teff$ (but the number of parameters is only limited by the dimensionality of the spectral grids). Once trained, the ANN layers, which act like a function that predicts a flux value at each point, can be used to quickly re-create an almost exact copy of the original spectrum. The typical error of the flux is $\lesssim 10^{-3}$, which corresponds to the interpolation error of $\lesssim0.1\%$ \citep{Kovalev2019}. Such "generic" spectra can be used to quickly create analogues of stellar spectra computed ab-initio. The input training grids of synthetic spectra must be computed on a random mesh and must be uniformly distributed in the full space of parameters. 

Here we use the Payne model trained on an 8-D grid ($\teff$, $\logg$, [Fe/H], micro-turbulence $\vmic$, $\vbrd$ (which accounts for macro-turbulence and projected rotation velocity v$\sin i$), [Mg/Fe], [Ti/Fe], [Mn/Fe]). The training grids \citep[see][for more details]{Kovalev2019} cover the entire parameter space of FGKM-type main-sequence stars, sub-giants and red giants, that is $4000 \le \teff \le 7000$ and $1.0 \le \logg \le 5.0$, and metallicity $-2.6 \le \text{[Fe/H]} \le +0.5$. $\vbrd$ is used because it is not possible to resolve the separate influence of macro-turbulence and rotation broadening components at the typical resolving power of observed stellar spectra \citep[see also][]{Mashonkina2017,Kovalev2019}. Both $\vmic$ and $\vbrd$ are randomly and evenly distributed in the parameter space, corresponding to typical values measured for FGK stars. $\vmic$ varies from 0.5 to 2 km$^{-1}$, whereas $\vbrd$ varies from 5 to 25 km$^{-1}$. The elements are chosen, because for them reliable NLTE atomic models were available to us at the stage of model grid development. The reference solar composition adopted in that grid was taken from \citet{Grupp2004a} and scaled-solar abundances were used for all elements, except Mg, Ti, Fe, and Mn, for which the meteoritic values from \citet{Grevesse1998} were used. The main advantage of this approach, compared to classical methods, is the significant improvement in performance. The standard method, which relies on symmetric grids with an equidistant step size is computationally very costly and forbids computing separate grids for all possible detailed abundances, while still keeping up with the rapid developments in atomic and molecular physics \citep[e.g.][]{Belyaev2020,DenHartog2021} and 3D NLTE radiative transfer methods \citep[e.g.][]{Bergemann2019, Gallagher2020, Bergemann2021}. The current grids allow us to determine [Mg/Fe], [Ti/Fe], [Mn/Fe], but it can be extended to an arbitrary number of chemical elements in future.

The physics of the input model spectra was extensively described in \citet{Kovalev2019}. In short, these are  synthetic spectra models computed using 1D NLTE radiative transfer for Fe, Mg, Ti, and Mn. The MAFAGS-OS \citep{Grupp2004a,Grupp2004b} 1D hydrostatic model atmospheres with opacity sampling (OS) were used, owing to their more extended parameter coverage (up to mid A-type) and denser grid sampling. We note, however, that extensive comparative tests were performed in different studies, showing that the MAFAGS-OS and MARCS model \citep{Gustafsson2008} provide very similar thermodynamic structures and results based on these models are virtually identical \citep{Bergemann2012, Bergemann2019}. Owing to the assumption of hydrostatic equilibrium, the transport of energy by convection has to be parametrised. In the MAFAGS-OS model, the latter is computed using the mixing-length theory \citep{1958Bohm,1968Cox,1991Canuto}. Microturbulence, the parameter that - for the lack of full 3D RHD treatment - approximates the effect of velocities on scales smaller than the photon mean free path, is a free parameter in the grid. We note, however, that the calculations of new multi-dimensional NLTE grids with average 3D models are already in progress, and the grids will make the use of ad-hoc adjustable parameter $\vmic$ obsolete.
\subsubsection{Model-data comparison}\label{subsec:spec_model_data_comp}
In the spectroscopic module, the SAPP relies on the gradient descent method, a standard first-order iterative optimisation algorithm that allows to locate the global minimum in the parameter space. As shown in \cite{Kovalev2019}, the main advantage of the gradient descent method is that it allows to greatly speed-up the spectroscopic analysis compared to other methods. The average run-time of this module is about $0.4$ seconds per star, and this procedure yields all spectroscopic quantities, including $\teff$, metallicity, detailed chemical abundances, etc.

Once the global minimum in the 8-D parameter space is found, we construct the spectroscopic PDF P(O$_{spec}|$\textbf{X}) by assuming a normal distribution with the 1 $\sigma$ uncertainty for $\teff^{spec}$, log$g_{spec}$, and [Fe/H]$_{spec}$ which are provided by the gradient descent method. Other types of spectroscopic PDFs were extensively studied in \citet{Schoenrich2014}. The other derived spectroscopic parameters (chemical abundances, $\vbrd$, $\vmic$) are kept at their best fit values. The discretisation and parameter space of the spectroscopic PDF follows the grid points defined by the tracks. We note, however, that the latter has no influence on the shape and amplitude of P(O$_{spec}|$\textbf{X}).

Hence, P(O$_{spec}$|\textbf{X}) for a given parameter space point $i$ is indexed over quantity $k_3$,

\begin{equation}
	\label{eqn:spec_gaussian_prod}
	P_i(O_{spec}|\text{\textbf{X}}) = \prod_{k_3}^{N_{k_3}}P(O_{k_3}|\text{X}_{i,k_3}),
\end{equation}

for each spectroscopic parameter O$_{k_3}$,

\begin{equation}
\label{eqn:spec_gaussian}
P(O_{k_3}|\text{X}_{i,k_3}) = G_j(\text{X}_{i,k_3} - O_{k_3},\sigma_{k_3})
\end{equation}

where O$_{spec}$ = $\teff^{spec}$, $\logg_{spec}$, [Fe/H]$_{spec}$ and \textbf{X}$_{i}$ = $\teff{}_i$, $\logg{}_i$, [Fe/H]$_i$ respectively.

For the case of covariance, equation \ref{eqn:cov_gaussian} is used instead of equation \ref{eqn:red_gaussian}.

\subsubsection{Error model}
\label{sec:errormask}

Since our model grids have only eight dimensions, we have implemented a more restrictive approach with respect to the statistical analysis of individual observations. \citet{Kovalev2019} used the full spectrum fit, however, this is sub-optimal, because different spectral lines correspond to the abundances of different chemical elements across the entire periodic table, and using low-dimensional training grids does not allow to account for the true chemical abundance patterns of the star, which are in most cases not scaled-solar \citep{Bergemann2014}. 

Therefore, in this work, instead of pre-selecting spectral lines by eye, as it has been common in previous literature studies, we resort to a robust statistical procedure. The procedure entails a comparison of the reference synthetic spectra of the benchmark stars (using the reference stellar parameters and abundances obtained as described in Sec. \ref{sec:benchmark_stars_parameters}) with the observed spectra that allows to find the wavelength regions that are poorly described by our synthetic models. These regions are not masked, but rather we evaluate the residuals between the model flux and observed flux at each wavelength point. These mono-chromatic residuals, which we refer to as the error model, serve as systematic uncertainties, so that the total combined error per wavelength point is determined by adding the error model in quadrature with the observed flux error.
\subsubsection{Correlation}
\begin{figure}
\centering
\vbox{
\includegraphics[scale=0.4]{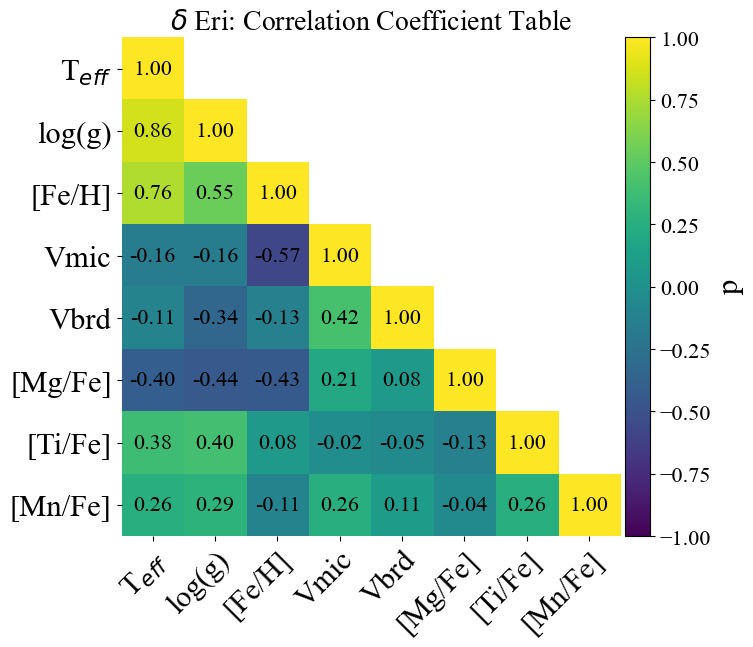}
\includegraphics[scale=0.4]{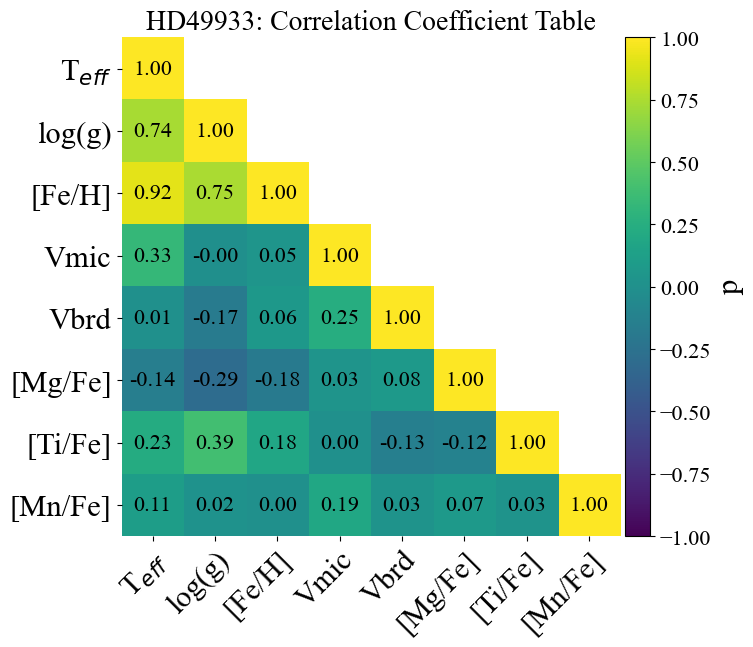}
}
\caption{Correlation coefficient tables split into two sub-figures. Each sub-figure has a colour scale that represents the correlation coefficient, which ranges from -1 (negatively correlated), 0 (no correlation), 1 (positively correlated). This number informs how each spectroscopic parameter is correlated with another. The top panel is the table for a K-type star ($\delta$ Eri) and the bottom panel is a table for a F-type star (HD49933). The vertical and horizontal axes for both depict the 8 spectroscopic parameters from SAPP's spectroscopy module.}
\label{fig:correlation_table}
\end{figure}

It shall be pointed out that some spectroscopic parameters are highly correlated with each other, owing to underlying physics of energy transfer in stellar atmospheres. Fig. \ref{fig:correlation_table} shows the covariances for all eight parameters in the spectroscopic module. The core parameters - $\teff$, $\logg$, and [Fe/H] show the most significant correlation, but also the $\alpha$-enhancement (e.g., [Mg/Fe] or [Ti/Fe]) and micro-turbulence correlate with metallicity in different regimes of parameter space. This is often the reason why alternative constraints on the core parameter space are necessary. The impact of covariance on the final results is discussed in Sect. \ref{subsec:bayes_covariance}.
\subsection{Infra-red flux method}
\label{sec:IRFM}
The Infra-Red Flux method (IRFM) is another efficient method to determine effective temperature and angular diameter of a star \citep[e.g.,][]{Blackwell1977,Casagrande2006,ghb09}. The IRFM does not depend much on the physics of stellar model atmospheres employed (see e.g., \citealt{Casagrande2006,C10}, for a discussion of H$^{-}$ opacity and the use of MARCS vs ATLAS9 models across FGK dwarfs, and \citealt{2001Asplund} and \citealt{2018Chiavassa} for an analysis of the use of 3D vs 1D models). An estimate of $\logg$ and [Fe/H] is also required, but has minimal impact on the methodology \citep[see e.g,][for a discussion]{Casagrande2006,ghb09}

However, the adopted interstellar reddening and absolute flux calibration can easily introduce systematic errors of order $\sim 100$~K \citep{Casagrande2006, Ruchti2013}. In the context of this work, the use of IRFM is limited to stars with reliable 2MASS photometry, which is about half of those in Table \ref{tab:ref_primary} due to their brightness and 2MASS saturation. For some of the remaining stars with good photometry in fewer bands, we tested the colour-$\teff$ relations derived from the same implementation of the IRFM \citep{C20}. These calibrations are available in up to 12 different colour combinations, with extra quality cuts based on photometry and realistic errors from a Monte Carlo approach\footnote{\url{https://github.com/casaluca/colte}}. The sensitivity of these relations to the adopted input parameters varies depending on the colour indices. 

In Sec. \ref{sec:IRFMres} we compare the results of $\teff^{\rm IRFM}$ with our spectroscopic $\teff$ values. This allows us to test the precision with which we can determine the $\teff$ by means of the IRFM method. This is important because we expect that this method, along with the Surface-Brightness Colour relationships (SBCR), will be the source of constraints on $\teff$ for those PLATO targets that do not have spectra.

\subsection{Surface-brightness colour relationships}
\label{sec:SBCR}

Surface-Brightness Colour relationships (SBCR) allow one to easily and independently estimate the limb-darkened angular diameter of the star. Combining the latter with the distance to the star, the linear radius can be computed. Furthermore, the combination with the bolometric flux yields the effective temperature.

Many SBCRs exist in the literature, but all of them are empirical and are calibrated on interferometric measurements (see \citealt{Nardetto18} and \citealt{Salsi20} for a short review) or on Galactic eclipsing binaries \citep{graczyk17, graczyk19}.
For example, \citet{Pietrzynski19} derived the distance to the Large Magellanic Cloud with a 1 $\%$ precision using eclipsing binaries as distance indicators. In order to achieve  such  precision, they used a dedicated SBCR based on the observations of 48 red clump stars with the PIONIER/VLTI instrument \citep{Gallenne18}. Assuming the expected 1$\%$ precision and accuracy of angular diameters expected from CHARA/SPICA (see Sect. \ref{sec:future}), we estimate that the method will provide the precision of 50 to 100K in $\teff$ and 1 to 2$\%$ in radii. 

The main advantage of the method is its simplicity: only two magnitudes (usually V and K) are required and it is not very sensitive to reddening, because the surface brightness and the colors have a similar sensitivity to extinction. For example, a difference of $0.2$ mag in the visual extinction translates into a difference of 2$\%$ on the radius and $60$ K on the temperature. For comparison, a 2$\%$ error on F$_{\text{bol}}$ translates into an error of $30$ K on the temperature. The main limitation of the method in the context of PLATO is likely its sensitivity to the uncertainty of the K magnitude. A $0.03$ mag error in the K band magnitude causes an 80 K uncertainty in $\teff$ and a 2 $\%$ uncertainty on the radius. The results also depend on the spectral type and luminosity class \citep{Salsi20}. Additionally, the method cannot be used for stars that show a significant activity level, have a companion(s), are flattened due to rotation \citep{challouf14, challouf15}, show signatures of wind and/or circumstellar shells.

In this work, we employ the SBCR relationships from \citet{Salsi2021}, which were derived for F5-K7 IV/V stars using the methodology described in \citet{Salsi20}. The relationships are based on a careful selection of the interferometric data, suitable sample selection (no activity), and a homogeneous set of 2MASS photometry. 

In Sect. \ref{sec:IRFMres}, we compare our SBCR estimates obtained with these relationships with the reference values. The bolometric fluxes, which are needed in the SBCR method to determine $\teff$ from the linear radius, are taken from \citet{Heiter2015}. However, it is expected that SAPP will provide bolometric fluxes as an output, from the photometric module. 

\subsection{Stellar evolution models}\label{sec:stellar_evo_models}
The SAPP code also includes a detailed grid of \texttt{GARSTEC} stellar evolution models \citep{Weiss2008}, which covers the mass range from $0.6$ to 5.0~$\msun$ with a step of $0.02$~$\msun$ and metallicity from $-2.50$ to $0.60$ with a step of $0.05$ dex. 

In short, the models were computed as follows. Stellar atmospheres are based on the T-$\tau$ VAL-C model \citep{Vernazza1981}, implemented with the analytic fit given in \citet{Sonoi2019}. This relation leads to stellar models that reproduce well the $\teff$ scale of RGB stars in the APOKASC sample \citep{Serenelli2017,2018Pinsonneault}. Convective overshooting is treated in \texttt{GARSTEC}  as a diffusion process parametrised with a coefficient $f$, which is fixed to 0.02 at all convective boundaries. This is roughly equivalent to an overshooting region extending over $0.2H_P$, with $H_P$ being the pressure scale height at the convective boundary. For small convective cores, $f$ is decreased linearly from 0.02 down to 0 for stellar masses from 1.4 down to 1.1~$\msun$. This prescription has been found to describe well results from binary stars \citep{Higl2018} and mid-age open clusters \citep{2020Semenova}. No convective core overshooting in the main sequence is used below 1.1~$\msun$. Mass loss is modelled with a Reimers law and $\eta= 0.2$. Microscopic diffusion, without taking into account radiative levitation, is included according to \citet{Thoul1994}. Its efficiency is suppressed linearly with stellar mass in the range $1.25-1.35~\msun$ and it is not included for higher masses. The reason is that microscopic diffusion and radiative levitation in the presence of very thin convective envelopes lead to theoretically large changes in surface abundances not supported by observations, likely due to the presence of a yet physically unidentified macroscopic mixing process. For lower masses, however, extra-mixing below the convective envelope is included following the prescription described in \citet{Vandenberg2012}, with a metallicity dependent efficiency adjusted to reproduce the solar lithium depletion and the depletion of lithium typical of the Spite plateau \citep[e.g.][]{Spite1982, Sbordone2010} stars. 

The relation between the abundances of metals and helium is assumed to follow a linear relation calibrated using the Standard Big Bang Nucleosynthesis (SBBN) helium value $Y_P=0.2485$ \citep{Steigman2007} and a solar model calibration, which lead to a slope on the enrichement law $\Delta= 1.14$ \citep{Serenelli2017}. More recent determinations of $Y_P$ lead to slightly lower values, ($Y_P=0.2470\pm0.0002$ and $Y_P=0.2446\pm0.0041$, consistent with each other, for SBNN and non-SBBN respectively \citealp{fields:2020}). The impact on the slope of the enrichment law is about 7\%, equivalent to that produced by a [Fe/H] error of 0.025~dex.

In addition, the models include synthetic photometry computed using ATLAS12/SYNTHE bolometric corrections\footnote{\url{http://waps.cfa.harvard.edu/MIST/model_grids.html\#bolometric}} for different passbands: UBVRI, 2MASS, Kepler, Gaia, Tess, Hipparcos, Tycho and PanStarrs. Bolometric corrections have so far been implemented corresponding to A$_{\rm v}$=0, so extinction has been applied a posteriori. This leads to errors of approximately a few hundredths of magnitude for Av $< 2$ mag.
\subsection{Numerical approach}\label{sec:interpolation}
The exact procedure is as follows. 

In this first step, $\teff$, $\logg$, and [Fe/H] are fixed to the best fit spectroscopic value determined using the gradient descent method. This is because this method is fast and it does not require probing the entire parameter space. In the second step, the SAPP collects stellar evolution tracks which have $\teff$, $\logg$, [Fe/H] centred on the first-guess parameters within a certain range defined by the user. For simplicity, we limit the sub-domain range to $\pm$ 300 K in $\teff$, $\pm$ 0.5 dex in $\logg$, $\pm$ 0.6 dex in $\feh$. Alongside these parameters collected, mass, radius, age, and luminosity are also tabulated. 

Second, for each point in the sub-domain of the stellar evolution models, $i$,the spectroscopic probability is calculated using equations \ref{eqn:spec_gaussian_prod},\ref{eqn:spec_gaussian}. 

Finally, this spectroscopic PDF defined on the sub-domain of stellar evolution models is folded with the PDFs computed from photometric, astrometric, and asteroseismic data. This procedure returns the full posterior PDF. 

The final estimates of all output parameters ($\teff$, $\logg$, [Fe/H], M, age, R, L) are calculated by fitting a Gaussian to the posterior PDF with mean $\mu$ and standard deviation $\sigma$. In other studies, \citet{Schoenrich2014}, the expectation values and their moments were used However, our extensive inspection of the posterior PDFs showed that they are rather symmetric and can all be well approximated by a Gaussian.
\section{Results}
\label{sec:results}

The results of our analysis of the benchmark stars will be discussed in detail below. We will begin with quantifying the accuracy and precision of our estimates of $\teff$, $\logg$, and [Fe/H] obtained using different spectroscopic approaches and different types of observational data (Sect. \ref{sec:constrained_spec}). We will then proceed to the analysis of using IRFM and SBCR to calculate $\teff$ and radius (Sect. \ref{sec:IRFMres}).  The influence of seismic priors on $\logg$ and so on the analysis of other stellar parameters will be the subject of Sect. \ref{sec:seismic_priors}. The results for masses, radii, and ages of stars obtained using the full Bayesian solution will be presented and discussed in Sect. \ref{sec:bayes_results}. The impact of including covariance from spectroscopy is detailed in Sect. \ref{subsec:bayes_covariance}. Finally, the analysis of medium resolution spectra is shown in Sect. \ref{subsec:med_spectra}.

\begin{figure*}[ht!]
\centering
\vbox{
\includegraphics[scale=0.35]{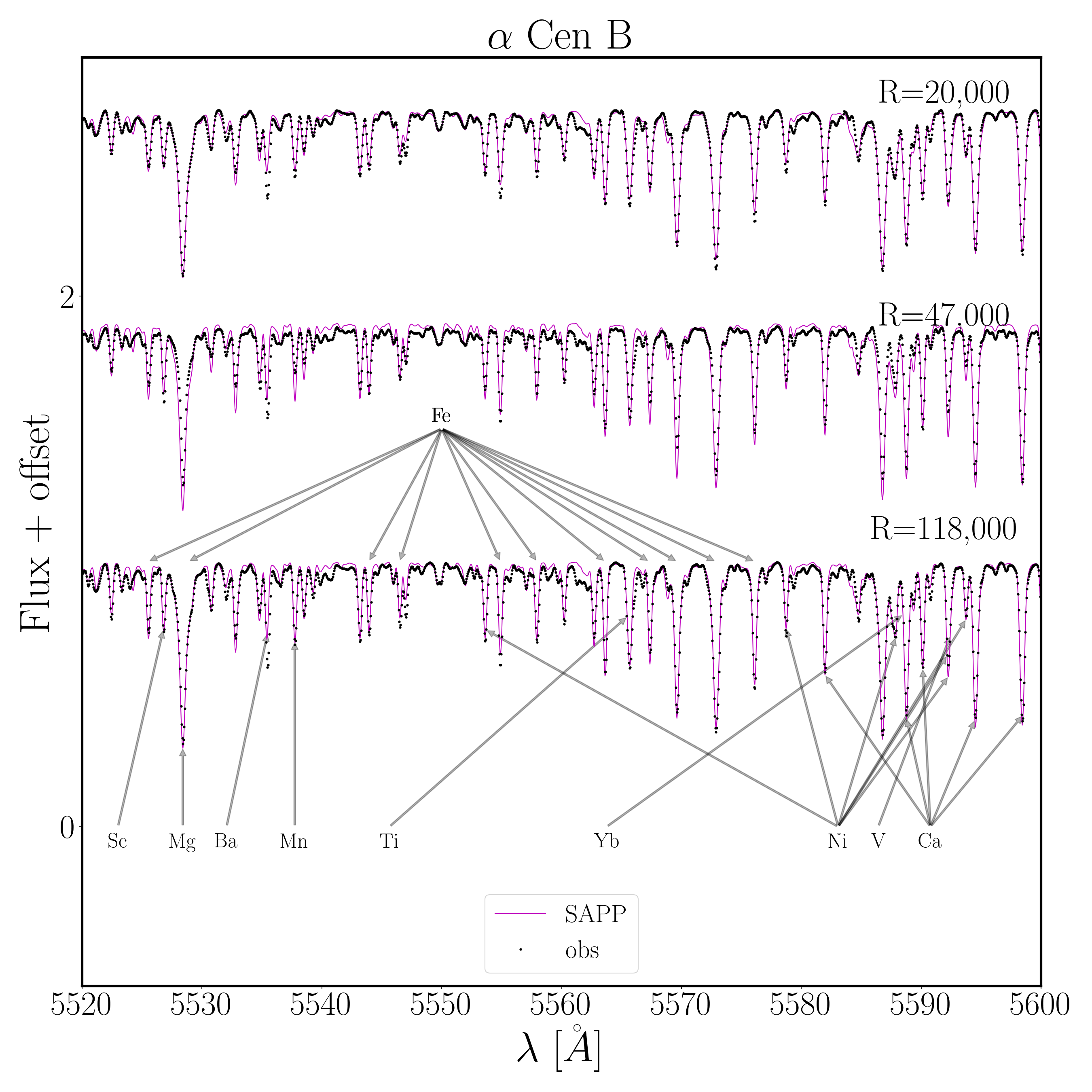}
}
\caption{Normalised Flux versus wavelength in angstroms. The observed spectra of $\alpha$ Cen B (black dots) is overlaid with SAPP's best fit model from the spectroscopy module (magenta solid line). This contains HARPS, UVES and HR10 spectra with original resolving powers R$\sim 118\,000$, $47\,000$ and $20\,000$ respectively. The UVES and HARPS spectra have been convolved down to that of HR10.}
\label{fig:spec1}
\end{figure*}

Figure \ref{fig:spec1} depicts three spectra (black dots) of $\alpha$ Cen B (one of the reference targets) from different telescopes compared to best fit spectra models from SAPP (magenta lines). A medium resolution spectra (HR10) with R = $20,\ 000$ and two high-resolution spectra (UVES, HARPS) with original R = $47,\ 000$ and $118,\ 000$ respectively. The high resolution spectra have been degraded to that of HR10 in order to demonstrate the performance of fitting different quality spectra of the same star. The original resolutions are annotated above each spectra as well as a sample of diagnostic lines which populate the HR10 wavelength range. This is just for an illustration, but the agreement for all other stars in the sample is equally good. The agreement between the observed data and the models is very good, which suggests that the SAPP in combination with available theoretical models can be used with confidence to analyse spectra of FGK-type dwarfs and sub-giants.

\subsection{Constrained versus unconstrained spectroscopic calculations}\label{sec:constrained_spec}
We begin with the analysis of the influence of the error model (see Sect. \ref{sec:errormask}) on the spectroscopic calculations. Hereafter, we refer to the results obtained using the error model as "constrained" analysis, and those obtained without the error model as an "un-constrained" analysis.
\begin{figure}[ht!]
\centering
\includegraphics[scale=0.15]{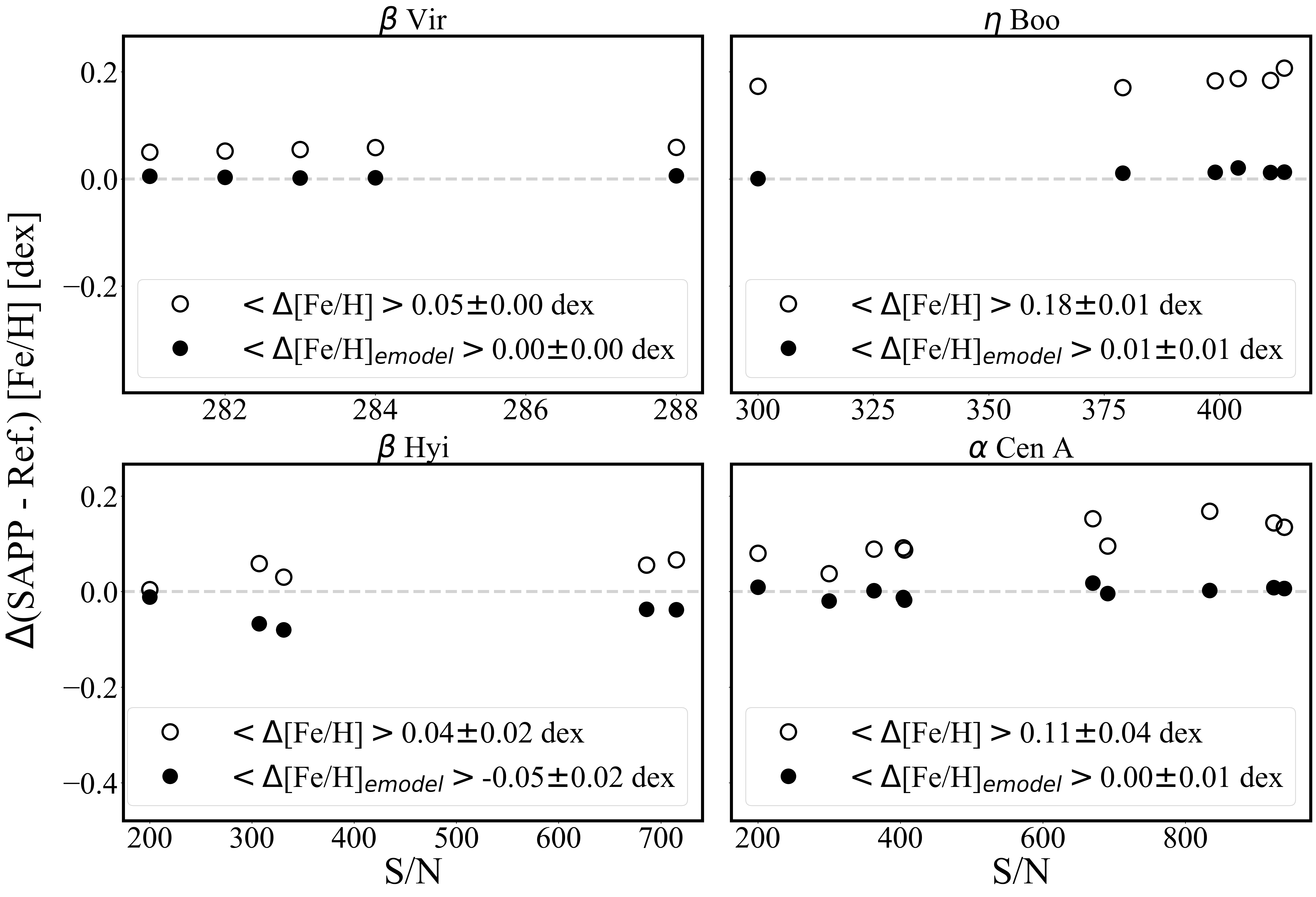}
\caption{Residuals between SAPP parameters and reference versus the signal-to-noise ratio of four stars: $\beta$~Vir, $\eta$~Boo, $\beta$~Hyi, and $\alpha$~Cen~A. Open circles represent using purely spectroscopic module (without error model), these have an average and 1$\sigma$ in the legend named $\left<\Delta\text{[Fe/H]}\right>$. Filled circles represent using spectroscopic module with the error model, these have an average and 1$\sigma$ in the legend named $\left<\Delta\text{[Fe/H]}_{emodel}\right >$.}
\label{fig:maskscatter}
\end{figure}

We find that the accuracy of stellar parameters, $\teff$, $\logg$, and [Fe/H], is significantly improved, if the error model is employed with respect to the reference stars. The scatter of the SAPP $\teff$ estimates decreases from $101$ K (unconstrained analysis) to $42$ K (constrained analysis) and the bias decreases from $109$ K (unconstrained analysis) to $-1$ K (constrained analysis). Even more impressive is the improvement in $\logg$ estimates. The systematic $\log g$ bias decreases from $-0.16$ dex (unconstrained analysis) to $-0.04$ dex (constrained analysis) and the scatter reduces enormously, from $0.13$ dex to $0.02$ dex, which makes our spectroscopic results competitive with other techniques, such as asteroseismology.  As Fig. \ref{fig:maskscatter} shows, in the unconstrained analysis the residuals are large and positive. The constrained analysis, in contrast, is much more successful, because it removes all systematic effects and returns metallicities with the precision of $0.02$ dex,this is likely due to the error model reducing the impact of features within the spectrum which are consistently causing the [Fe/H] to be initially overestimated. Not surprisingly, the constrained analysis also improves the results for the detailed chemical composition of stars. The abundances of Ti, Mg and Mn are in excellent agreement with the reference values, with mean differences of $0.02 \pm 0.04$ dex, $0.04 \pm 0.04$ dex, and $-0.04 \pm 0.03$ dex, respectively.

It should be stressed, however, that the literature values of[Fe/H] and any other element abundances cannot be treated as 'reference' quantities, because there is no model-independent (fundamental) way to determine the detailed chemical composition star. All methods depend on models of stellar atmospheres, atomic, and molecular physics, and, therefore, on the level of physical complexity of the systems. Therefore, most differences between our values and the reference metallicities stem from the different radiative transfer methods and input physics employed in the spectroscopic calculations. 

\subsection{IRFM and SBCR modules}\label{sec:IRFMres}
\begin{figure}
\centering
\includegraphics[scale=0.26]{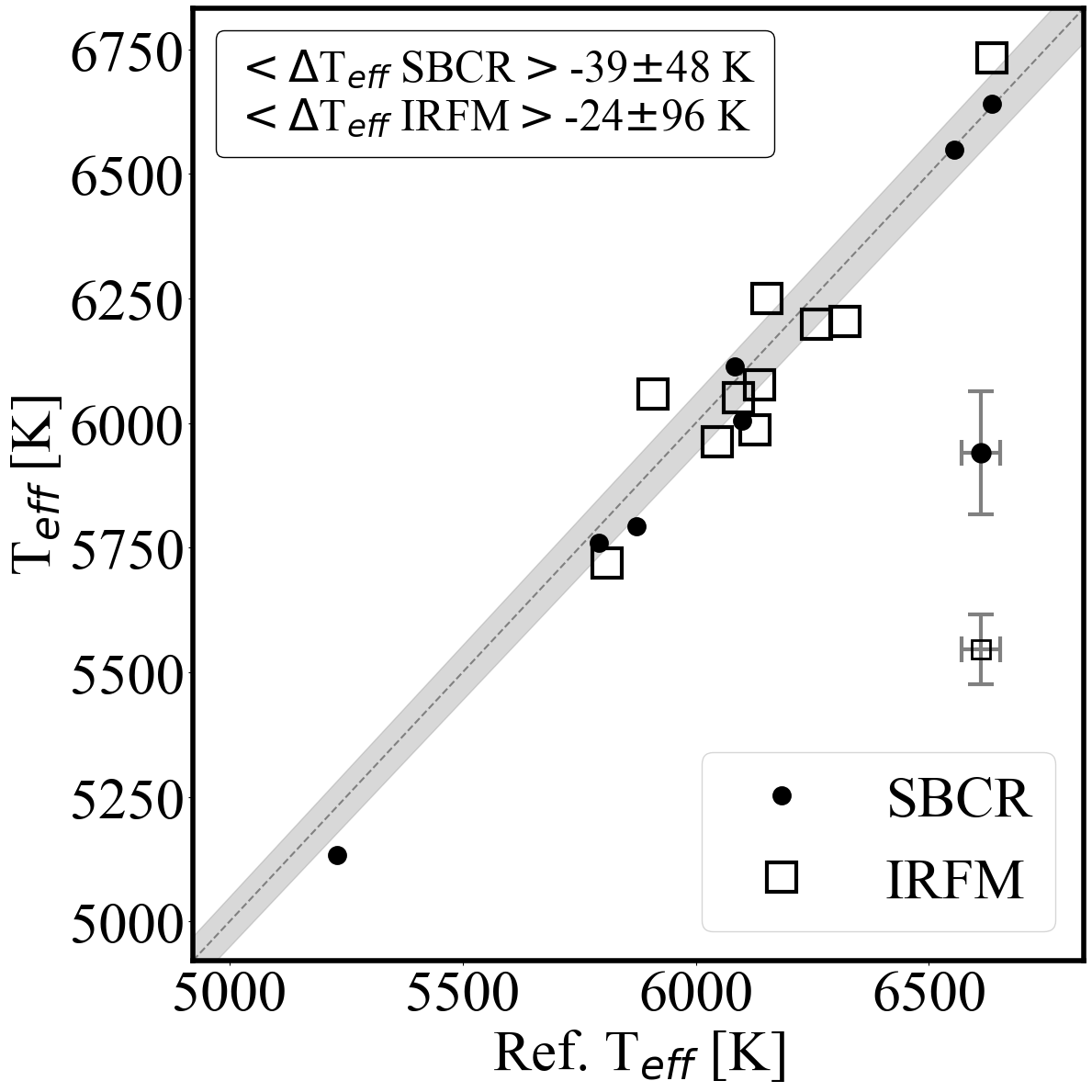}
\includegraphics[scale=0.25]{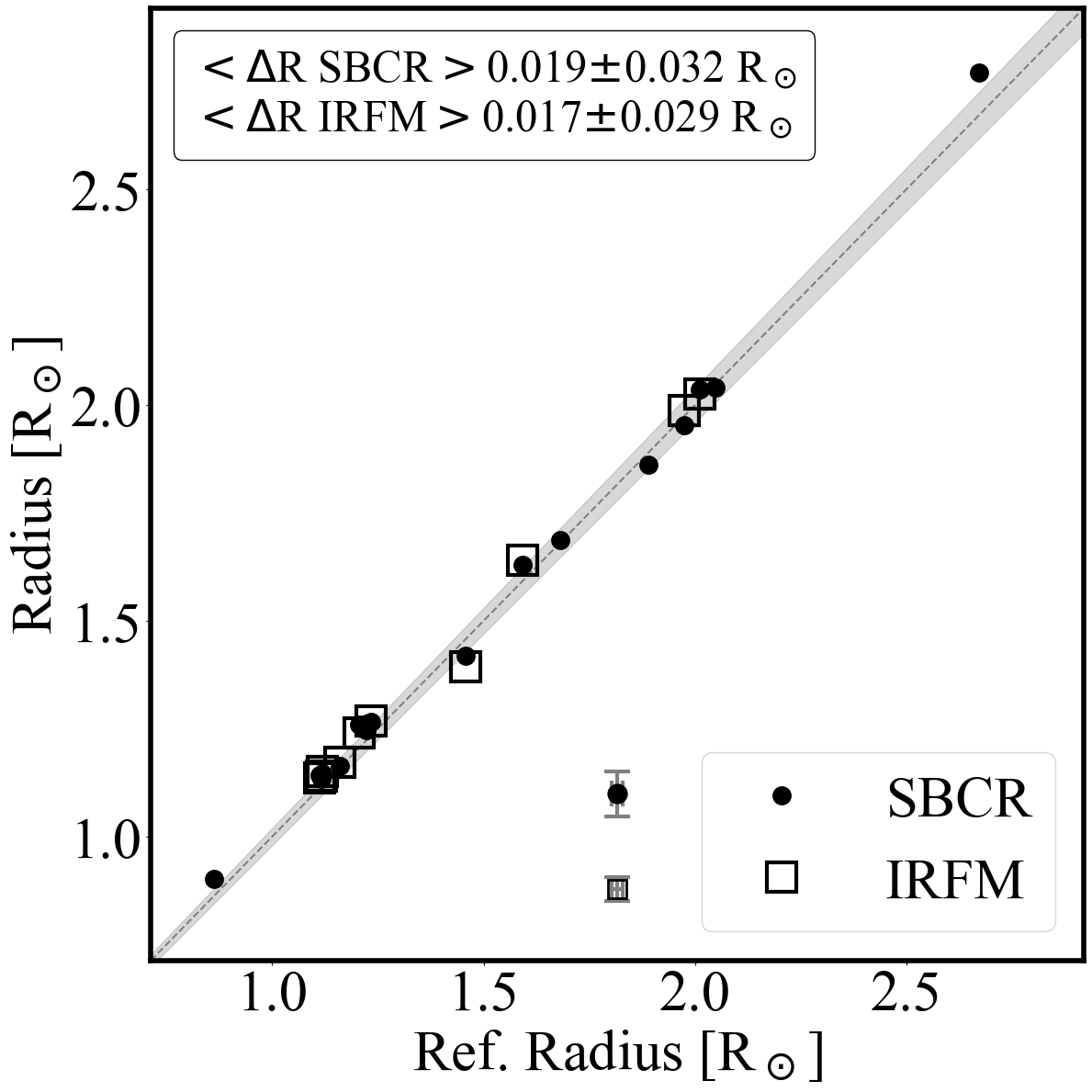}
\caption{Comparison of effective temperatures (top panel) and radii (bottom panel) derived from the SBCR (solid circles) and IRFM (open squares) method to Reference values. The dark grey line corresponds to the 1:1 line, the light grey shaded region represents our adopted threshold of 2$\%$. The average uncertainty for IRFM, SBCR (vertical) and Reference (horizontal) is annotated in the middle right for $\teff$ and bottom right for radii. The annotation in the upper left represents the average difference between SBCR, IRFM and Reference with 1 $\sigma$ scatter about said average respectively.
}
\label{fig:SBCRIRFM}
\end{figure}
In this section, we explore whether constraints on $\teff$ from the IRFM and SBCR methods could help improve the estimation of the effective temperatures. 

Figure \ref{fig:SBCRIRFM} (top panel) shows our estimates of $\teff$ obtained using the IRFM and SBCR method in comparison with the reference $\teff$ measurements. The results based on both methods are sensitive to the quality of photometry and to extinction, therefore we limit this comparative analysis to stars with the 2MASS flags "A" or "B", which corresponds to the photometric errors below 0.05 mag in the J, H, and K$_s$ bands. Generally, we find that both methods provide an excellent constraint on the $\teff$ of a star, with the average bias of only $\sim$ $-32$ K and the scatter of $\sim$ 73 K around the reference values. The uncertainties of the IRFM estimates are of the order $\sim 70$ K. As to SBCR, the uncertainty on the effective temperatures is $\sim 123$ K, but it ranges from $98$ to $138$ K depending on the quality of photometry.

Figure \ref{fig:SBCRIRFM} (bottom panel) compares the resulting SBCR and IRFM radii of stars with the reference values. The uncertainties indicated in the figure are the quadratic sum of the individual uncertainties. The SBCR uncertainties, of the order $\sim 0.052$ $\rsun$ are dominated by the RMS error of the method, with the precision of $\sim$ 3.1 to $4.5\%$. The IRFM radii are slightly more precise, with $\sigma \sim 0.027 \rsun$. 

\begin{figure*}
\vbox{
\hbox{
\begin{tabular}{@{}c@{}}
\large (a) \\
\includegraphics[scale=0.19]{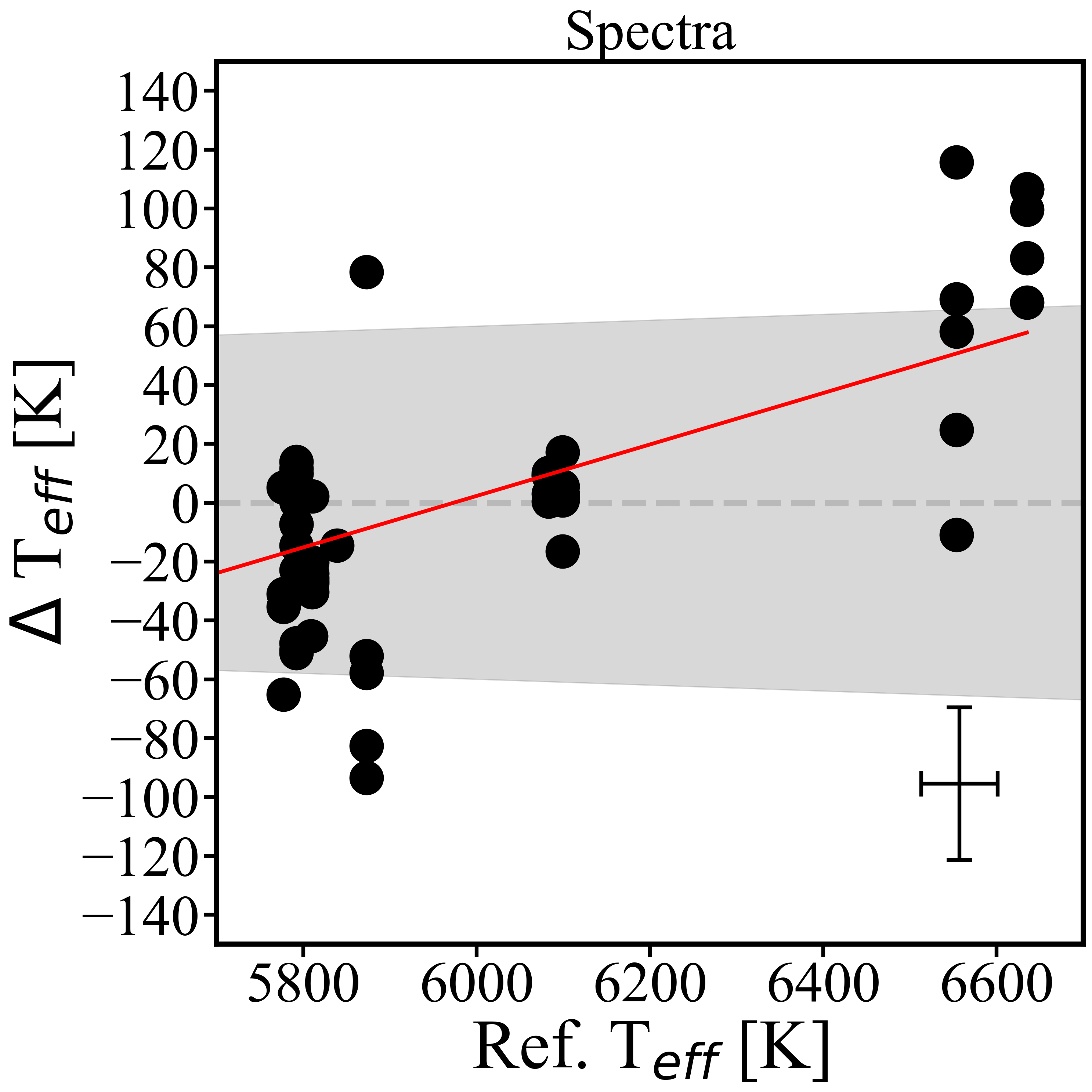}
\end{tabular}
\begin{tabular}{@{}c@{}}
\large (b) \\
\includegraphics[scale=0.19]{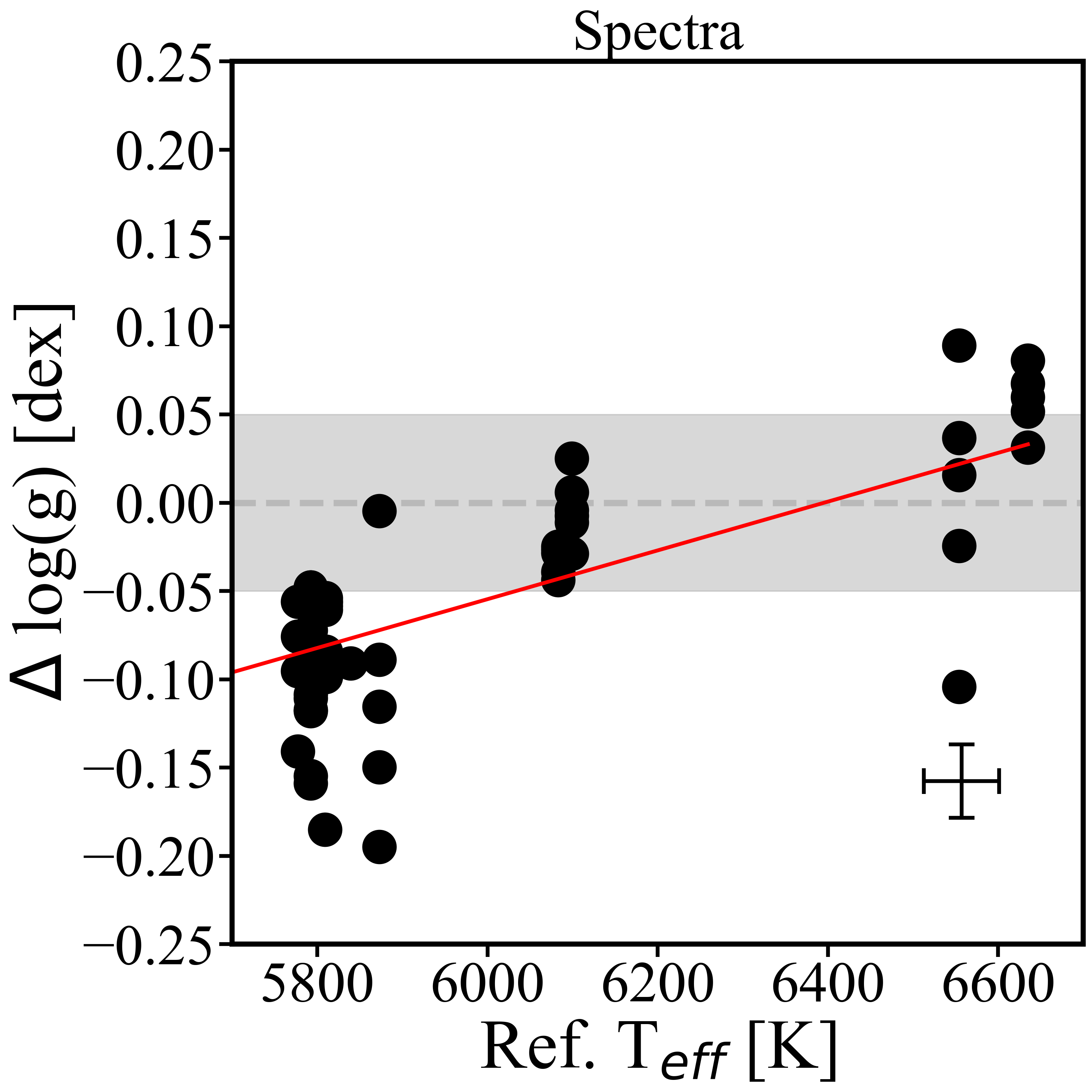}
\end{tabular}
\begin{tabular}{@{}c@{}}
\large (c) \\
\includegraphics[scale=0.19]{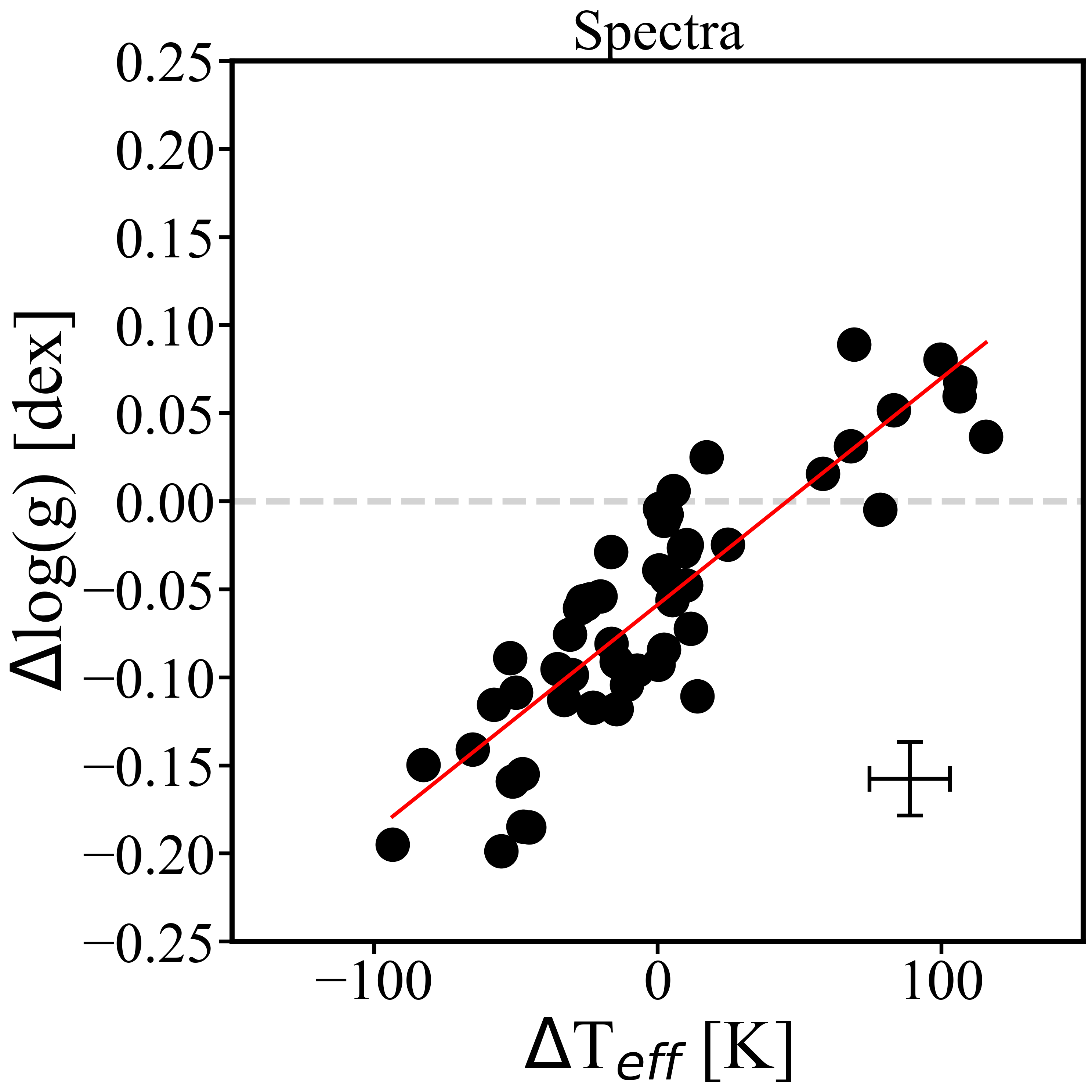}
\end{tabular}
}
\hbox{
\begin{tabular}{@{}c@{}}
\large (d) \\
\includegraphics[scale=0.19]{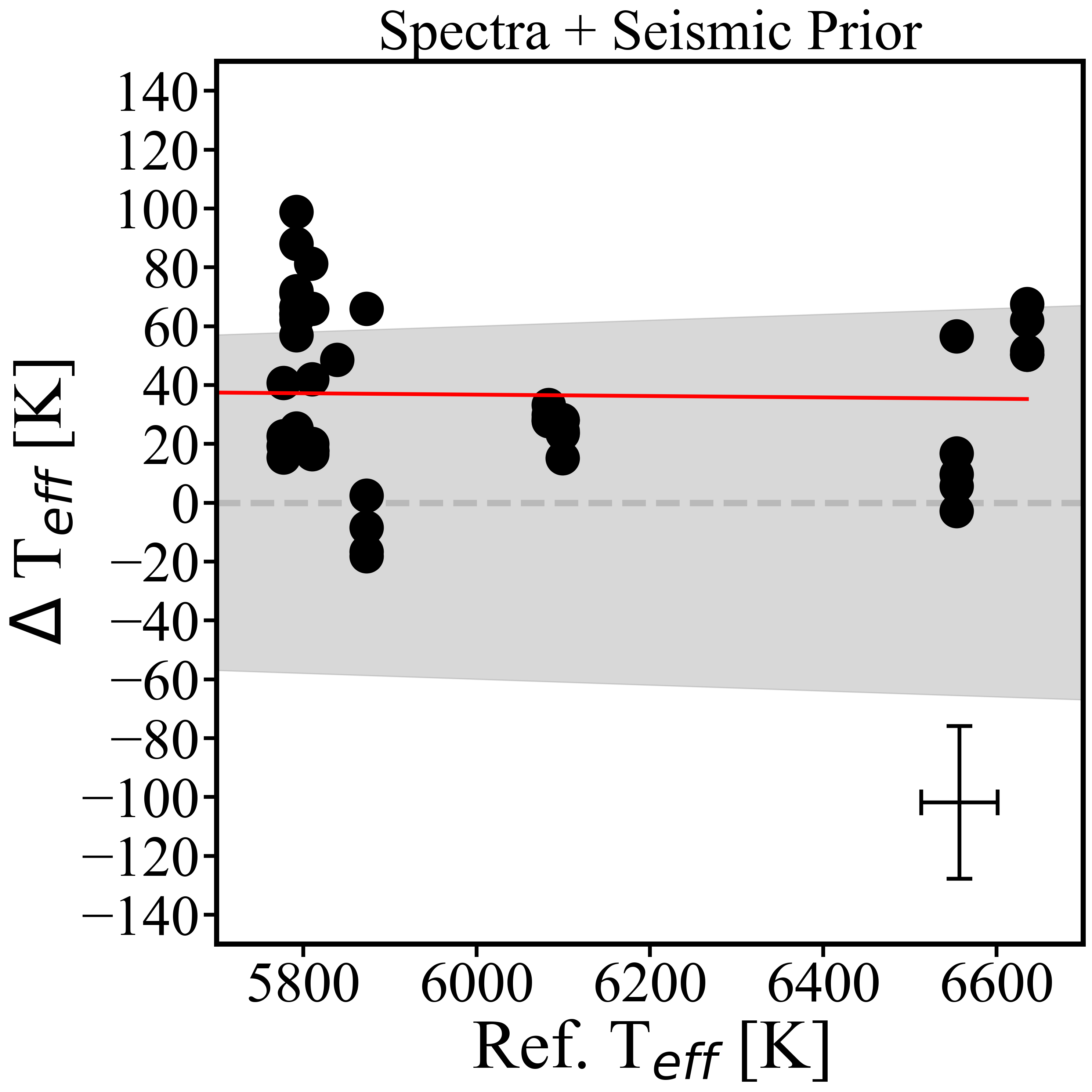} \\
\end{tabular}
\begin{tabular}{@{}c@{}}
\large (e) \\
\includegraphics[scale=0.19]{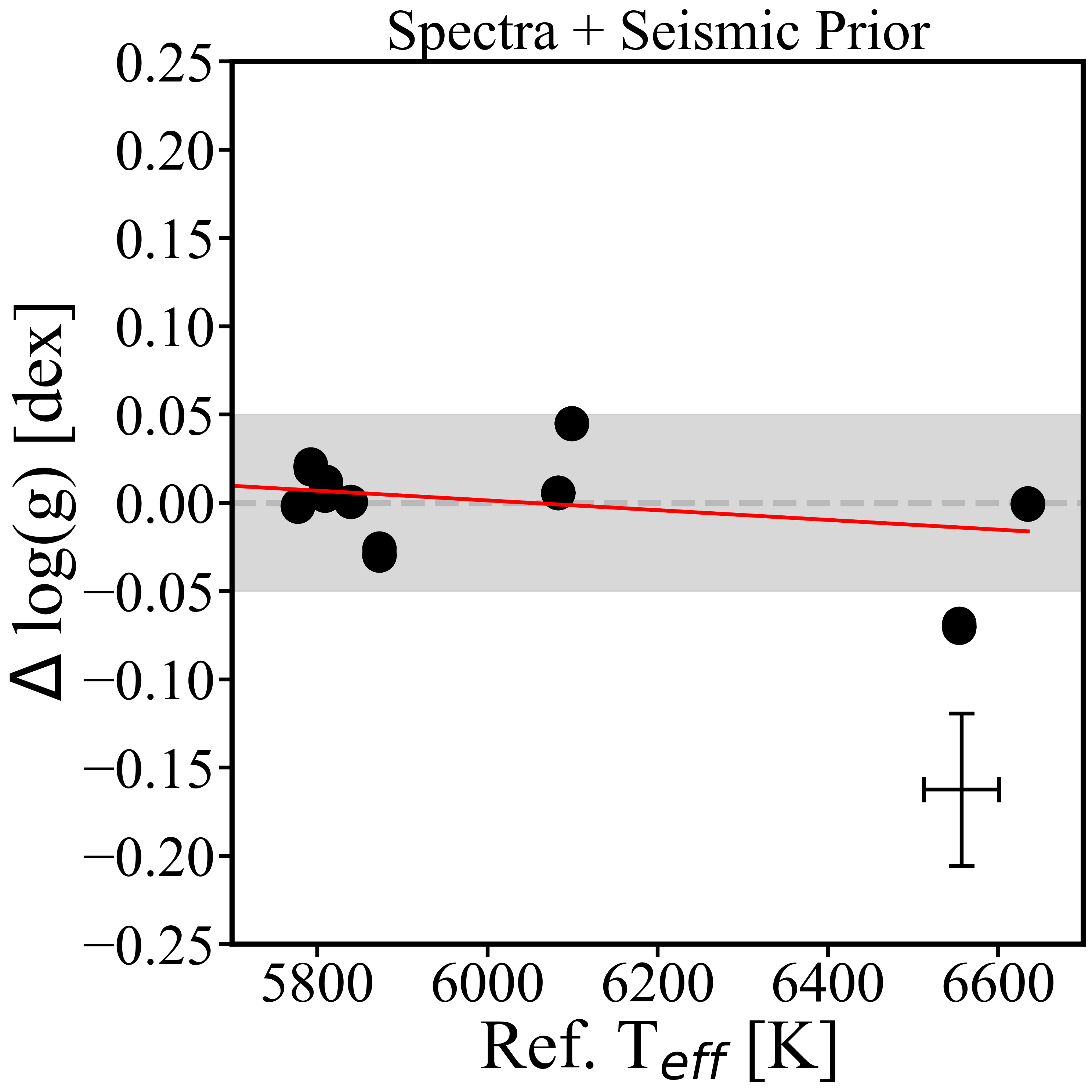} \\
\end{tabular}
\begin{tabular}{@{}c@{}}
\large (f) \\
\includegraphics[scale=0.19]{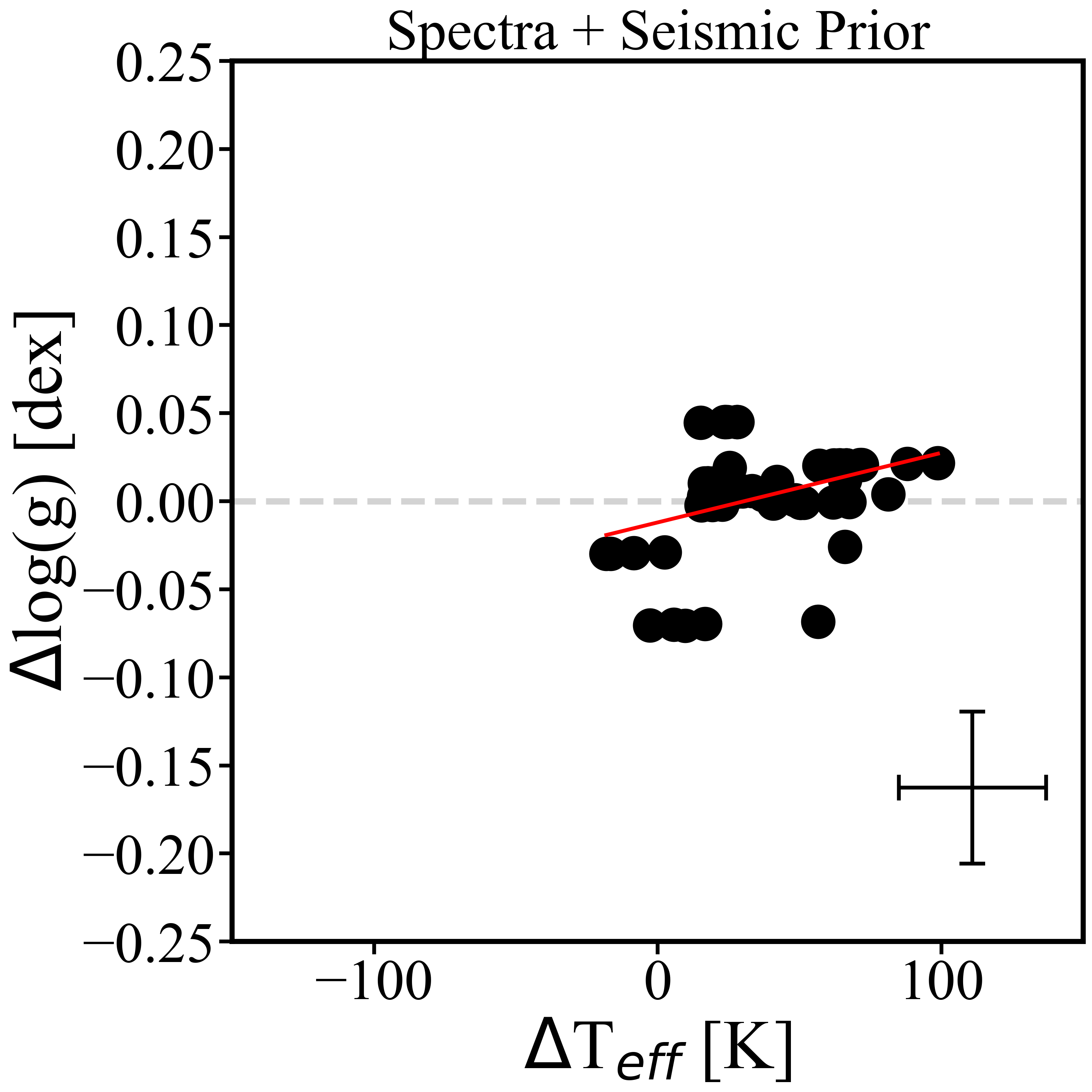} \\
\end{tabular}
}
}
\caption{SAPP spectroscopic parameters $\teff$ and $\logg$ of the benchmark stars which have interferometric $\teff$ reference values above 5500 K (this includes two KIC stars 16 Cyg A and B). The vertical axis represents the residuals between $\teff$, $\logg$ and their corresponding reference values. The horizontal axis represents the reference parameters. The red line is a linear regression of the data points. The top panel contains results produced by SAPP's constrained spectroscopy module, and the bottom panel contains results produced by SAPP's constrained spectroscopy module with $\nu_{max}$ prior applied. The grey shaded region is our tolerance of 1$\%$ for $\teff$ and 0.05 dex for $\logg$. The average uncertainty in each panel is located in the bottom right. For each star, several spectra were analysed (see Sect. \ref{sec:observations})}.
\label{fig:fstar_seismic_improve}
\end{figure*}
We conclude that both methods, SBCR and IRFM, require accurate and precise 2MASS photometry (with errors less than $0.05$ mag) in order to achieve the PLATO space mission requirements. Also extinction may influence the results. Currently, both methods may be used as auxiliary methods to constrain $\teff$ and radii, foremost as priors on spectroscopy. However, their use as standalone modules will require an improvement in the parameter coverage and reference parameters of the calibration samples. In particular, we anticipate that with the data from the upcoming CHARA/SPICA facility, the SBCR method will be improved significantly to reach the desired precision of $1-2\%$ on the radii.
\subsection{Influence of seismic priors on stellar parameters}
\label{sec:seismic_priors}

In the subsequent analysis, we investigate whether the use of asteroseismic data improves the constraints on stellar parameters, compared to the approach when only stellar spectra are used. So far, most studies employing asteroseismic constraints on $\logg$ have reported an improvement in the accuracy of results  \citep{Bruntt2012,Buchhave2015,Nissen2017}.

We use two asteroseismic quantities: $\numax$ -- the frequency of the maximum oscillation power -- and $\dnu$ --  the large frequency separation \citep[e.g.][]{Serenelli2017}. These seismic data are used as a prior in two methods. The first approach is to estimate a $\logg$ value from the asteroseismic PDF, which utilizes both quantities and then fix it in the spectroscopic analysis. We note that instead of fixing the surface gravity, a prior on $\logg$ can be imposed, however, for the benchmark stars this alternative approach yields the same result. This is due to the very small, of the order $0.01$ to $0.02$ dex, uncertainties on the seismic $\logg$ values. We also explore a different approach, in which the final solution is obtained using an iterative algorithm, with $\logg$ constrained via the empirical $\numax$ $= f(\teff, \logg)$ relationship \citep[e.g.][]{Belkacem2012}. In short, using the observed stellar and solar $\nu_{max}$ values, we input an initial guess of $\teff$ from spectroscopy, the resulting $\logg$ value from the equation \ref{eqn:nu_max} is then fixed in the spectroscopic analysis. The re-calculated set of parameters give a new estimate of $\teff$, which is used in the same equation, and the loop continues until the $\teff$ estimate does not change by more than $10$ K. The choice of this convergence criterion is not critical at this stage and the convergence is usually very fast (Fig. \ref{fig:teff_iteration} in Appendix). Such an approach has been adopted, for instance, by \citet{Lund2019}. Figure \ref{fig:fstar_seismic_improve} shows our results obtained without and with a $\logg$ prior, using the iterative solution with $\logg$.

\begin{figure*}
\centering
\hbox{
\begin{tabular}{@{}c@{}}
\large (a) \\
\includegraphics[scale=0.19]{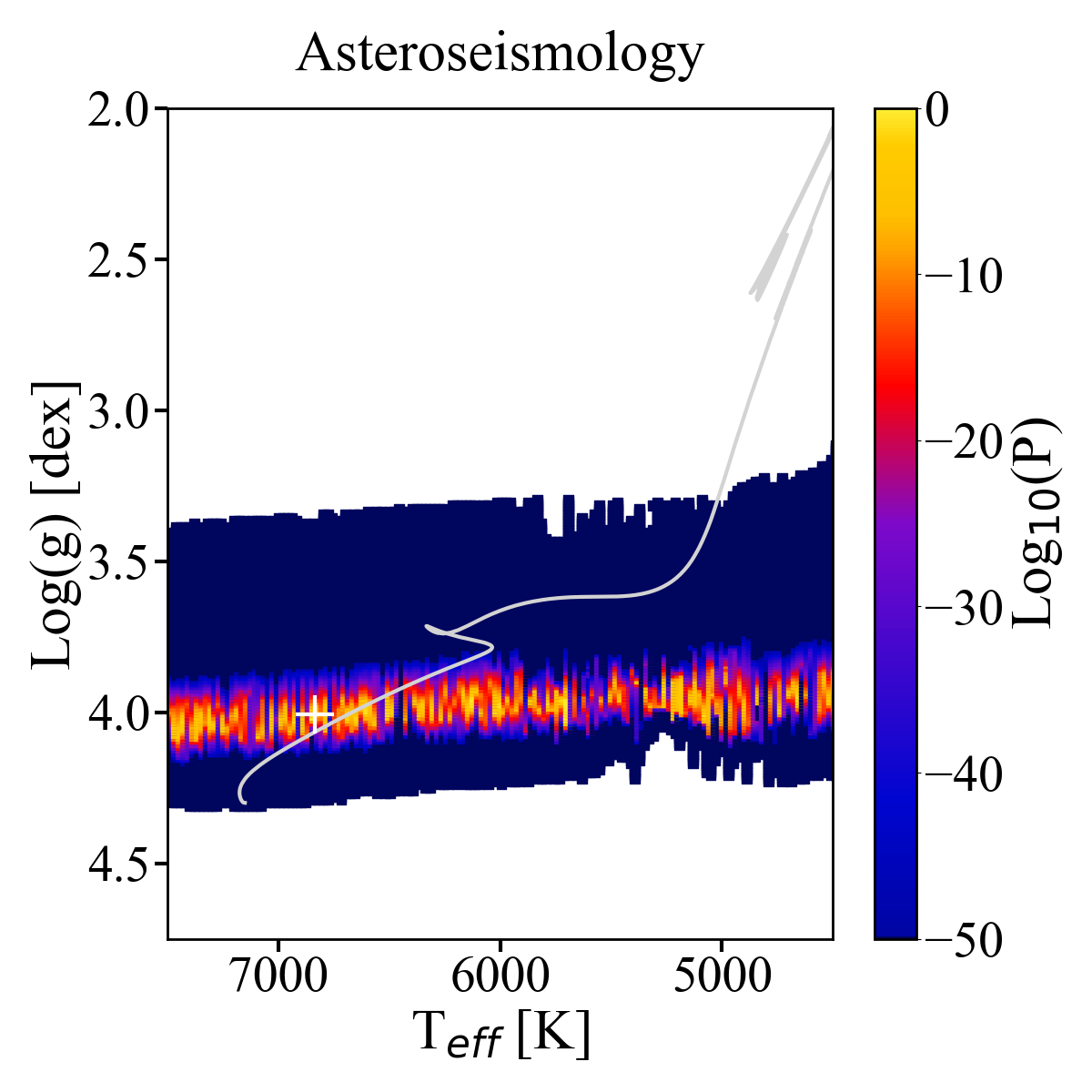}
\end{tabular}
\begin{tabular}{@{}c@{}}
\large (b) \\
\includegraphics[scale=0.19]{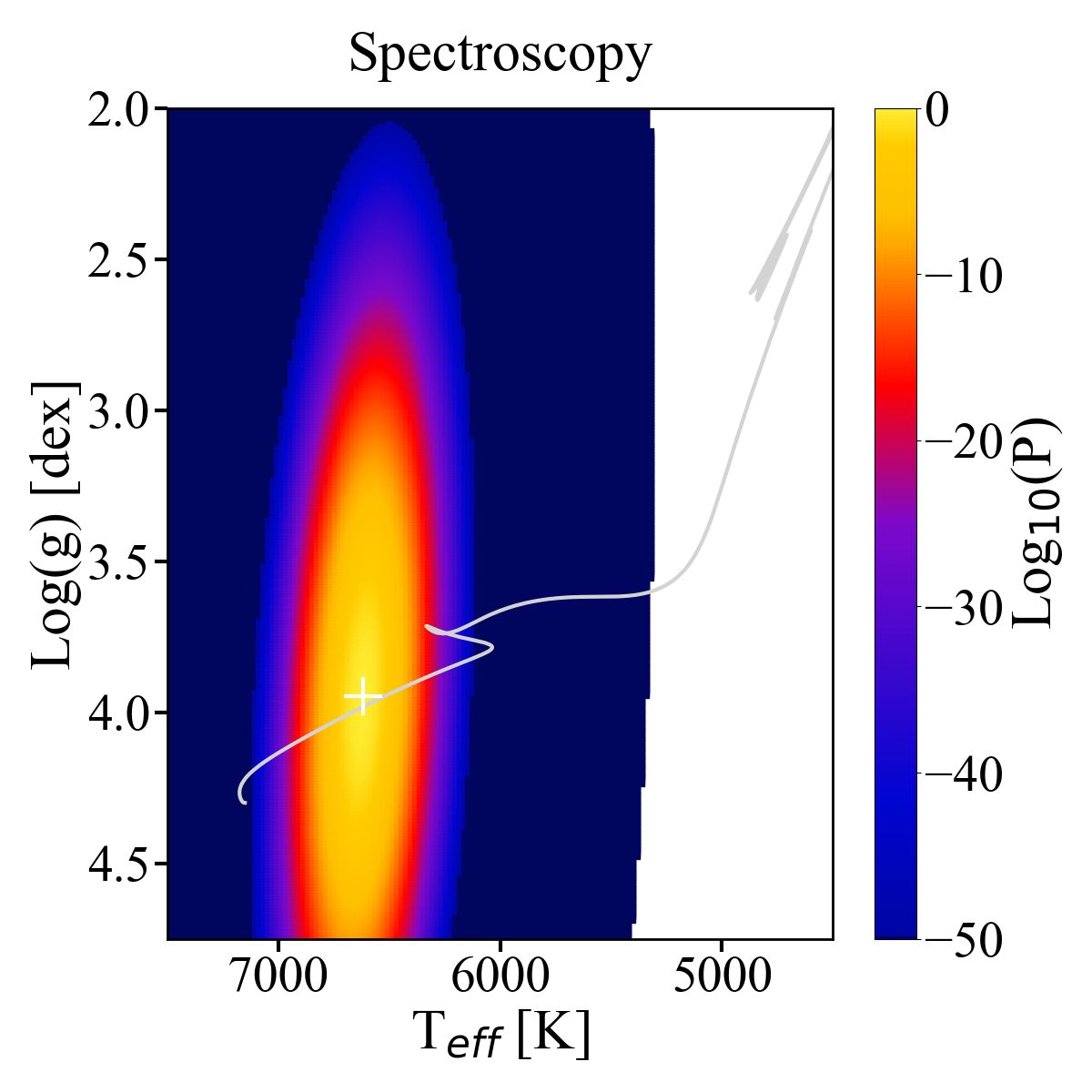}
\end{tabular}
\begin{tabular}{@{}c@{}}
\large (c) \\
\includegraphics[scale=0.19]{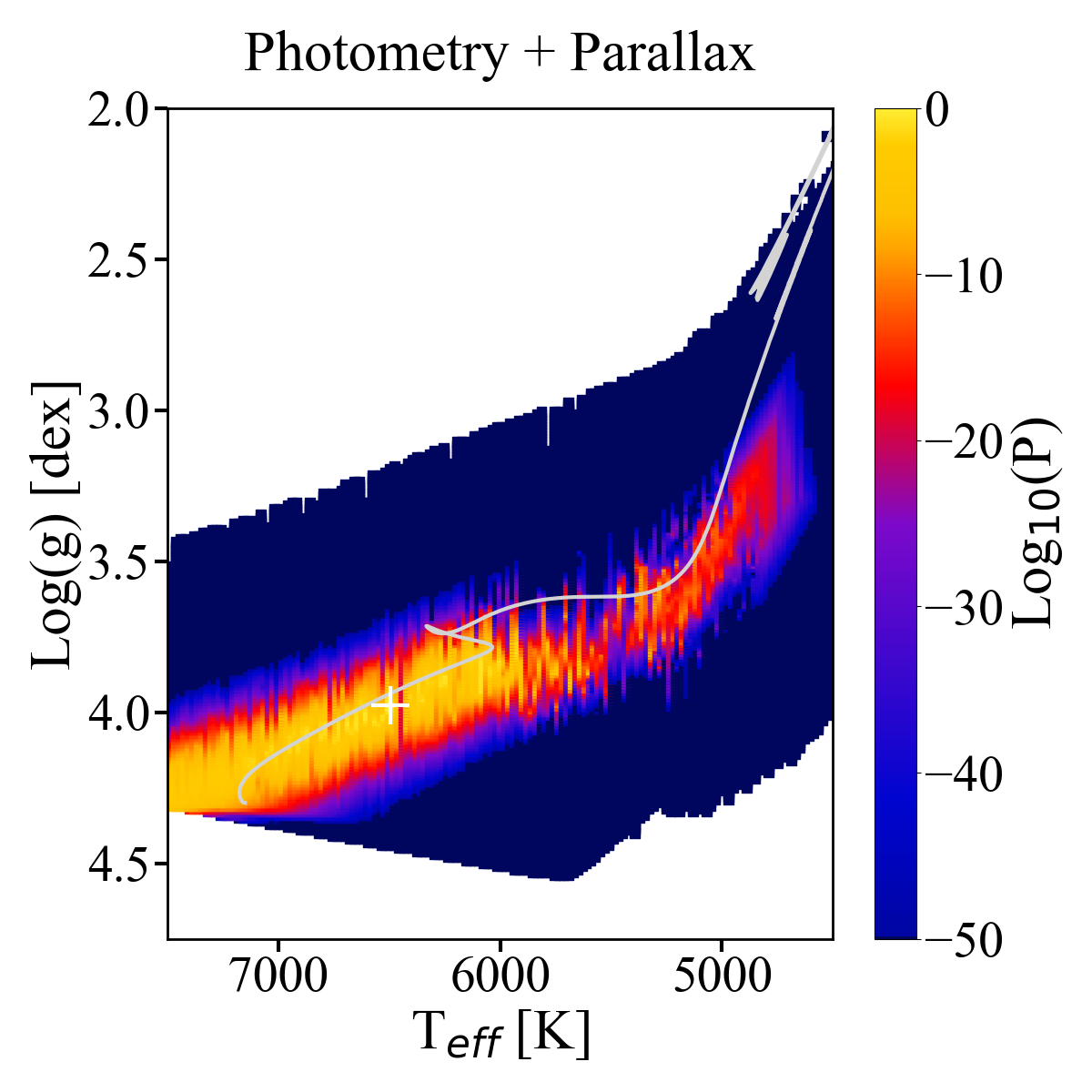}
\end{tabular}
}

\hbox{
\begin{tabular}{@{}c@{}}
\large (d) \\
\includegraphics[scale=0.19]{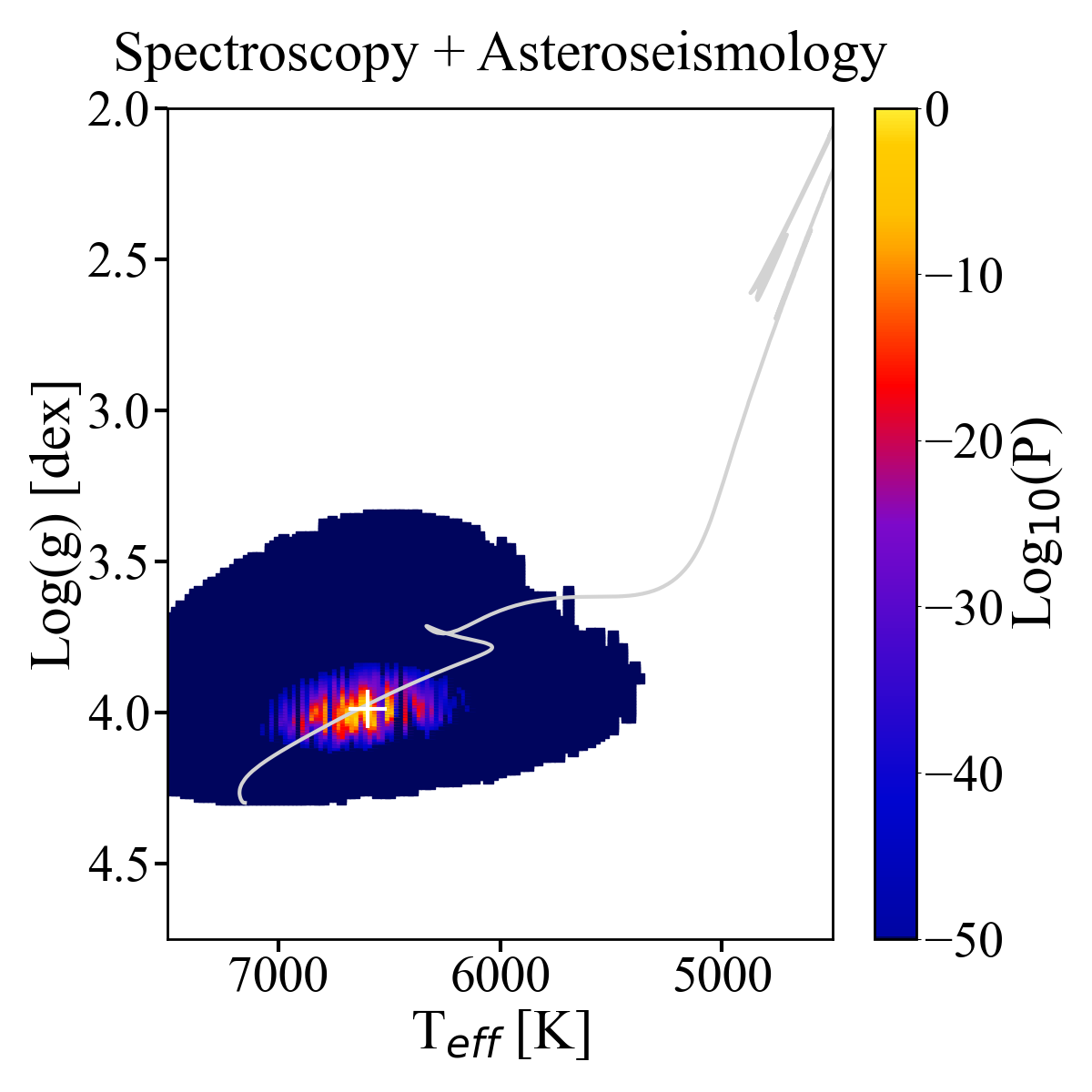} \\
\end{tabular}
\begin{tabular}{@{}c@{}}
\large (e) \\
\includegraphics[scale=0.19]{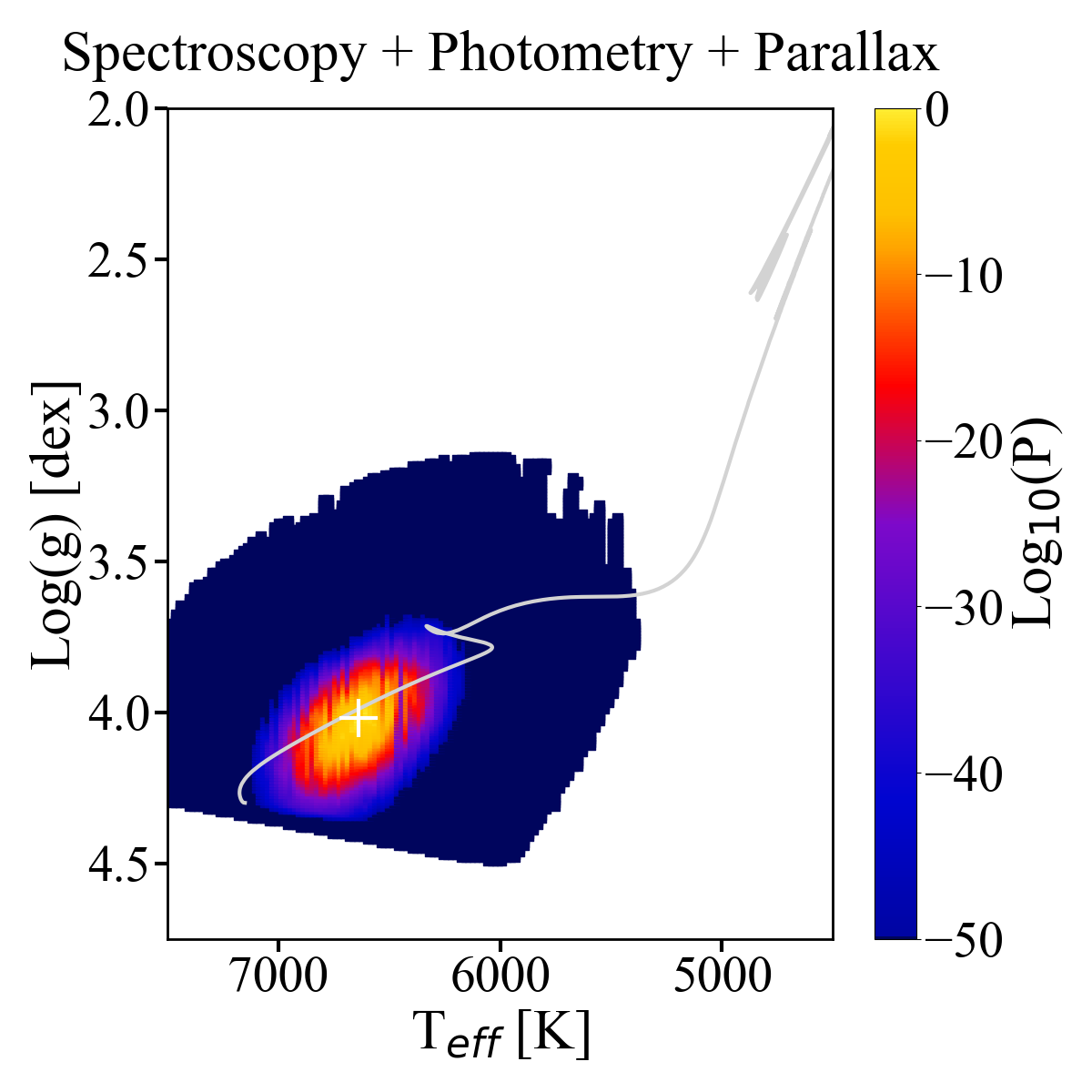} \\
\end{tabular}
\begin{tabular}{@{}c@{}}
\large (f) \\
\includegraphics[scale=0.19]{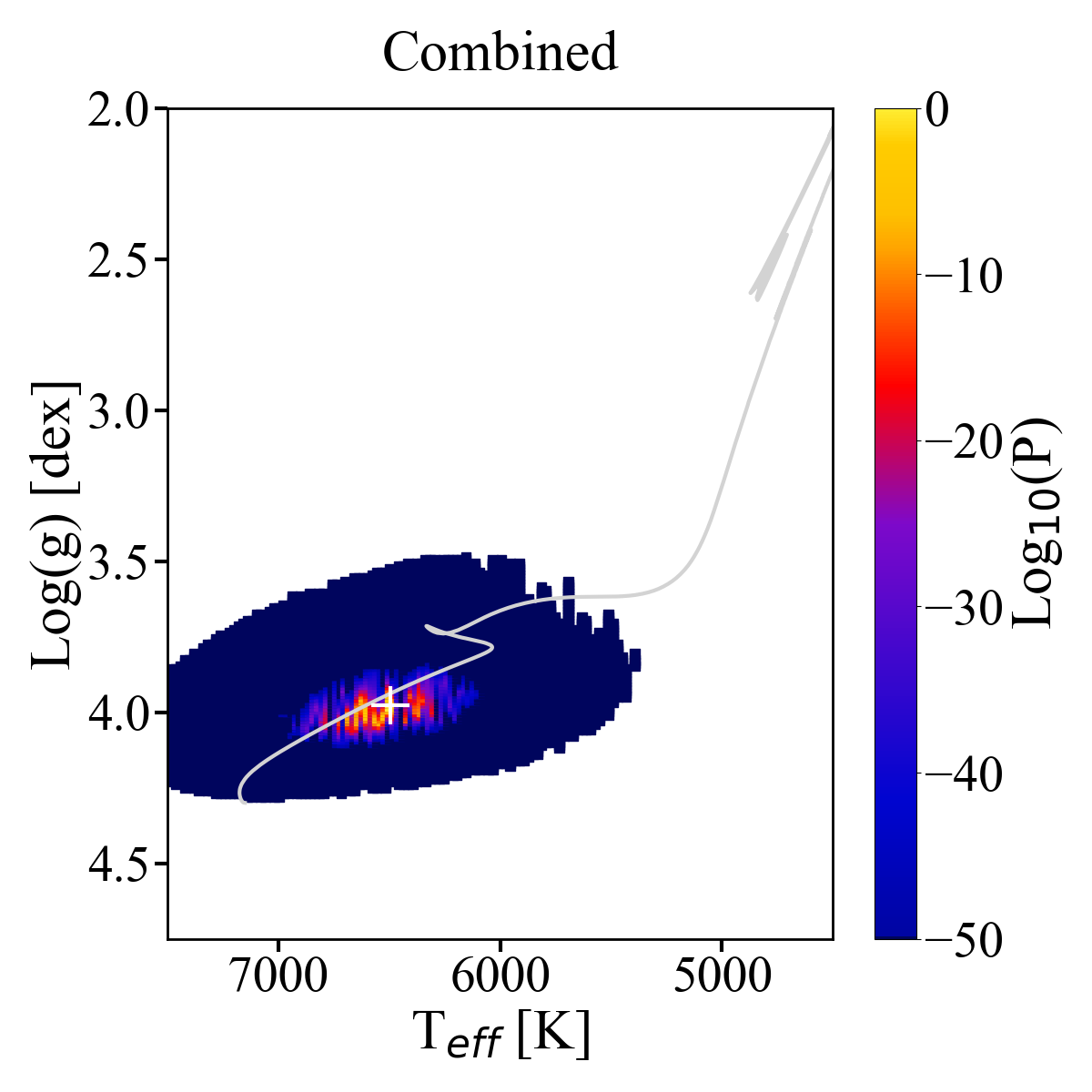} \\
\end{tabular}
}
\caption{Five sub-figures depicting a PDF calculated for Procyon from different modes of SAPP a) Asteroseismology, b) Spectroscopy c) Photometry and Parallax, d) Spectroscopy and Asteroseismology, e) Spectroscopy and Photometry (with parallax) and f) Combined (Bayesian scheme). The horizontal axes is effective temperature, the vertical axes is surface gravity and the colour bar is the logarithm of probability. Each sub-figure PDF is sliced in [Fe/H] dimension. The grey solid line represents a stellar evolution track with 1.5 M$_\odot$ and Z = Z$_\odot$ (the Pre-Main Sequence is not included). The white cross represents the best fit value for the given PDF.}
\label{fig:Procyon_quad_fig}
\end{figure*}

Overall, it is clear that the use of seismic prior on $\numax$ improves the accuracy of $\teff$ and $\logg$ estimates (compare panels (a)-(d) and (b)-(e)), resolving the notorious problem of systematically under- or over-estimated temperatures and surface gravities \citep{2014Mortier,2019Tsantaki}. The iterative solution improves not only the accuracy (central values), but also the precision of the results: the scatter of $\teff$ residuals is reduced by a factor of two, from $\pm 42$ K to $\pm 21$ K. The uncertainties of individual values, however, increase slightly when a seismic prior is applied, which is due to the error in $\nu_{max}$ being taken into account. Without the seismic prior, the uncertainties are purely statistical in nature. The most obvious improvement concerns the stars with $\teff \gtrapprox 6200$ K: Procyon and HD 49933. For these F-type stars, the spectroscopic analysis without any prior returns a very problematic (and well-known) systematic bias in $\teff$ and $\logg$, which manifests itself in the unfortunate correlation between the residuals of the both parameters (Fig. \ref{fig:fstar_seismic_improve}, panel c). In other words, the $\teff$ over-estimated by $100$ K also implies that the $\logg$ estimate is $\sim +0.1$ dex too high. The causal connection between the residual of $\teff$ and that of $\logg$ cannot be established, as both parameters are highly correlated (Fig. \ref{fig:correlation_table}). Remarkably, including the seismic priors resolves the problem: the $\teff$ and $\logg$ values are now in a good agreement with the reference values, with systematic bias on the order of $40$ K between the two quantities.

Accurate surface gravity estimates are essential to determine chemical abundances from gravity-sensitive spectral lines, such as Mg triplet lines at 5100 $\AA$, Ca near-IR triplet lines, and the majority of diagnostic lines of singly-ionised elements \citep{Gehren2004, Lind2012, Bergemann2017}.

We therefore conclude that, similar to $\teff$, the seismic prior offers a clear improvement in surface gravity and helps to break the degeneracy between the influence of $\log g$ and $\teff$ for stars hotter than the Sun.
\subsection{Bayesian solution}\label{sec:bayes_results}
In this section, we describe the results obtained, when all stellar parameters are determined consistently using the Bayesian approach (Sect. \ref{sec:Bayes_approach}), employing photometric (magnitudes), astrometric (parallaxes) , spectroscopic, and asteroseismic ($\Delta\nu$, $\nu_{max}$) data. As emphasised in Sect. \ref{sec:intro}, we want to leverage the exquisite statistical capabilities of Bayesian inference methods and therefore we also include stellar structure models, which gives us the ability to determine radii, masses, and ages of stars in the consistent homogeneous framework. This also allows us to quantify the correlations between various parameters and to explore the influence of uncertainties in the fundamental parameters of stars: their initial mass and detailed chemical composition.

Figure \ref{fig:Procyon_quad_fig} depicts the PDFs for Procyon from different modes of SAPP with the vertical axes as $\logg$ and the horizontal axes as $\teff$. Each PDF shows the likelihood landscape sliced in the [Fe/H] dimension with respect to their maximum probability. The colour scale is the logarithm of the corresponding probability, over-plotted is an evolution track with M = 1.5 M$_\odot$ and Z = Z$_\odot$, the best-fit value is plotted as a white cross. The differences in Figs. \ref{fig:Procyon_quad_fig}d,e, and f are subtle, (d) represents the combination of the Spectroscopy PDF and Asteroseismology PDF, this is analogue to our constrained spectroscopy + $\nu_{max}$ prior results. Fig. \ref{fig:Procyon_quad_fig} (e) represents the PDF resulting from the combination of Spectroscopy and Photometry with parallax included, and (f) represents the PDF resulting from the full Bayesian scheme. The probability range of 10$^{-50}$ to 1 allows us to see the detail of the likelihood landscape and how it changes with various combinations.
\begin{figure*}
\hbox{
\begin{tabular}{@{}c@{}}
\large (a) \\
\includegraphics[scale=0.2]{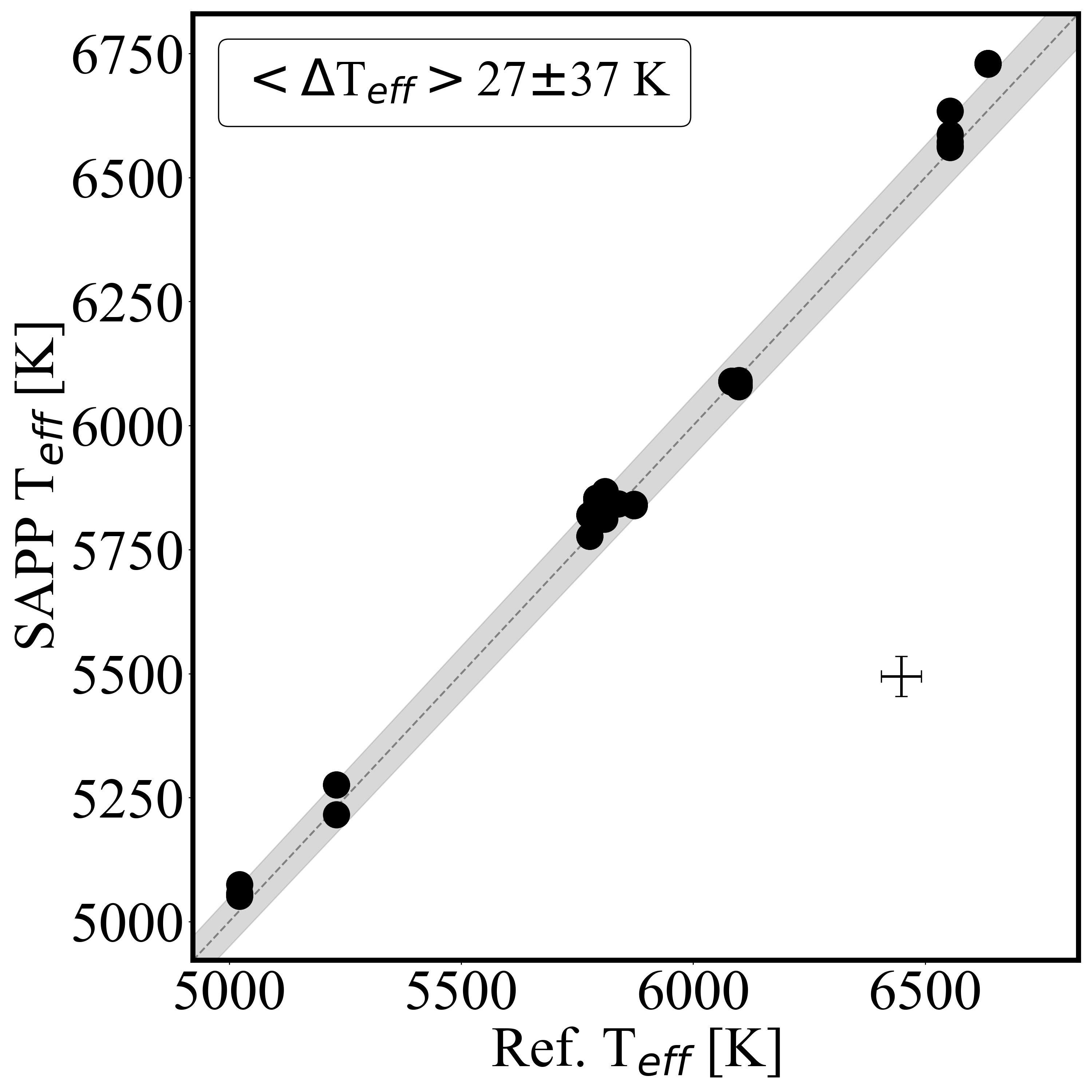} 
\end{tabular}
\begin{tabular}{@{}c@{}}
\large (b) \\
\includegraphics[scale=0.2]{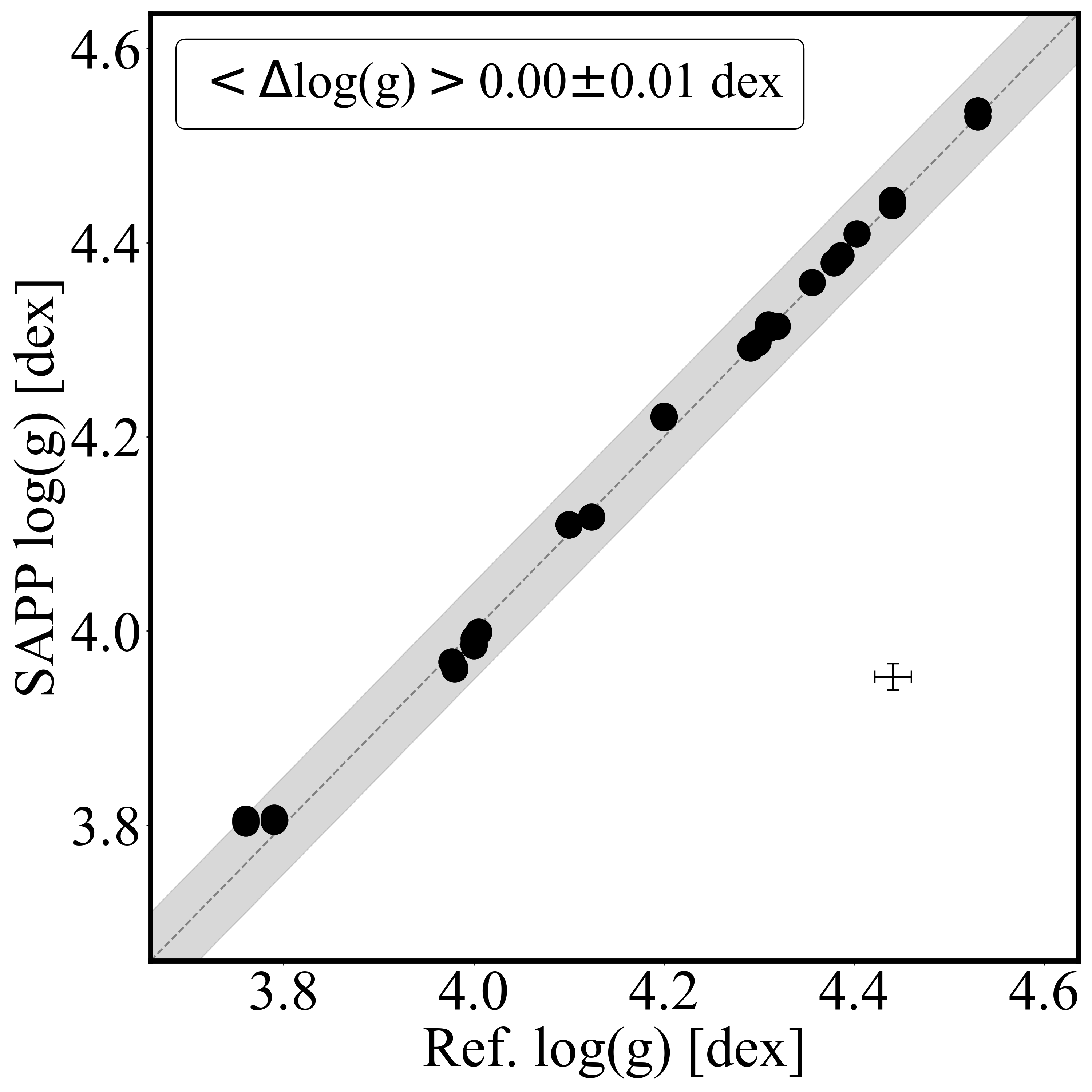}
\end{tabular}
\begin{tabular}{@{}c@{}}
\large (c) \\
\includegraphics[scale=0.2]{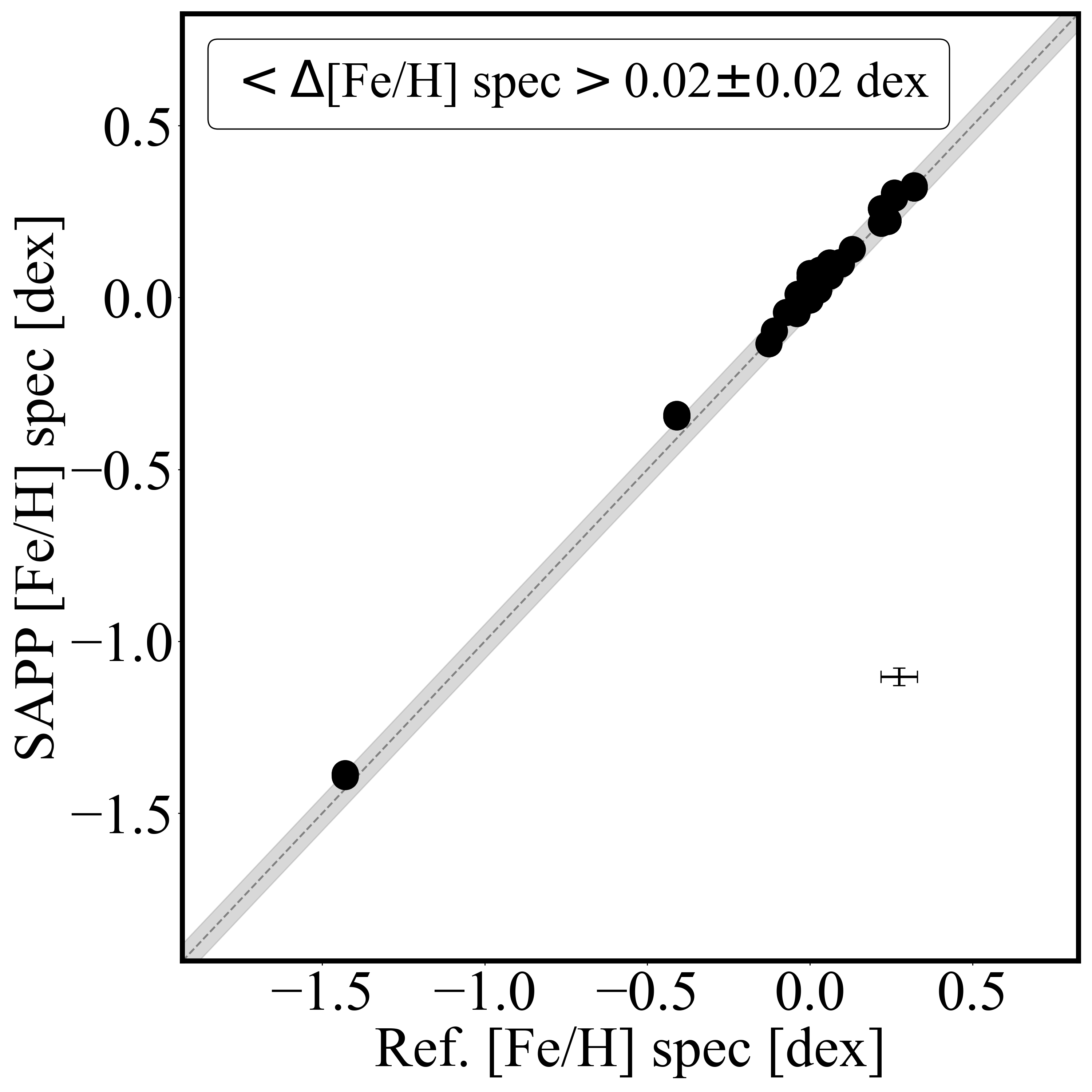} 
\end{tabular}
}
\hbox{
\begin{tabular}{@{}c@{}}
\large (d) \\
\includegraphics[scale=0.2]{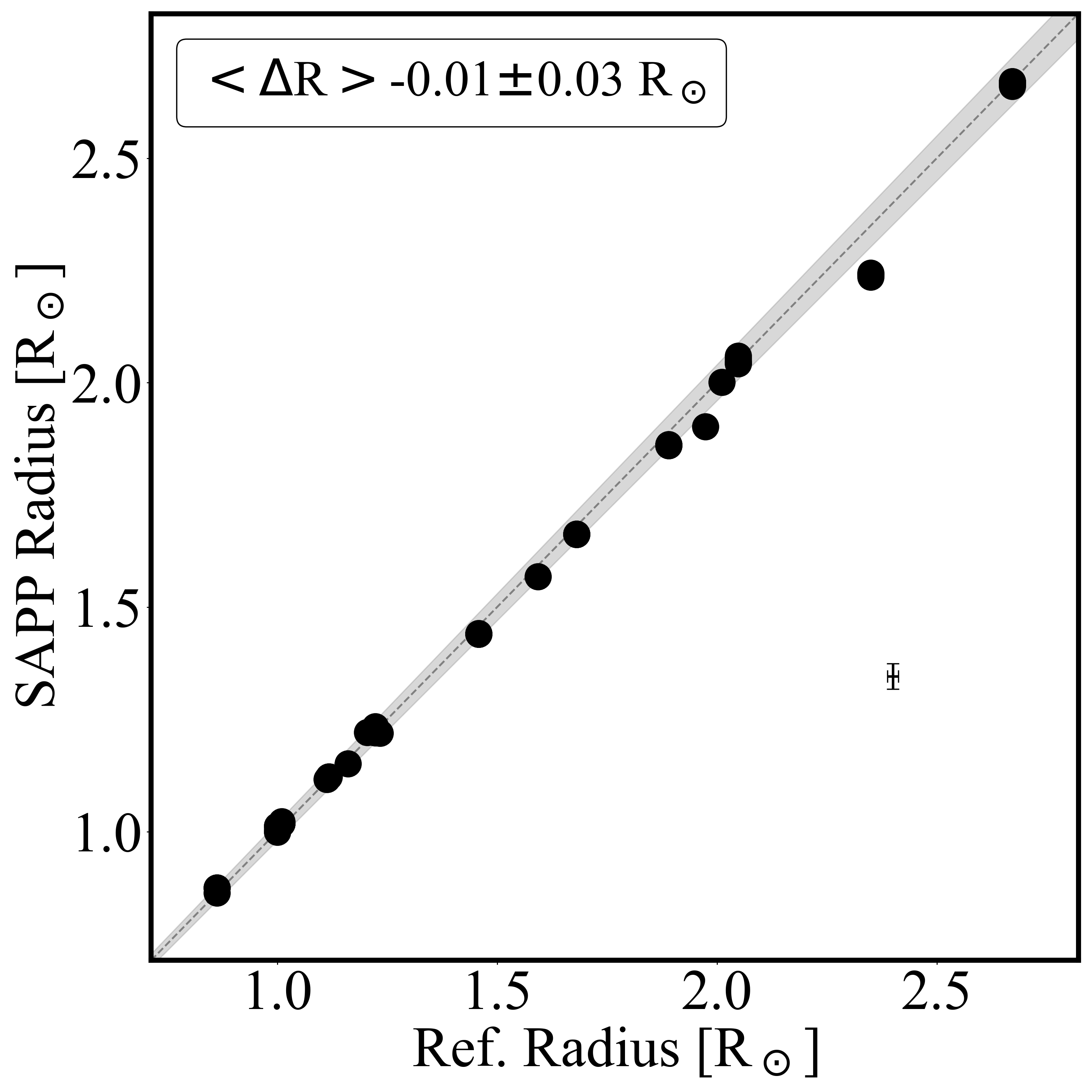} \\
\end{tabular}
\begin{tabular}{@{}c@{}}
\large (e) \\
\includegraphics[scale=0.2]{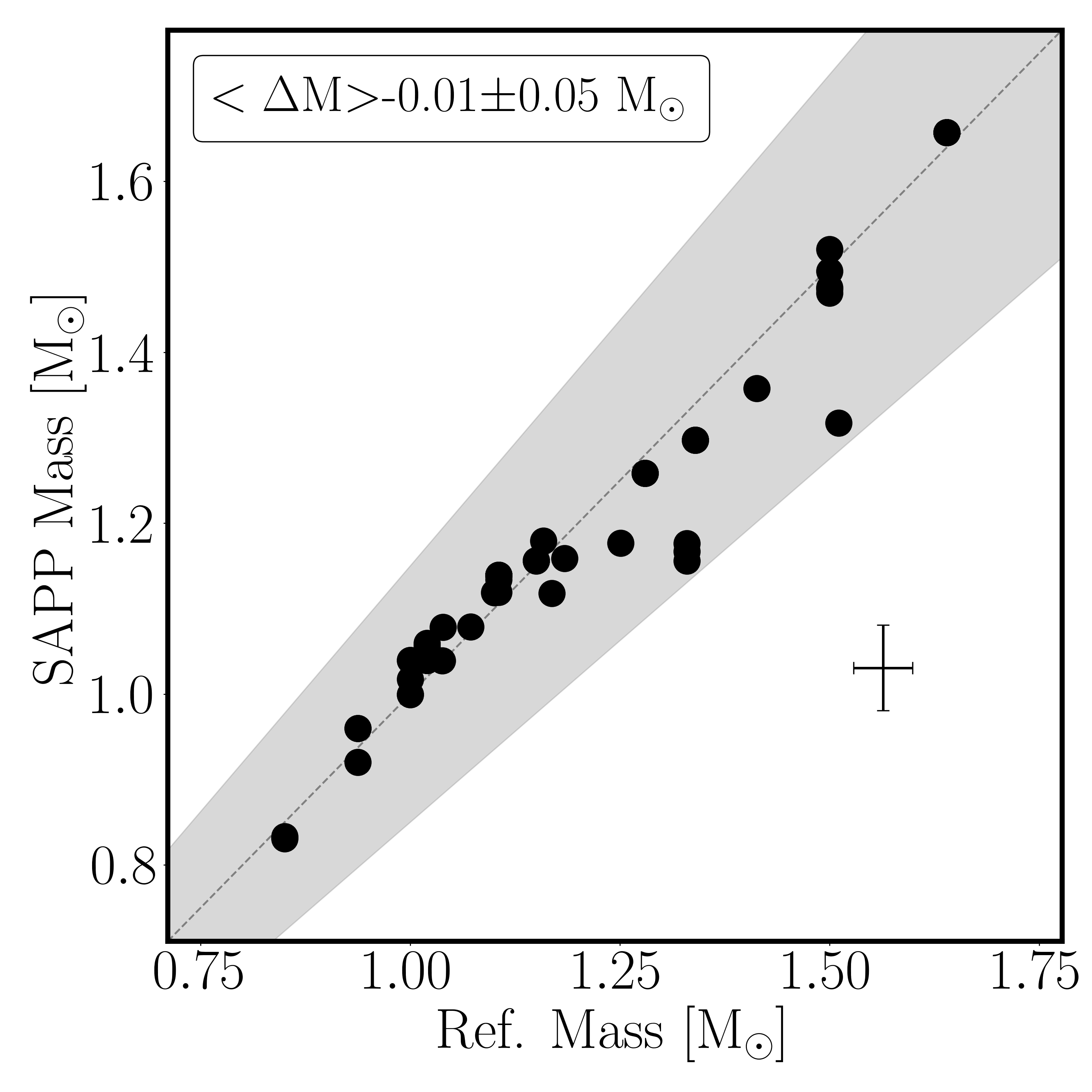} \\
\end{tabular}
\begin{tabular}{@{}c@{}}
\large (f) \\
\includegraphics[scale=0.2]{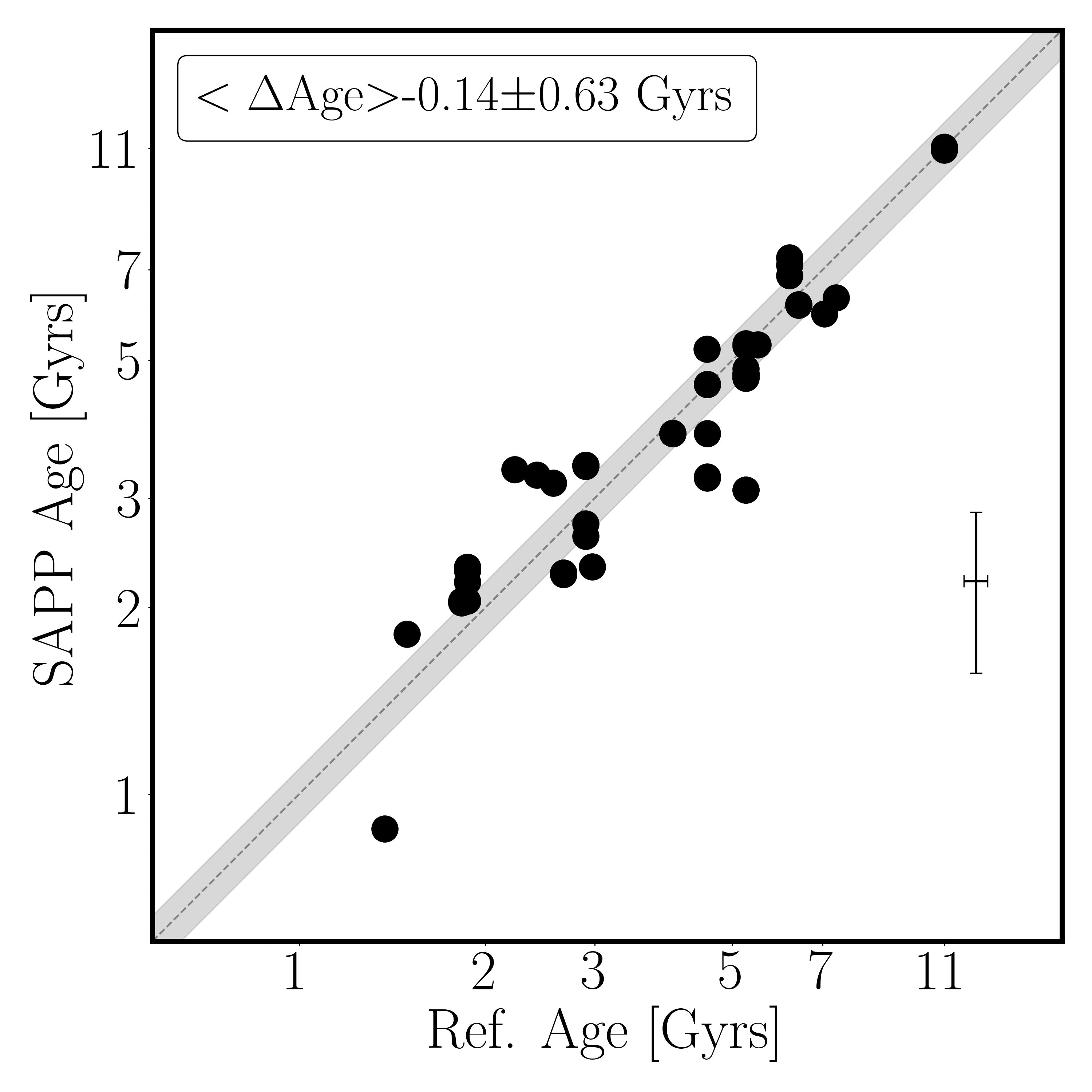} \\
\end{tabular}
}
\caption{SAPP stellar parameters from the full Bayesian scheme versus the reference parameters. The annotations represent the average difference between the SAPP parameters with covariance and the reference parameters respectively. The uncertainty represents the 1 $\sigma$ scatter about said average. The gray shaded region represents the desirable uncertainties of stellar parameters for PLATO: 1$\%$ for $\teff$, 0.05 dex for $\logg$ and [Fe/H], 15 $\%$ for mass, 10 $\%$ for age, 2 $\%$ for radius. The vertical error bars represent the average 1 $\sigma$ uncertainties.}
\label{fig:finalvalues}
\end{figure*}

Our final results for the Bayesian scheme are provided in Table  \ref{tab:bayesian_primary} and they are compared with the reference quantities in Figure \ref{fig:finalvalues}. The average uncertainty for each panel is located in the bottom right. The systematic uncertainties of SAPP are determined from the average differences calculated and shown in the annotations of each sub-figure. Therefore, the final values presented in table \ref{tab:bayesian_primary} have the statistical and systematic uncertainties combined. For the stars which have multiple observation spectra, the final values are averaged and the uncertainties are propagated thoroughly. Clearly, our results for all stellar parameters are in excellent agreement with other estimates. The $\teff$ estimates are accurate to $27 (\rm{syst.}) \pm 37 (\rm{stat.})$ K ($\sim 0.5$ \%), whereas the error on  $\logg$ and [Fe/H] does not exceed $0.00 (\rm{syst.}) \pm 0.01 (\rm{stat.})$ dex and $0.02(\rm{syst.}) \pm 0.02 (\rm{stat.})$ dex, respectively. Also the estimates of radii, masses, and ages of the benchmark stars are consistent with the reference values. Radii and masses are best determined, with the statistical uncertainty of only $0.03~$R$_\odot$  and $0.05$~M$_\odot$ respectively. Ages are determined with a precision of $0.63$ Gyr and a small bias of $\sim -0.14$ Gyr. It shall be stressed, however, that ages are highly model-dependent quantities, and their determination, in turn, relies on the quality of atmospheric parameters ($\teff$, [Fe/H], $\alpha$-enhancement...). Also the reference values are highly heterogeneous, and were determined by different methods and models. Therefore a disagreement (however small) is not surprising and may simply indicate that different types of stellar evolution models (or isochrones) and/or different input values for atmospheric parameters of stars were employed in the literature studies that we use for comparison with our results. 

Comparing our final Bayesian results with the literature estimates \citep[e.g.][]{Sahlholdt2019, Das2019, Howes2019}, we can conclude that the SAPP code provides robust estimates of fundamental stellar parameters, comparable or even superior to other methods. In \cite{Das2019}, the precision of parameters is $10$\% on mass and 10-25\% on age. \citet{Sahlholdt2019} find more conservative uncertainties, of the order $\sim 15 - 40$\% on age. \citet{Howes2019} did not use asteroseismic information, and their estimates are strongly dependent on the evolutionary stage of a star, with typical uncertainties of $15\%$ on age at the turn-off and RGB, but reaching up to $50\%$ on the main-sequence and sub-giant branch (see their Fig. 18).  Our results are qualitatively similar to these estimates, if we were to limit the input data to observed spectra, photometry, and parallaxes, but the use of asteroseismology greatly improves the precision of M, R, and $\tau$, allowing to constrain their values to a precision of a few ($1$ to $5\%$) percent for masses and radii, and $15\%$ for ages. Whereas in PLATO (see section \ref{sec:BO-SAPP}), these quantities will not be determined using the SAPP, our results suggest that the SAPP delivers robust and accurate estimates of atmospheric and fundamental parameter of stars, making the code useful for the characterisation of stars observed within other ongoing and forthcoming programs, such as WEAVE, 4MOST, and SDSS-V.
\subsection{Spectroscopy parameter covariance impact on bayesian scheme}
\label{subsec:bayes_covariance}

The current formulation of spectroscopy produces a PDF that can reliably be combined with other modules. In this section, we explore the influence of using the spectroscopic covariance (Eqn. \ref{eqn:cov_gaussian} in Sect. \ref{sec:Bayes_approach}) in the full Bayesian calculations.

We have also compared the results for stellar parameters computed with and without the inclusion of spectroscopic covariance. Fig. \ref{fig:covariance} (see the Appendix \ref{app:covariance}) confirms that the differences between the results obtained using the two approaches are small. The only somewhat significant deviation can be seen in $\teff$ and $\tau$, which scatter around $\sim$ $\pm$ $40$ K and $0.67$ Gyrs, respectively. We do not detect any significantly large systematic bias associated with the assumption of independent likelihoods.

The inclusion of covariance is formally correct with respect to statistical analysis, as otherwise the assumption is that the core parameters and their errors are independent. It is also not a significant problem in terms of calculation overheads, as the covariance matrix is always available as a by-product of spectroscopic calculations in the SAPP (Sect. \ref{subsec:spec_model_data_comp}). On the other hand, it is clear that the influence of using the covariance is small in the parameter space of FGK-type stars. We can therefore conclude that it is not critical to include the spectroscopic covariance in the full Bayesian analysis, as long as one does not require the precision of better than $1$ percent in the astrophysical characterisation of stars.
%
%
%
%
%
\subsection{Analysis of medium-resolution spectra}
\label{subsec:med_spectra}

\subsubsection{Gaia-ESO benchmark stars}

\begin{figure*}
\centering
\vbox{
\includegraphics[scale=0.4]{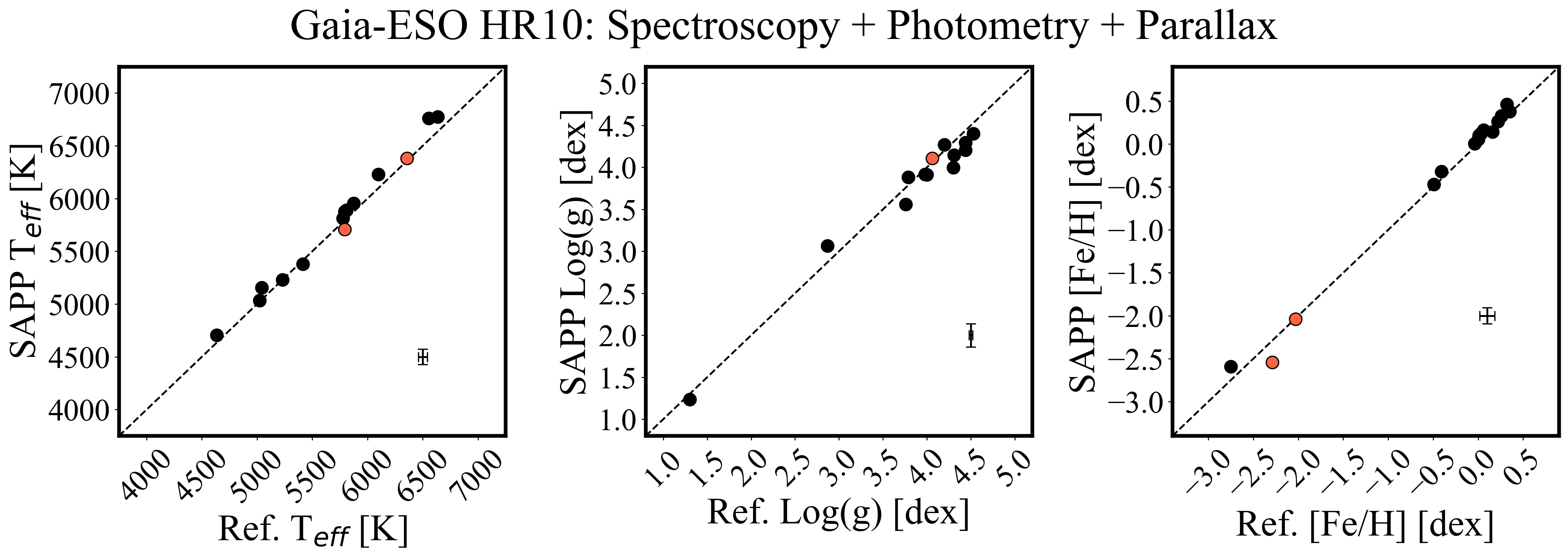}
\includegraphics[scale=0.4]{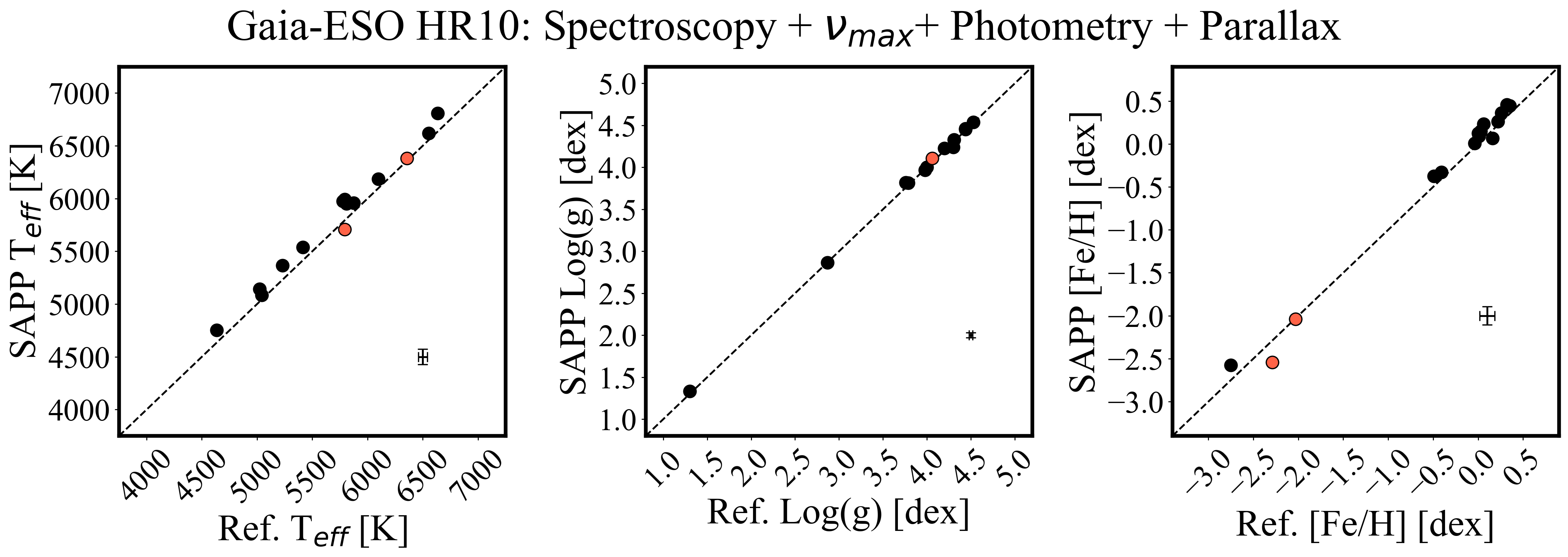}}
\caption{SAPP best-fit parameters of Gaia-ESO HR10 spectra from Bayesian scheme, with resolution R = 20,000 and coverage $\lambda\subset$[5300,5600] $\AA$. In total there are 17 
benchmark stars depicting a different parameter ($\teff$ $\logg$, [Fe/H] and [Mg/Fe]) derived from SAPP (vertical axes) versus the given reference parameter (horizontal axes). The panels represents data determined using the Bayesian scheme without asteroseismology (upper) and with asteroseismology (lower). The red data points represent stars without any asteroseismic parameters.}
\label{fig:HR10_bmk_test}
\end{figure*}

We present our results obtained from the analysis of medium-resolution HR10 spectra of the Gaia-ESO benchmark stars and compare them with the independently-determined stellar parameters (see Sect. \ref{sec:benchmark_stars_parameters}). The GIRAFFE HR10 spectra cover a narrow wavelength range from 5300 to 5600 $\AA$ at the resolving power of $20\,000$. The median S/N is in the range from $70$ to over $2\,000$. The HR10 spectra have lower sampling than the degraded UVES and HARPS spectra analysed in Sect. \ref{sec:bayes_results}, this allows the difference of instrument effects to be explored. These stars also include one FGK sub-giant (HD 140283), two red giants ($\xi$ Hya, HD 122563), an F dwarf (HD 84937), and two G dwarfs ($\mu$ Ara, $\tau$ Cet). These targets are important because red giants highlight the extent of SAPP's applicability towards metal-poor stars outside of PLATO's core program. In Fig. \ref{fig:HR10_bmk_test} a total of $17$ benchmark stars are analysed, where the SAPP estimates of $\teff$, $\logg$, [Fe/H] are compared to the reference values, respectively. The average uncertainty values are shown in the bottom right corner. The upper panel depicts SAPP data derived without asteroseismology data, and the lower panel with asteroseismology. The red data points represent stars which have no asteroseismic information, therefore are identical in both panels.

We find that the analysis of medium-resolution stellar spectra  with the SAPP also yields reliable results. If asteroseismic constraints are used, we obtain the average $\teff$ bias of about $\sim 101$ K and a dispersion of $76$ K, very precise $\log g$ estimates on the order of $0.012$ dex, and metallicities accurate to $\approx 0.05$ dex. Also the Mg abundances can be determined with the uncertainty of $\approx\ 0.01 \pm 0.08$ dex. Interestingly, the $\teff$ value obtained from the HR10 spectra alone are superior to those obtained using $\numax$. This could possibly be the consequence on non-trivial relationships between the physics adopted in the spectroscopic module (Sect. 3.4.1), which is based on MAFAGS-OS atmospheric models. The model atmospheres adopt the mixing length that is empirically constrained to reproduce observed Balmer line profiles (given a pre-defined $\teff$ scale), and may, therefore by coincidence, compensate for the deficits of the model atmospheres, yet at the expense of surface gravity accuracy.  Nonetheless, our results in both cases reinforce the evidence that medium-resolution optical spectra, with $R \approx 20,000$, are well suited for the analysis of fundamental parameters and composition of PLATO stars. Spectra of such quality will be obtained in future with 4MOST \citep{Bensby2019, dejong2019} and WEAVE medium-resolution spectrographs \citep{Dalton2014}. Therefore it can be foreseen that the characterisation of PLATO targets can indeed be done at the required level of accuracy and precision, if 4MOST and WEAVE medium-resolution spectra are available.

\begin{figure*}
\centering
\includegraphics[scale=0.3]{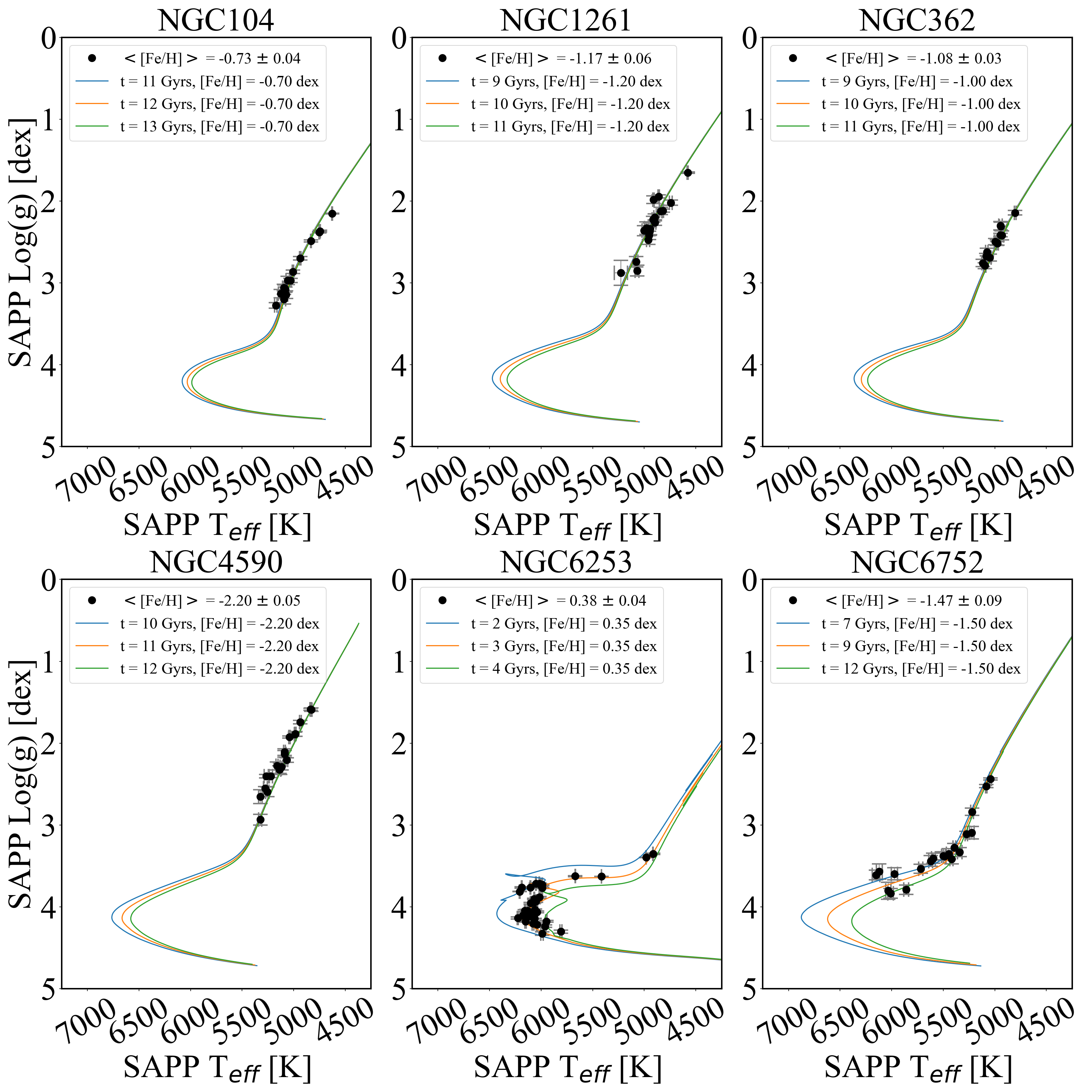}
\caption{$\teff$-$\logg$ diagram of the Gaia-ESO clusters. Only data analysed in this work with the SAPP are shown. In each panel, the black solid circles are a star with corresponding error bars in $\teff$ and $\logg$. There are three isochrones varying in age (blue youngest, orange median, green oldest) which have a metallicity close to the average [Fe/H] of the cluster.}
\label{fig:cluster_HRD}
\end{figure*}

\begin{figure*}
\centering
\includegraphics[scale=0.3]{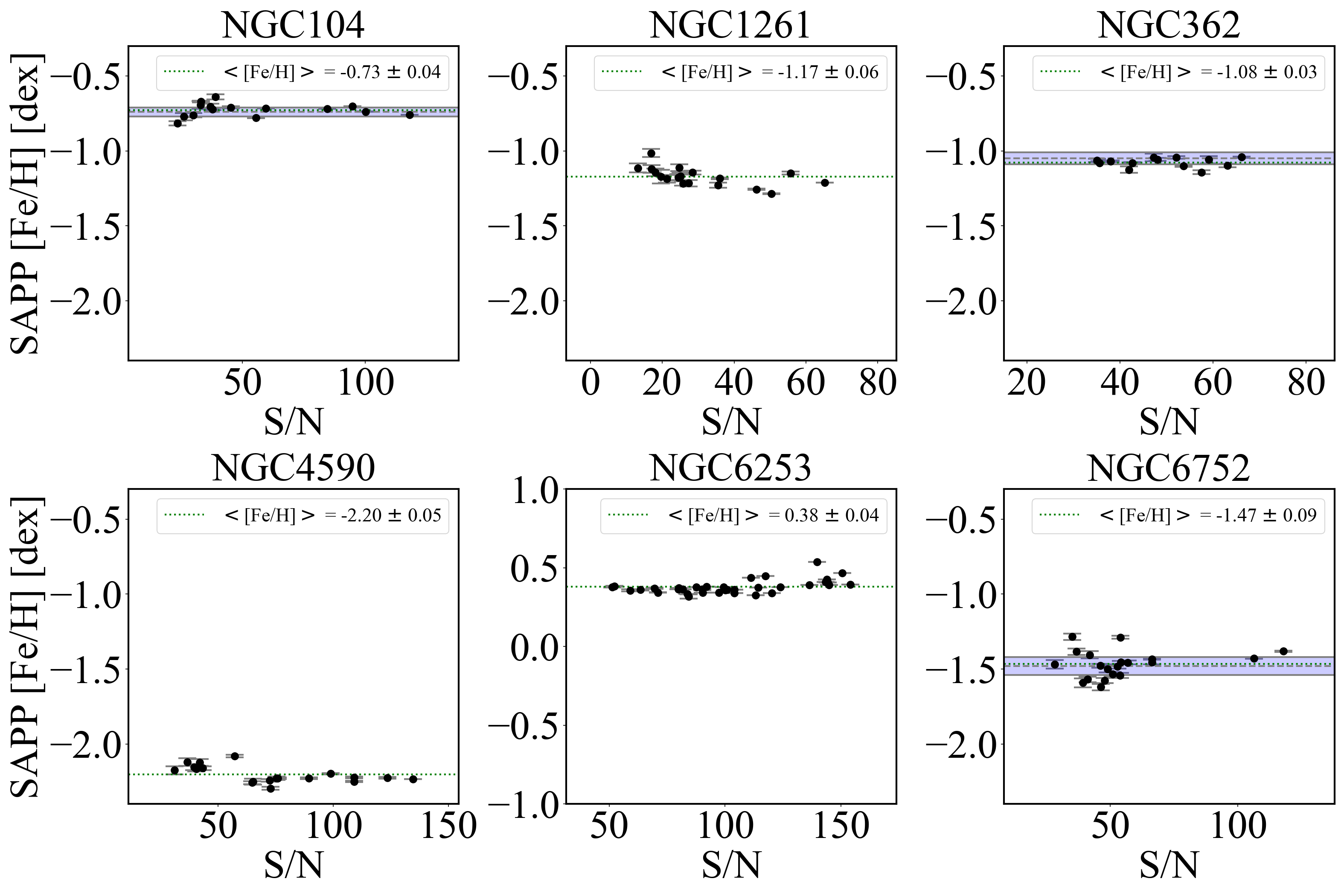}
\caption{SAPP metallicities [Fe/H] of stars in clusters versus the spectral SNR. In each panel, the average (NLTE) [Fe/H] and 1$\sigma$ scatter is annotated, the former being shown as the green dotted line. The grey dashed line and corresponding blue shaded region represents NLTE [Fe/H] estimates and their errors from \cite{Kovalev2019} respectively.}
\label{fig:cluster_feh_scatter}
\end{figure*}
\subsubsection{Gaia-ESO open and globular clusters}

In Figures \ref{fig:cluster_HRD} and \ref{fig:cluster_feh_scatter} we present the results from our analysis of six open and globular clusters\footnote{Cluster RV values from Gaia EDR3 were used to differentiate non-cluster members; here we adopted a typical threshold of $5 \sigma$ in RV for all clusters, except NGC 6352, for which the threshold value of $3$ kms$^{-1}$ was assumed}, for which Gaia-ESO HR10 spectra are available in the public Gaia-ESO data release\footnote{\url{http://archive.eso.org/scienceportal/home?data_collection=GAIAESO&publ_date=2020-12-09}}. None of the clusters were processed using the Asteroseismic module in SAPP, therefore only photometric, astrometric and spectroscopic data was analysed in this section. In Fig. \ref{fig:cluster_HRD}, we also show several isochrones (derived from the GARSTEC stellar evolution tracks described in Sect. \ref{sec:stellar_evo_models}) with varying ages and similar metallicity respectively. These isochrones were not fit to the data, and are only displayed to guide the eye. The unique value of clusters is not just because they, to a first order, represent mono-age and mono-metallic stellar populations \citep{Magrini2017, Bastian2018}, but also because homogeneous observations of stars at different evolutionary stages are available. For NGC 6253 and NGC 6752, we can directly test the metallicities in red giants against our estimates obtained for main-sequence and sub-giant stars. Additionally, literature values of metallicity are available for the turnoff or sub-giant stars in NGC 104 \citep{Carretta2004} and NGC 6253 \citep{Maderak2015}, and we use these estimates to validate our results.

Generally, the results for all six Gaia-ESO clusters are very robust (Fig. \ref{fig:cluster_HRD}), with the $\teff$ and $\logg$ values being in good agreement with the isochrones. Also the intracluster metallicity variation (Fig. \ref{fig:cluster_feh_scatter}) is very small, consistent with expectations from  previous literature estimates of metallicity. In particular, three of the clusters with NLTE metallicites were analysed in \cite{Kovalev2019} and their corresponding average [Fe/H] agree very well with our study. Furthermore, our metallicity estimates for NGC 104 and NGC 6253 are consistent with values based on non-RGB stars from \citet[][]{Carretta2004} and \citet[][]{Maderak2015}, who obtain [Fe/H] $=-0.67 \pm 0.01 \pm 0.04$ dex and [Fe/H] $=+0.445 \pm 0.014$ dex, respectively. However, it should be noted that NLTE effects are particularly large in the atmospheres of red giants at low [Fe/H] typical of globular clusters \citep{Kovalev2019}, and most literature estimates use 1D LTE models, which is why a comparison with other studies is not meaningful. There is no systematic bias or correlation of metallicity with stellar parameters ($\teff$ or $\logg$) or the SNR. That  suggests that our abundances are robust and do not depend on the quality of the data, at least for stellar spectra with SNR $>20$. The typical [Fe/H] dispersion is about $0.03$ dex (NGC 362, NGC 104, NGC 6253) to $0.06$ dex (NGC 1261). Only for NGC 6752, the dispersion is a bit larger, $\sigma \rm{[Fe/H]} \approx 0.09$ dex. Whether some of this dispersion is caused by the presence of sub-giants is not clear yet \citep[e.g.][]{Geller2017,Giesers2019}.

\section{BO-SAPP and DO-SAPP}
\label{sec:BO-SAPP}

In the context of PLATO, the SAPP will operate in two modes: the version that will run During Operations (DO-SAPP) and the version that will be used before the mission to provide an initial characterization of all targets (Before Operations, BO-SAPP). Spectroscopy will be activated in both modes. However because of scientific and runtime considerations, it is the only module that will be activated in DO-SAPP, which overall minimizes SAPPs complexity. Thus, DO-SAPP will only rely on stellar spectra and $\logg$ priors available from the seismic analysis of PLATO light-curves or from the granulation signal. Also, PLATO Follow-Up (FU) spectra and PLATO photometry will be used, if necessary. Unlike the current implementation of the SAPP, which uses the MAFAGS-OS model atmospheres, PLATO modules will use the MARCS 1D and Stagger 3D models combined with a newly developed NLTE version of the Turbospectrum code \citep{Plez2012} as inputs. 

In BO-SAPP, all modules combine together into SAPP’s Bayesian framework to produce best-fit parameters $\teff$, $\logg$, [Fe/H], chemical abundances, V$_{mic}$, Luminosity, etc. Currently, the code makes use of photometry, spectra, and parallaxes, whereas IRFM and SBCR techniques are used to constrain $\teff$ independently and to validate the spectro-photometric results. We furthermore plan to include the interferometry data.Stellar evolution models will not be used, to minimise overlap with other modules of the PLATO Stellar Analysis System (SAS) system. Furthermore, we anticipate that once the Gaia RVS spectra are available, these will be used to complement spectra from instrument archives and surveys. The details of combining different observed spectra, such as, e.g. Gaia RVS, SDSS-V, and 4MOST data, will be presented in a separate study, however, the main advantage of the SAPP is that \textit{same} physical models and the same statistical approach will be used for the analysis of all different kinds of spectroscopic observations to maximise the consistency of the results. Photometric data will be sourced from 2MASS, SPHEREx, and other surveys. 

SAPP’s parameters will be used by different PLATO work-groups to determine  radii, mass, and ages of stars, as well as astrophysical parameters of their orbiting planets. Therefore, very stringent requirements are posed on the accuracy of the atmospheric characterization. For example, a $2 \%$ error in $\teff$ translates into a $3$ percent error in radius and a 5 $\%$ error in mass for a typical FGK-type un-evolved star \citep{Serenelli2017}. Owing to the complexity of PLATO algorithms and the details of work organisation within different PLATO packages, the exact scope of both branches of the SAPP pipeline may change slightly in future. 
\section{Future developments}
\label{sec:future}
The future improvements of the code will involve the transition to 3D convective stellar model atmospheres, but also updates to the SBCR relationship, and a module to analyse M-dwarfs based on optical and infra-red spectra. 3D models are needed, because they provide a physically-realistic description of convective and radiative energy transport in late-type stars \citep[e.g.][]{2005A&A...429..335V,2009LRSP....6....2N, 2012JCoPh.231..919F}, and thereby remove the need for ad-hoc adjustable parameters (like the mixing-length, micro-turbulence, and macro-turbulence), which are used in 1D hydrostatic models. Besides, the most accurate solar photospheric abundances require 3D NLTE modelling from first principles \citep[e.g.][]{Asplund2005, Asplund2009, Lind2017,Bergemann2019, 2019A&A...630A.104A, Gallagher2020}.

In terms of 3D modelling, the Stagger grid \citep{Magic2013a,Magic2013b} will be used in the PLATO consortium to obtain the non-seismic stellar parameter determination, asteroseismic surface effect corrections, limb darkening, and stellar convective noise. The existing STAGGER grid has currently about $200$ models with ${\rm T_{eff}}$ from 3500 to 7000 K (step $500$ K), $\logg$ from 1.5 to 5 dex (step of 0.5 dex), metallicity from $-$4 to $+0.5$ dex. The grid contains also models for specific benchmark stars which are out of the nodes of the grid. The grid is currently being refined to $250$ K step in ${\rm T_{eff}}$ and the models are run for longer time sequences to have better relaxation and more snapshots per model, i.e. more reliable time average. These models will be used to compute 3D NLTE synthetic grids computed using MULTI3D \citep{Bergemann2021} with NLTE-Turbospectrum (Gerber et al. in prep.), and they will supersede the currently used 1D NLTE grids based on the MAFAGS-OS models (Sect. \ref{subsec:spectroscopic_models}).

Further, we are planning to improve the calibration relationships for the SBCR module (Sec. \ref{sec:SBCR}). We are planning to use the data collected with the CHARA/SPICA instrument\footnote{First light expected in 2022.} \citep[][]{spicajosa,spicaspie}. With the 300m baselines of the CHARA array \citep{chara} and the optical regime of the SPICA instrument (6500 - 8500 $\AA$), an angular resolution of 0.2 mas could be reached, yielding angular diameters with a precision of about 1\% for several hundreds of FGK-type stars.
\section{Conclusions}\label{sec:Conclusion}

In this study, we present the SAPP pipeline for the astrophysical characterisation of FGK-type stars. The code is based on the method of Bayesian inference and it is capable of combining various sources of observational information, including but not limited to spectroscopy, photometry, parallaxes, and asteroseismology, along with their uncertainties. Thus, the code avoids the assumption of Gaussian uncertainties and uncorrelated variables, which is commonly used in astronomical literature. 

We tested the pipeline on a sample of well-studied FGK-type stars, most of them included in the set of Gaia benchmark stars \citep{Jofre2018}, and we complemented them with a metal-poor TESS sub-giant $\nu$ Ind \citep{Chaplin2020} and with stars from the Kepler legacy sample \citep{Nissen2017}. High- and medium-resolution spectra ($R > 40,000$ and $R = 20,000$ respectively) with high-S/N (S/N > $300$) spectra for these stars are available from the ESO archives. The combined sample includes $27$ FGK-type stars that cover a broad range in $\teff$, $\logg$, and metallicity. The majority of our benchmarks are main-sequence stars, but there are several sub-giants. The reference parameters of these targets were adopted from recent literature studies that employed asteroseismology, interferometric angular diameters, and spectro-photometry.

We find that spectroscopic data provides most the reliable information about atmospheric parameters of stars. From spectra alone, we can determine $\teff$, $\logg$,  metallicity, and individual abundances of $\alpha$- and iron-group elements with a precision of a $50$ to $100$ K ($\teff$) and $0.05$ to $0.1$ dex in other parameters. These parameters can be recovered from the medium-resolution (R $=20,000$) spectra with narrow wavelength coverage in the optical. Higher-resolution or wide-band spectra would be needed to obtain reliable abundances of carbon, oxygen or rare chemical elements, such as Li or neutron-capture species.

We also find that the asteroseismic prior offers a clear improvement in the accuracy and precision of $\teff$ and $\logg$ estimates. This can be achieved by either using the asteroseismic scaling relationships and iterating with the spectroscopic module, or by using asteroseismic constraints as an independent likelihood in the PDF. The combined approach allows us to achieve $1 \%$ accuracy in the estimate of $\teff$ for all stellar types relevant to PLATO (FGK-type). M-type stars are relevant to the context of PLATO, however will be tested in future studies. The strictly spectroscopic analysis without a seismic prior yields highly biased parameters for hotter stars, with $\teff$ and $\logg$ estimates severely over-estimated. This bias is the consequence of the progressive loss of important lines of neutral species, which makes it difficult to break the degeneracies between $\logg$ and $\teff$. The SBCR and IRFM modules provide useful constraints on $\teff$, however, they are currently not competitive with spectroscopy, as the precision of both methods is not better than 100 K in $\teff$. On the other hand, both modules deliver robust estimates of stellar radii, with the accuracy of $3$ to $5 \%$ in our parameter space. With the upcoming CHARA/SPICA data we expect that the SBCR method can be improved further to reach the precision of $1$ to $2 \%$ in radii.

Comparing our results computed using the SAPP pipeline with the independent values, we find that $\teff$ and $\logg$ estimates are recovered to better than 1$\%$, assuming the validity of the absolute scale of the interferometric $\teff$ and asteroseismic $\logg$ estimates. Our NLTE estimates of metallicity ([Fe/H]), Mg, Ti, and Mn abundances agree with the NLTE estimates based on very high-resolution (HARPS, UVES) spectra to better than $0.04$ dex. This precision and accuracy will help PLATO to attain its goals of 2 $\%$ in R, 15 $\%$ in mass and 10 $\%$ in age \citep{Goupil2017,Serenelli2017}. Analysis of medium-resolution HR10 spectra of stars with asteroseismic data showed that we obtain reliable results within 5$\%$ for $\teff$ and $\logg$ within 1$\%$. NLTE [Fe/H] are accurate to within 0.1 dex with [Mg/Fe] accurate to -0.03$\pm$0.07 dex. SAPP was tested on spectra of stars in six open and globular clusters for an independent assessment of the pipeline. Out of the six, three have been analysed by \cite{Kovalev2019} for the determination of NLTE [Fe/H] and abundances, which agree very well with our study. Our results of the six clusters in general are robust and present little scatter with respect to stellar atmospheric parameters ($\teff$, $\logg$, [Fe/H]). Spectra of this resolution and quality are expected from future facilities, such as 4MOST \citep[e.g][]{Bensby2019,dejong2019} and WEAVE \citep[e.g.][]{Dalton2014}.

Finally, we note that although in this work our focus is on PLATO, the SAPP pipeline is versatile enough to provide astrophysical parameters for other types of stars, such as red giants, M dwarfs, etc. \citet{Kovalev2019} presented comprehensive tests and validation of the NLTE spectroscopic module on high- and medium-resolution spectra of evolved and un-evolved stars across a broad range of metallicities $-2.5 \lesssim$ [Fe/H] $\lesssim +0.3$, effective temperatures $4000 \lesssim \teff \lesssim 7000$ K, and surface gravities $1.2 \lesssim \logg \lesssim 4.6$ dex. The advantages of combining  spectra, photometry, astrometry, and stellar models to derive ages and masses of evolved stars were carefully explored by \citet{Serenelli2013} and \citet{Schoenrich2014}. We therefore conclude that the SAPP code delivers robust estimates of stellar astrophysical parameters, which makes the code useful for the analysis of low-mass stars observed with different large-scale spectroscopic surveys, such as 4MOST and WEAVE \citep[e.g.][]{Dalton2014,Bensby2019}.
\begin{table*}
\small
\begin{center}
\setlength{\tabcolsep}{0.005\linewidth}
\caption{Final estimated parameters of benchmark stars}
\label{tab:bayesian_primary}
\begin{tabular}{l c c c c c c c c c r}
\hline
\hline
\noalign{\smallskip} Star ID  & $\teff$  & $\log g$ &  [Fe/H] & Mass & Age & Radius & Luminosity & [Mg/Fe] & [Ti/Fe] & [Mn/Fe] \\ 
              & K  & dex & dex & M$_\odot$ & Gyrs & R$_\odot$ & L$_\odot$ & dex  & dex & dex\\
\noalign{\smallskip}
\hline
\noalign{\smallskip}
18 Sco & 5833 $\pm$ 37 & 4.44 $\pm$ 0.01 & 0.06 $\pm$ 0.02 & 1.04 $\pm$ 0.05 & 3.21 $\pm$ 0.63 & 1.02 $\pm$ 0.03 & 1.08 $\pm$ 0.16 & -0.01 $\pm$ 0.03 & 0.03 $\pm$ 0.02 & -0.02 $\pm$ 0.03\\
$\alpha$ Cen A & 5852 $\pm$ 37 & 4.31 $\pm$ 0.01 & 0.30 $\pm$ 0.02 & 1.14 $\pm$ 0.05 & 4.73 $\pm$ 0.63 & 1.23 $\pm$ 0.03 & 1.60 $\pm$ 0.16 & -0.04 $\pm$ 0.03 & -0.03 $\pm$ 0.02 & -0.02 $\pm$ 0.03\\
$\alpha$ Cen B & 5246 $\pm$ 37 & 4.53 $\pm$ 0.01 & 0.24 $\pm$ 0.02 & 0.94 $\pm$ 0.05 & 4.18 $\pm$ 0.64 & 0.87 $\pm$ 0.03 & 0.51 $\pm$ 0.16 & 0.09 $\pm$ 0.03 & 0.08 $\pm$ 0.02 & -0.00 $\pm$ 0.03\\
$\beta$ Hyi & 5840 $\pm$ 37 & 3.96 $\pm$ 0.01 & -0.04 $\pm$ 0.02 & 1.16 $\pm$ 0.05 & 6.15 $\pm$ 0.63 & 1.86 $\pm$ 0.03 & 3.62 $\pm$ 0.16 & 0.01 $\pm$ 0.03 & -0.03 $\pm$ 0.02 & -0.11 $\pm$ 0.03\\
$\beta$ Vir & 6089 $\pm$ 37 & 4.11 $\pm$ 0.01 & 0.22 $\pm$ 0.02 & 1.30 $\pm$ 0.05 & 3.81 $\pm$ 0.63 & 1.66 $\pm$ 0.03 & 3.41 $\pm$ 0.16 & -0.13 $\pm$ 0.03 & -0.13 $\pm$ 0.02 & -0.16 $\pm$ 0.03\\
$\delta$ Eri & 5061 $\pm$ 37 & 3.80 $\pm$ 0.01 & 0.10 $\pm$ 0.02 & 1.17 $\pm$ 0.05 & 7.10 $\pm$ 0.64 & 2.24 $\pm$ 0.03 & 2.96 $\pm$ 0.16 & 0.14 $\pm$ 0.03 & -0.01 $\pm$ 0.02 & -0.08 $\pm$ 0.03\\
$\eta$ Boo & 6082 $\pm$ 37 & 3.80 $\pm$ 0.01 & 0.32 $\pm$ 0.02 & 1.66 $\pm$ 0.05 & 2.27 $\pm$ 0.63 & 2.67 $\pm$ 0.03 & 8.75 $\pm$ 0.16 & 0.04 $\pm$ 0.03 & -0.09 $\pm$ 0.02 & -0.06 $\pm$ 0.03\\
HD 49933 & 6730 $\pm$ 37 & 4.22 $\pm$ 0.01 & -0.34 $\pm$ 0.02 & 1.26 $\pm$ 0.05 & 2.04 $\pm$ 0.63 & 1.44 $\pm$ 0.03 & 3.82 $\pm$ 0.16 & 0.06 $\pm$ 0.03 & 0.00 $\pm$ 0.02 & -0.17 $\pm$ 0.03\\
Procyon & 6583 $\pm$ 47 & 3.99 $\pm$ 0.01 & 0.04 $\pm$ 0.03 & 1.49 $\pm$ 0.05 & 2.23 $\pm$ 0.63 & 2.05 $\pm$ 0.03 & 7.09 $\pm$ 0.24 & 0.02 $\pm$ 0.04 & -0.06 $\pm$ 0.02 & -0.16 $\pm$ 0.03\\
Sun & 5803 $\pm$ 37 & 4.44 $\pm$ 0.01 & 0.02 $\pm$ 0.02 & 1.02 $\pm$ 0.05 & 3.89 $\pm$ 0.63 & 1.01 $\pm$ 0.03 & 1.03 $\pm$ 0.16 & -0.01 $\pm$ 0.03 & 0.01 $\pm$ 0.02 & -0.03 $\pm$ 0.03\\
KIC 10162436 & 6289 $\pm$ 37 & 3.97 $\pm$ 0.01 & -0.04 $\pm$ 0.02 & 1.36 $\pm$ 0.05 & 3.17 $\pm$ 0.63 & 2.00 $\pm$ 0.03 & 5.63 $\pm$ 0.16 & -0.03 $\pm$ 0.03 & 0.02 $\pm$ 0.02 & -0.05 $\pm$ 0.03\\
KIC 10644253 & 6111 $\pm$ 37 & 4.41 $\pm$ 0.01 & 0.14 $\pm$ 0.02 & 1.18 $\pm$ 0.05 & 0.88 $\pm$ 0.63 & 1.12 $\pm$ 0.03 & 1.58 $\pm$ 0.16 & -0.03 $\pm$ 0.03 & 0.02 $\pm$ 0.02 & -0.02 $\pm$ 0.03\\
KIC 12069424 & 5842 $\pm$ 37 & 4.29 $\pm$ 0.01 & 0.10 $\pm$ 0.02 & 1.08 $\pm$ 0.05 & 6.31 $\pm$ 0.63 & 1.23 $\pm$ 0.03 & 1.58 $\pm$ 0.16 & 0.03 $\pm$ 0.03 & 0.03 $\pm$ 0.02 & -0.02 $\pm$ 0.03\\
KIC 12069449 & 5811 $\pm$ 37 & 4.36 $\pm$ 0.01 & 0.06 $\pm$ 0.02 & 1.04 $\pm$ 0.05 & 5.95 $\pm$ 0.63 & 1.12 $\pm$ 0.03 & 1.28 $\pm$ 0.16 & 0.03 $\pm$ 0.03 & 0.04 $\pm$ 0.02 & -0.00 $\pm$ 0.03\\
KIC 12258514 & 6026 $\pm$ 37 & 4.12 $\pm$ 0.01 & 0.02 $\pm$ 0.02 & 1.18 $\pm$ 0.05 & 5.30 $\pm$ 0.63 & 1.57 $\pm$ 0.03 & 2.91 $\pm$ 0.16 & -0.01 $\pm$ 0.03 & 0.01 $\pm$ 0.02 & -0.02 $\pm$ 0.03\\
KIC 3427720 & 6086 $\pm$ 37 & 4.39 $\pm$ 0.01 & 0.00 $\pm$ 0.02 & 1.12 $\pm$ 0.05 & 2.33 $\pm$ 0.63 & 1.12 $\pm$ 0.03 & 1.55 $\pm$ 0.16 & 0.00 $\pm$ 0.03 & 0.02 $\pm$ 0.02 & -0.02 $\pm$ 0.03\\
KIC 6106415 & 6002 $\pm$ 37 & 4.30 $\pm$ 0.01 & 0.01 $\pm$ 0.02 & 1.08 $\pm$ 0.05 & 5.21 $\pm$ 0.63 & 1.22 $\pm$ 0.03 & 1.74 $\pm$ 0.16 & -0.07 $\pm$ 0.03 & 0.02 $\pm$ 0.02 & -0.07 $\pm$ 0.03\\
KIC 6225718 & 6207 $\pm$ 37 & 4.31 $\pm$ 0.01 & -0.10 $\pm$ 0.02 & 1.12 $\pm$ 0.05 & 3.34 $\pm$ 0.63 & 1.22 $\pm$ 0.03 & 1.98 $\pm$ 0.16 & -0.04 $\pm$ 0.03 & 0.02 $\pm$ 0.02 & -0.11 $\pm$ 0.03\\
KIC 7940546 & 6305 $\pm$ 37 & 4.00 $\pm$ 0.01 & -0.13 $\pm$ 0.02 & 1.32 $\pm$ 0.05 & 3.27 $\pm$ 0.63 & 1.90 $\pm$ 0.03 & 5.14 $\pm$ 0.16 & -0.01 $\pm$ 0.03 & 0.01 $\pm$ 0.02 & -0.10 $\pm$ 0.03\\
KIC 9139151 & 6130 $\pm$ 37 & 4.38 $\pm$ 0.01 & 0.10 $\pm$ 0.02 & 1.16 $\pm$ 0.05 & 1.81 $\pm$ 0.63 & 1.15 $\pm$ 0.03 & 1.68 $\pm$ 0.16 & -0.07 $\pm$ 0.03 & 0.01 $\pm$ 0.02 & -0.04 $\pm$ 0.03\\
$\nu$ Ind & 5361 $\pm$ 37 & 3.44 $\pm$ 0.01 & -1.39 $\pm$ 0.02 & 0.83 $\pm$ 0.05 & 10.98 $\pm$ 0.64 & 2.87 $\pm$ 0.03 & 6.12 $\pm$ 0.17 & 0.27 $\pm$ 0.03 & 0.27 $\pm$ 0.02 & -0.30 $\pm$ 0.03
\end{tabular}
\tablefoot{The results of our full Bayesian analysis for the primary parameters $\teff$, $\logg$, and [Fe/H]. The Star IDs correspond exactly to the stars in table \ref{tab:ref_primary}. For each parameter, the uncertainty corresponds to the total error, computed by combining the statistical and systematic uncertainty in quadrature.}
\end{center}
\end{table*}
\begin{acknowledgements}
This work presents results from the European Space Agency (ESA) space mission PLATO. The PLATO payload, the PLATO Ground Segment and PLATO data processing are joint developments of ESA and the PLATO Mission Consortium (PMC). Funding for the PMC is provided at national levels, in particular by countries participating in the PLATO Multilateral Agreement (Austria, Belgium, Czech Republic, Denmark, France, Germany, Italy, Netherlands, Portugal, Spain, Sweden, Switzerland, Norway, and United Kingdom) and institutions from Brazil. Members of the PLATO Consortium can be found at \url{https://platomission.com/.} The ESA PLATO mission website is \url{https://www.cosmos.esa.int/plato.} We thank the teams working for PLATO for all their work.
We thank P. E. Nissen for providing the reduced spectra of the Kepler Legacy stars. MRG, MB, JG, and MK are supported by the Lise Meitner grant from the Max Planck Society BP is  partially supported by CNES, the Centre National d'\'{E}tudes Spatiales. We acknowledge support by the Collaborative Research centre SFB 881 (projects A5, A10), Heidelberg University, of the Deutsche Forschungsgemeinschaft (DFG, German Research Foundation). We thank C. Aerts for providing helpful comments to this study.
This work has made use of data from the European Space Agency (ESA) mission {\it Gaia} (\url{https://www.cosmos.esa.int/gaia}), processed by the {\it Gaia} Data Processing and Analysis Consortium (DPAC, \url{https://www.cosmos.esa.int/web/gaia/dpac/consortium}). Funding for the DPAC has been provided by national institutions, in particular the institutions participating in the {\it Gaia} Multilateral Agreement. AS acknowledges support from MICINN grant PID2019-108709GB-I00. TM acknowledges financial support from Belspo for contract PRODEX PLATO mission development. UH acknowledges support from the Swedish National Space Agency (SNSA/Rymdstyrelsen). SM has been supported by the J{\'a}nos Bolyai Research Scholarship of the Hungarian Academy of Sciences, and by the {\'U}NKP-20-5 New National Excellence Program of the Ministry for Innovation and Technology. NN acknowledges Anthony Salsi for useful discussions. V.A. is supported by FCT - Funda\c{c}\~ao para a Ci\^encia e Tecnologia (FCT) through national grants: UID/FIS/04434/2019; UIDB/04434/2020; UIDP/04434/2020. V.A. also acknowledges the support from FCT through Investigador FCT contract nr.  IF/00650/2015/CP1273/CT0001. M.T. acknowledges the funding from MIUR Premiale 2016: MITIC. LC is the recipient of the ARC Future Fellowship FT160100402. MB is supported through the Lise Meitner grant from the Max Planck Society. We acknowledge support by the Collaborative Research centre SFB 881 (projects A5, A10), Heidelberg University, of the Deutsche Forschungsgemeinschaft (DFG, German Research Foundation).  This project has received funding from the European Research Council (ERC) under the European Union’s Horizon 2020 research and innovation programme (Grant agreement No. 949173).

\end{acknowledgements}

\bibliographystyle{aa}
\bibliography{references_clean}
\appendix



\section{Iterative seismic prior - ceres}
\label{app:teff_iteration}
\begin{figure}[ht!]
    \centering
    \includegraphics[scale=0.25]{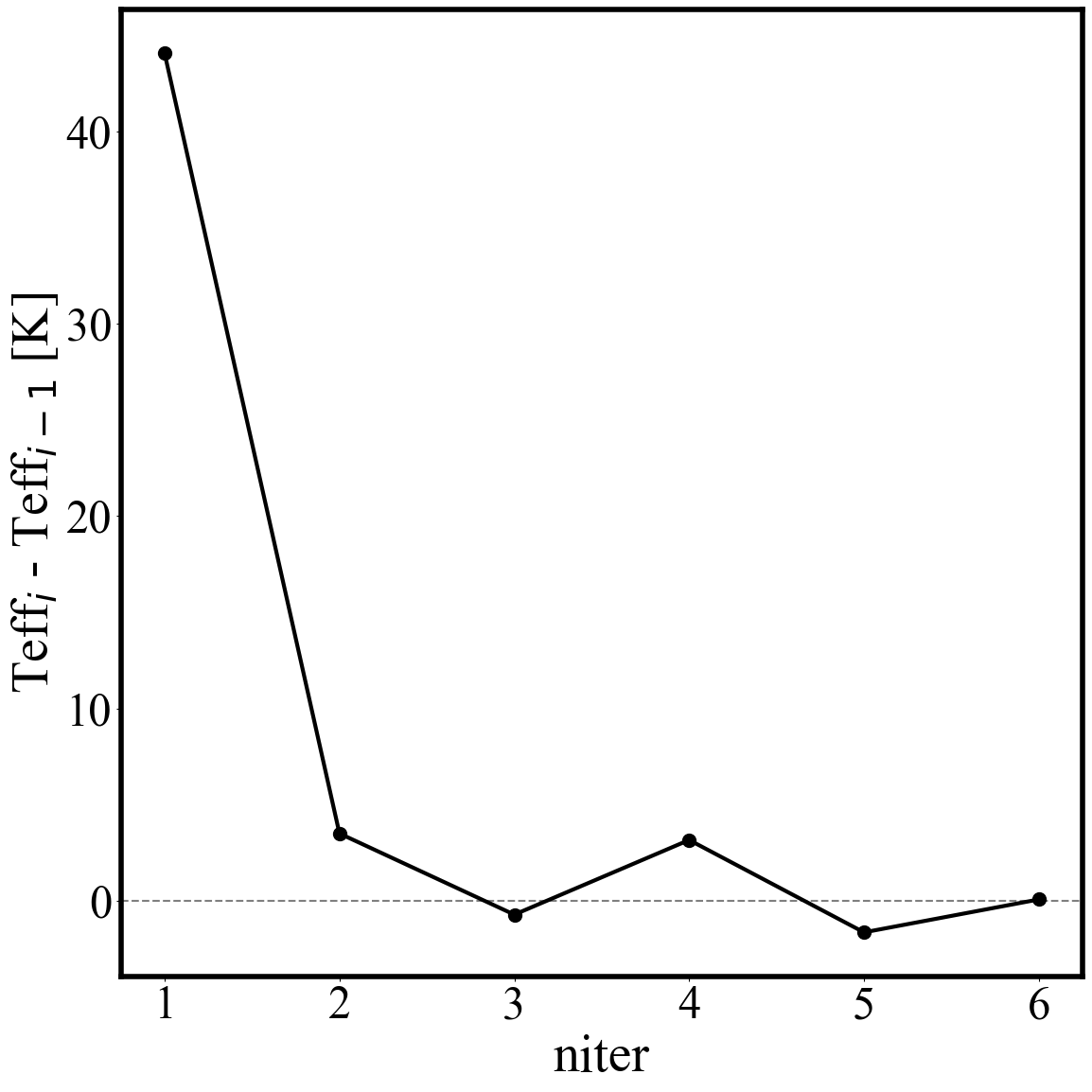}
    \caption{Difference between effective temperature estimations per iteration versus number of iterations. This is the iterative method described in section \ref{sec:seismic_priors} for the Sun, specifically Ceres HARPS spectra.}
    \label{fig:teff_iteration}
\end{figure}

Figure \ref{fig:teff_iteration} shows the iterative process described in Sect. \ref{sec:seismic_priors} for a spectra of the Sun from Ceres taken by HARPS. Whereby the process determines a new $\logg$ given a spectral $\teff$ and asteroseismic $\numax$ using equation \ref{eqn:nu_max}. Once a $\logg$ is calculated, it is fixed and the spectroscopy module is re-run, therefore determining a new $\teff$ which is used in the same equation. The change in effective temperature is tracked and represents the vertical axis, the number of iterations represents the horizontal axis.

\section{Pre-processing of spectra}
\subsection{Continuum normalisation}
\label{subsec:cont_normalisation}

Typically when spectra are taken, depending on the facility the data is run through a reduction pipeline which cleans the data of atmospheric interference such as tellurics (these aren't always caught), blaze functions, cosmic rays and other features. Some pipelines also perform continuum normalisation on the spectra, however, these aren't always \textit{Full}y successful. Meaning, the spectra collected is not ready for science. It must be further processed by the studies intending to analyse the data. We found a plethora of these issues for different spectra collected. The Gaia-ESO iDr5 fit files were somewhat continuum normalised but not \textit{Full}y, the given continuum was closer to 5 percent above unity. HARPS and UVES spectra were not continuum normalised at all but also suffered from different contaminants. They both had to be corrected for tellurics and treated for cosmic rays. 

\subsection{Contaminant treatment}

We mask the pixel location of the tellurics by inspecting the solar spectra for each given facility and then used those masks for any spectra taken by those telescopes. The masks were via "Noising-up" the errors i.e. we increase the error at the telluric pixel and 1.7 $\AA$ around it such that the fitting routine does not mistake it for a line. The cosmic ray treatment is simple, we apply a sigma clipping routine to the spectra pre-normalisation. The spectra are clipped to 2.5 sigma, no less otherwise vital information is lost. This is very important to treat as the cosmic ray contamination appears as an emission line, thus it will affect the continuum normalisation routine, especially ours since we use linear fitting. 

\subsection{Normalisation routine}

\begin{figure}[!ht]
    \centering
    \includegraphics[scale=0.35]{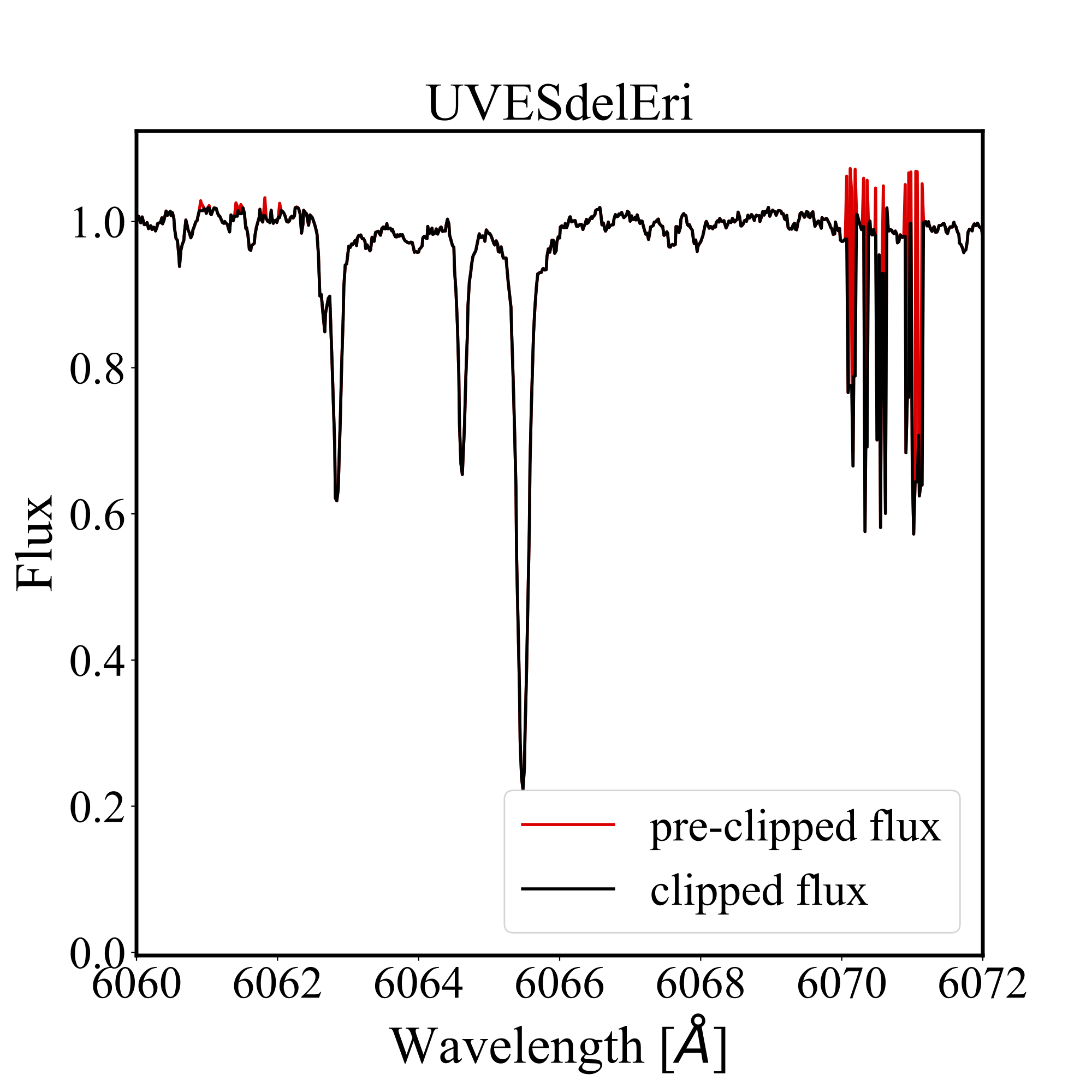}
    \caption{Comparing UVES spectra of the star $\delta$ Eri before and after sigma clipping with $\sigma = 2.5$}
    \label{fig:deleri_sigclip}
\end{figure}

\begin{figure}[!ht]
    \centering
    \includegraphics[scale=0.3]{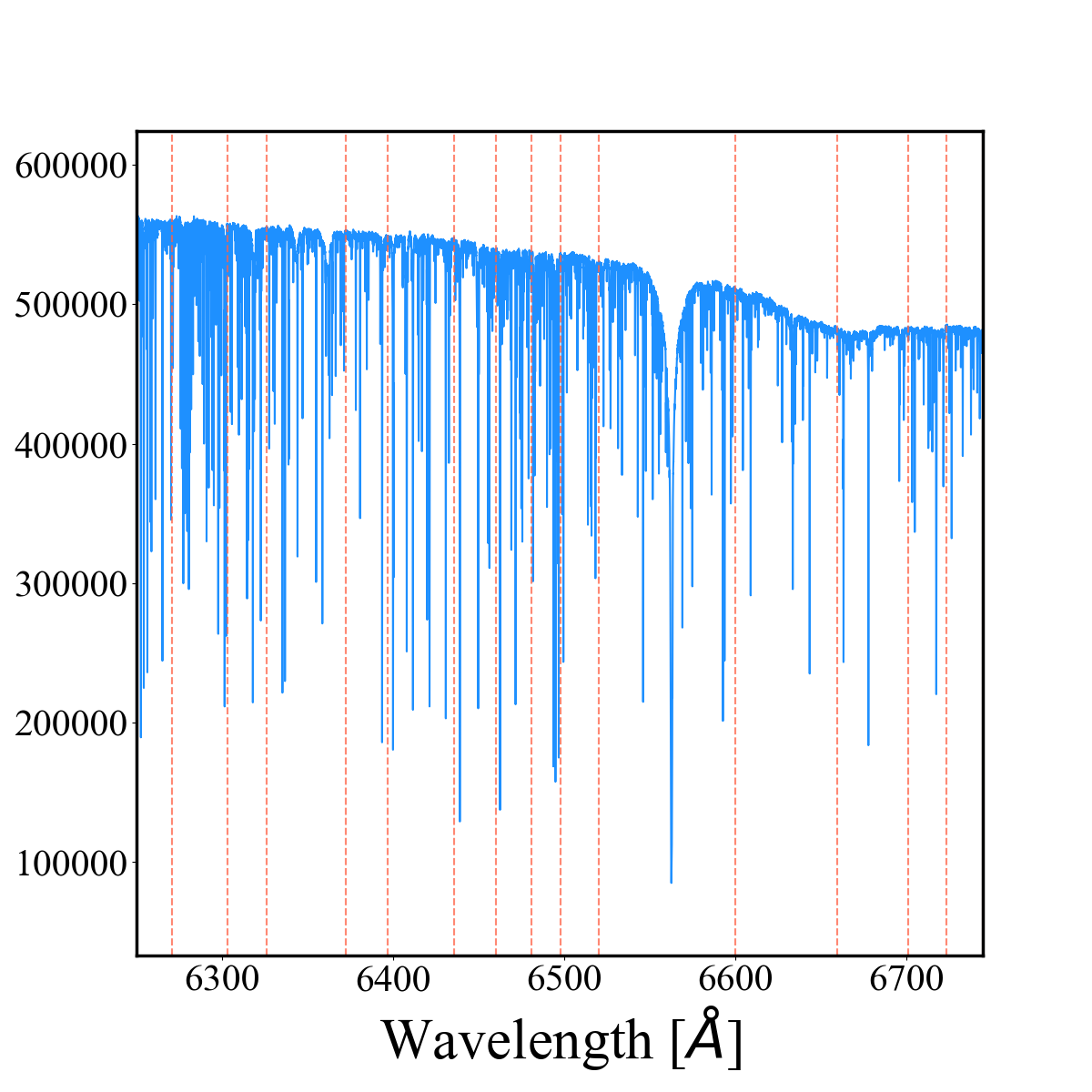}
    \caption{18 Sco HARPS-North spectra (blue solid line) from 6250 $\AA$ to 6745 $\AA$. The red-dashed lines over-plotted represent wavelength segments defined within the continuum normalisation routine.}
    \label{fig:contnorm_zones}
\end{figure}

\begin{enumerate}
    \item Search through spectra for negative fluxes, zero value or NaN pixels (this can happen due to reduction pipelines/bad pixels on CCD). If any bad pixels are found, remove said pixels from spectra.
    \item Sigma clip the data to remove cosmic ray emissions. See fig. \ref{fig:deleri_sigclip} 
    \item Search through the spectra for any significant gaps (lost data)
    \item Split spectra into individual segments. We developed our own list of zones to split the data in by care\textit{Full}y inspecting the wings of strong lines that appear in the spectra. See fig. \ref{fig:contnorm_zones}
    \item Divide the entire spectra by its median value. We found this aids in the normalisation routine by bringing the overall spectra closer to unity. For example if most of the points are above 1, dividing by the median will reduce the spectra down, if most of the points are below 1, dividing by the median will increase the spectra towards 1.
    \item Find a linear regression fit for each segment. Divide the segment by said fit.
    \item For each segment, split data in half (the split point actually is the pixel closest to unity within 2.5 $\AA$ of the middle pixel), divide by the median and find a linear fit for each half. Divide each half by its given linear fit.
    \item Ignore the previous step if the lines are deemed as broad. For example, H$_\alpha$ line would be too broad for this secondary normalisation to work.
    \item Stitch each normalised segment together and output the continuum normalised spectra. 
\end{enumerate}

\subsection{Radial-velocity correction}
\label{subsec:rv_correction}

After continuum normalising spectral data and treating contaminants, a further reduction step is required, radial velocity correction. The radial-velocity of a star represents the relative motion of the star with respect to the observer at the time of observation. The motions doppler shift the light and either blue shift or red shifts it. What we observe is the spectra is shifted along its wavelength scale by some amount. 

This greatly affects analysis of the spectra of the star, specifically the line profiles of the models will not match the lines from the observations. This will result in a high $\chi^2$ value for many pixels and thus produce an erroneous "best-fit" model with incorrect parameters. We decided to write our own RV correction procedure as it is relatively simple and allows us to verify the RV value given typically with the spectral data (sometimes it is not given). To shift the spectra to rest frame with a RV value known, the following equation is required,

\begin{equation}
\label{eqn:rv_shift}
    \lambda_{0} = \lambda_{obs}/(\Delta RV/c + 1)
\end{equation}

Our procedure took direct inspiration from PyAstronomy RV correction module, \textit{PyAstronomy.pyasl.crosscorrRV}, which is a collection of astronomy packages \citep{pya}. The procedure we follow involves testing a range of RV values, doppler shift the observed spectrum and compare it to a model which is in the rest frame. As all the stars in our benchmark sample are on the main-sequence and so sub-giant/dwarfs, we only require one model for RV correcting our spectra, a Main-Sequence star such as our Sun. A spectral model of R $\approx$ 500,000 was created using Turbospectrum \citep{Plez2012} with the classical parameters of the Sun (see table .\ref{tab:ref_primary}). This model works as a template because despite some changes in the HR Diagram, all the stars' line profiles should approximately match the location of the Solar line profiles. Therefore, one linearly shifts the observed spectrum and see how well the lines match. This is done via Cross-Correlating, a method whereby we measure the similarity of two spectra (observed and model) as a function of the displacement of one relative to the other. To calculate the Cross-Correlation exactly, the procedure does the following: 

\begin{enumerate}
    \item Create a list of plausible $\Delta$RV values, for example -50 to 50 kms$^{-1}$ with a resolution of 0.05 kms$^{-1}$.
    \item For each $\Delta$RV value, interpolate the model template spectra onto the observation spectra wavelength frame. This can be done by making $\lambda_{obs}$ the subject of equation \ref{eqn:rv_shift}, thus $\lambda_{0,shift}$ = $\lambda_0$ $\times$ (RV/c + 1) where $\lambda_0$ in this case is the original model wavelength (i.e. rest frame). Interpolating ensures the model wavelength scale sampling matches the observation scale exactly.
    \item Calculate the Cross-Correlation value using the following equation, 
    $$ \text{CC} = \sum\limits_{i} \text{flux}[i]_{\text{obs}}\times \text{flux}[i]_{0,\text{shift}}, $$
    where $i$ represents the wavelength pixel. This total number is the CC value given the $\Delta$RV value used. 
    \item Calculate CC for all $\Delta$RV values and locate where the maxima occurs, this will indicate the $\Delta$RV value required for Radial-Velocity correction.
\end{enumerate}

\begin{figure*}[ht!]
\includegraphics[scale=0.4]{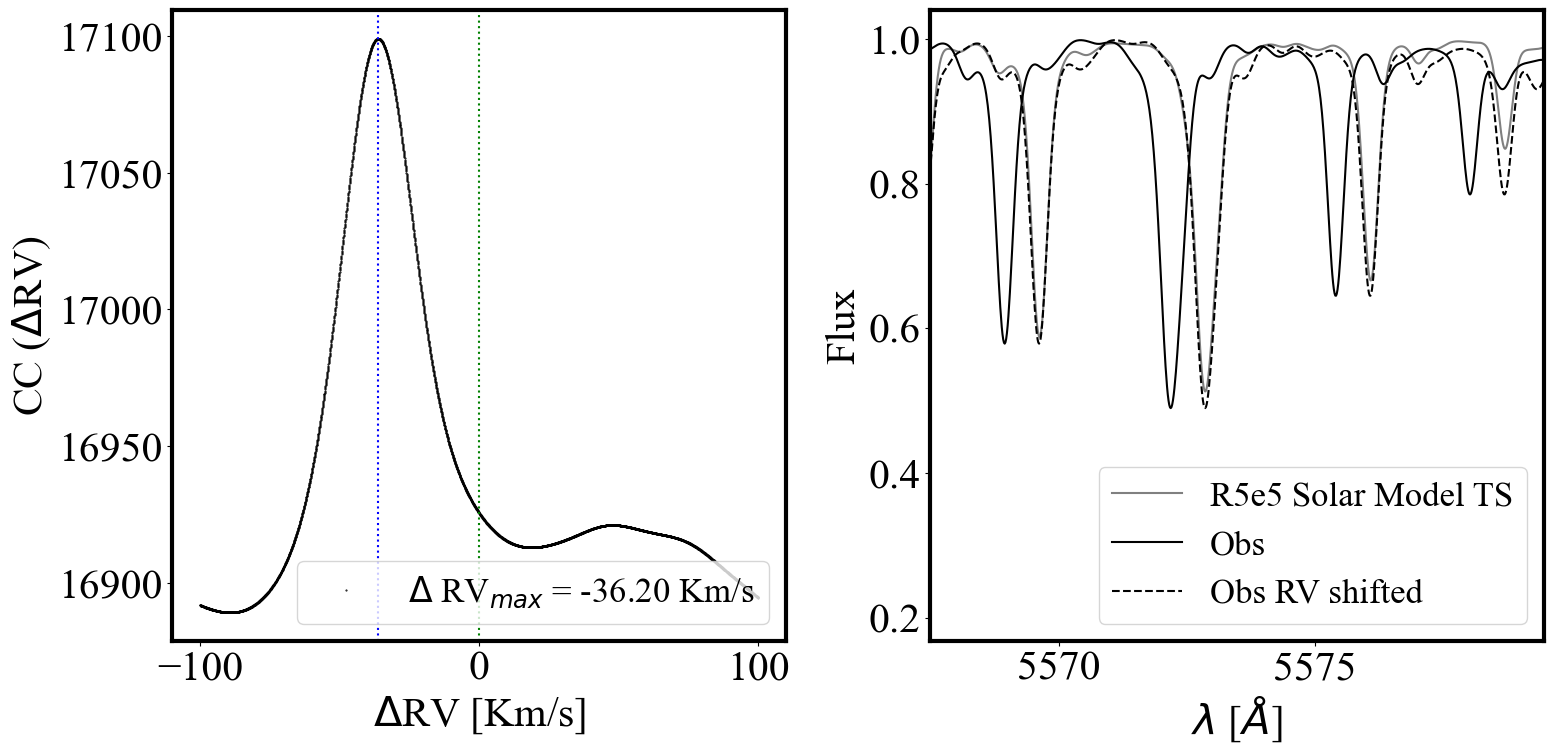}
\centering
\caption{Two subfigures representing the radial-velocity correction process for a UVES spectra of Alpha Cen A convolved to HR10 resolution. Sub-figure on the left depicts the Cross-Correlation value versus the doppler rv shift for a given model and observation comparison. The Sub-figure on the right shows the radial-velocity corrected spectra (black solid line) and the un-corrected spectra (black dashed line) versus the high resolution Solar model from Turbospectrum (grey solid line).}
\label{fig:rv_corrections}
\end{figure*}

Figure \ref{fig:rv_corrections} depicts two sub-figures with the results of calculating the RV correction of UVES $\alpha$ Centauri A spectra. The first sub-figure plots a variable called "CC" (Cross-Correlation) versus RV correction in kms$^{-1}$, here we see the peak of the CC is at $\Delta$RV = -36.2 kms$^{-1}$. The second sub-figure shows three spectra, the black solid line is the observed spectra pre-shifted, the black dashed line is the observed spectra RV shifted and the grey solid line is the Solar model.

\newpage
\section{Impact of spectroscopic covariance on bayesian scheme}
\label{app:covariance}

\begin{figure*}
\centering
\hbox{
\begin{tabular}{@{}c@{}}
\large (a) \\
\includegraphics[scale=0.2]{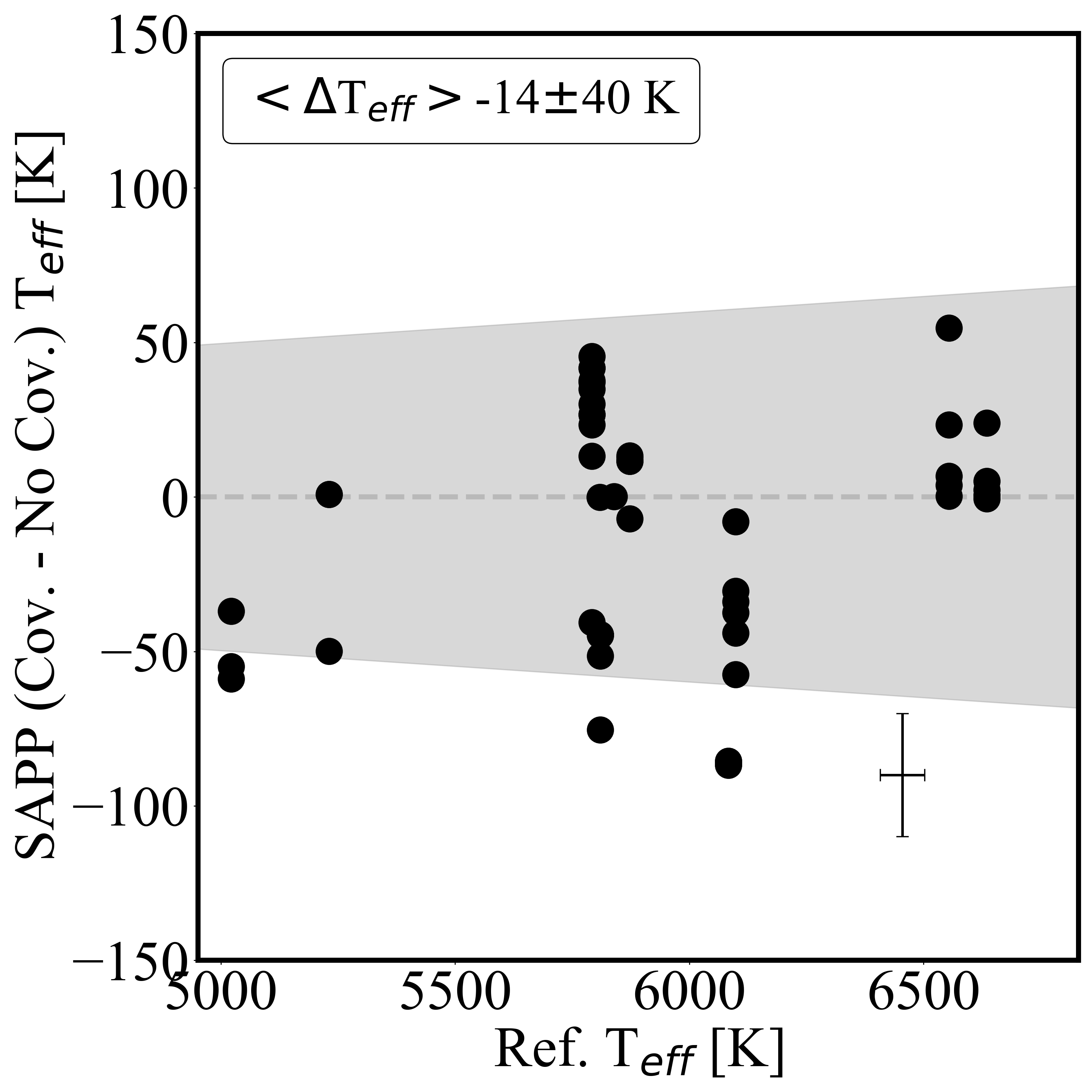}
\end{tabular}
\begin{tabular}{@{}c@{}}
\large (b) \\
\includegraphics[scale=0.2]{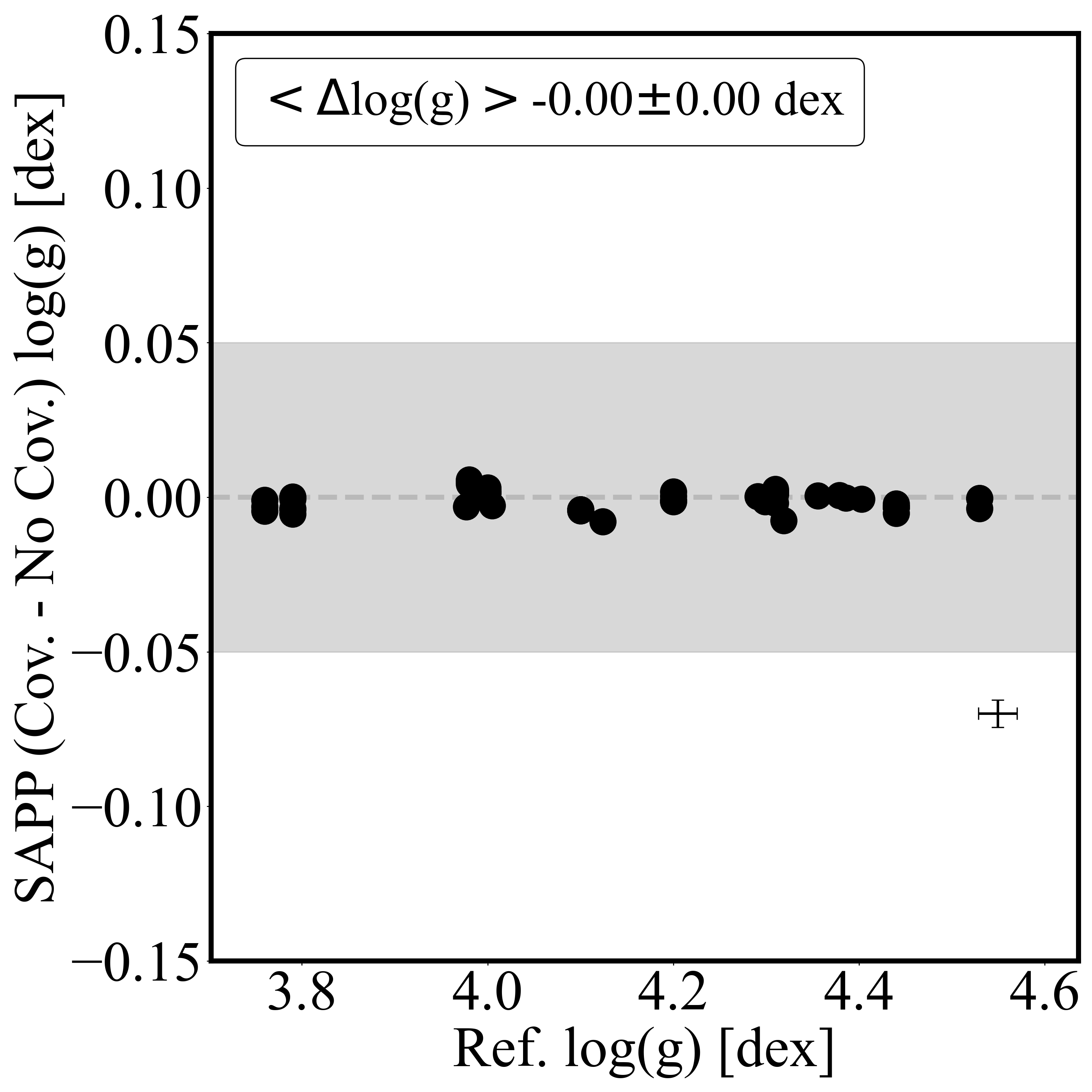}
\end{tabular}
\begin{tabular}{@{}c@{}}
\large (c) \\
\includegraphics[scale=0.2]{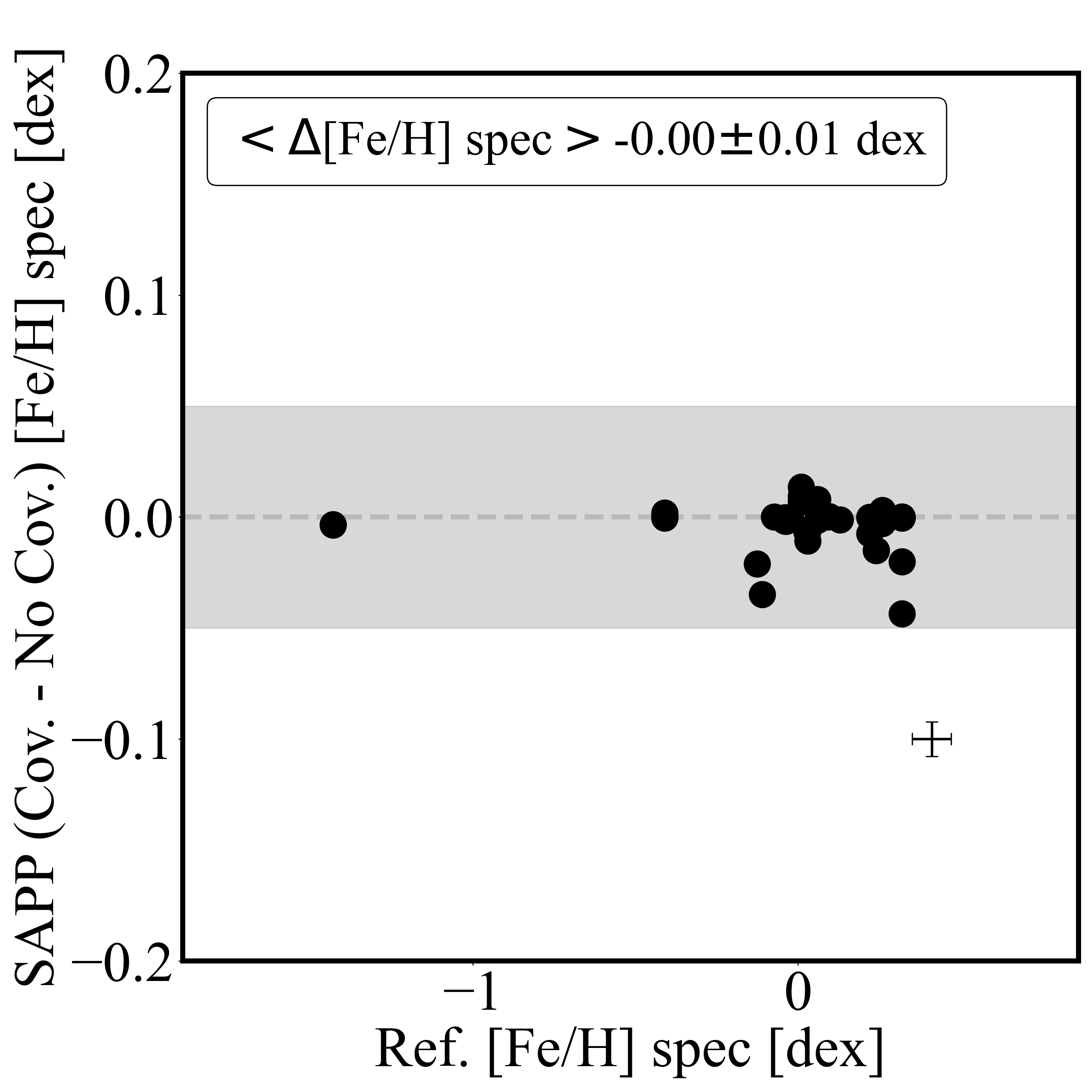}
\end{tabular}
}
\hbox{
\begin{tabular}{@{}c@{}}
\large (d) \\
\includegraphics[scale=0.2]{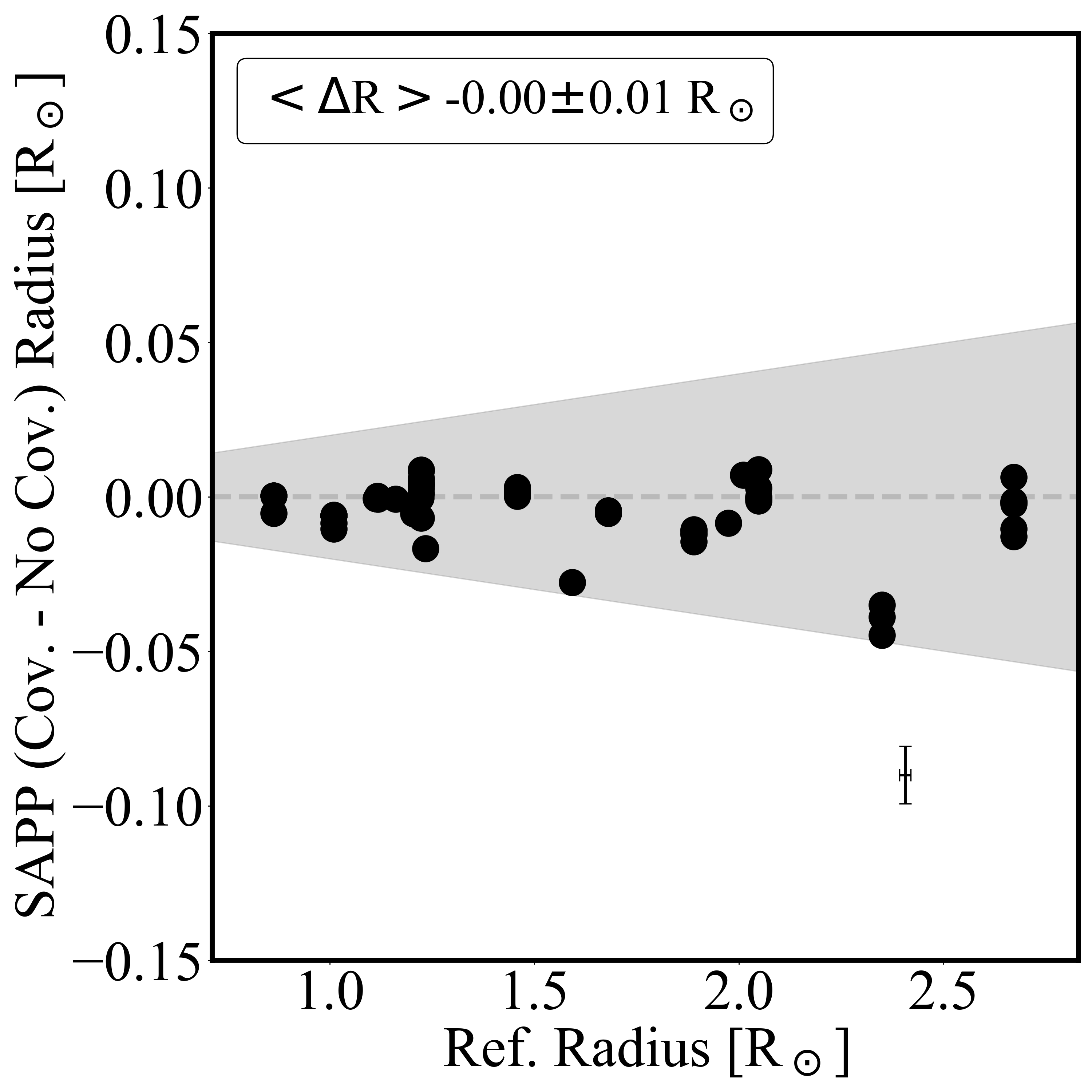} \\
\end{tabular}
\begin{tabular}{@{}c@{}}
\large (e) \\
\includegraphics[scale=0.2]{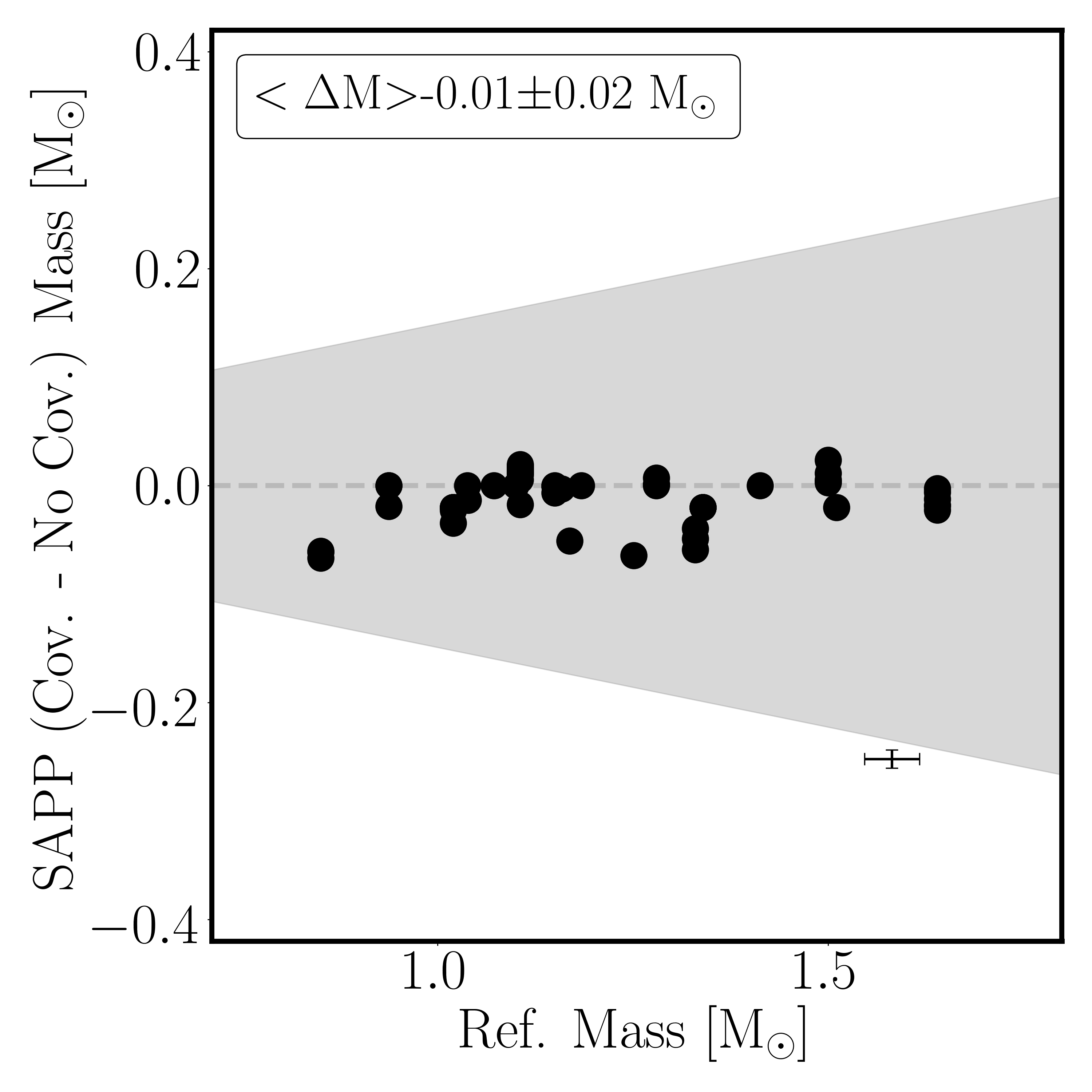} \\
\end{tabular}
\begin{tabular}{@{}c@{}}
\large (f) \\
\includegraphics[scale=0.2]{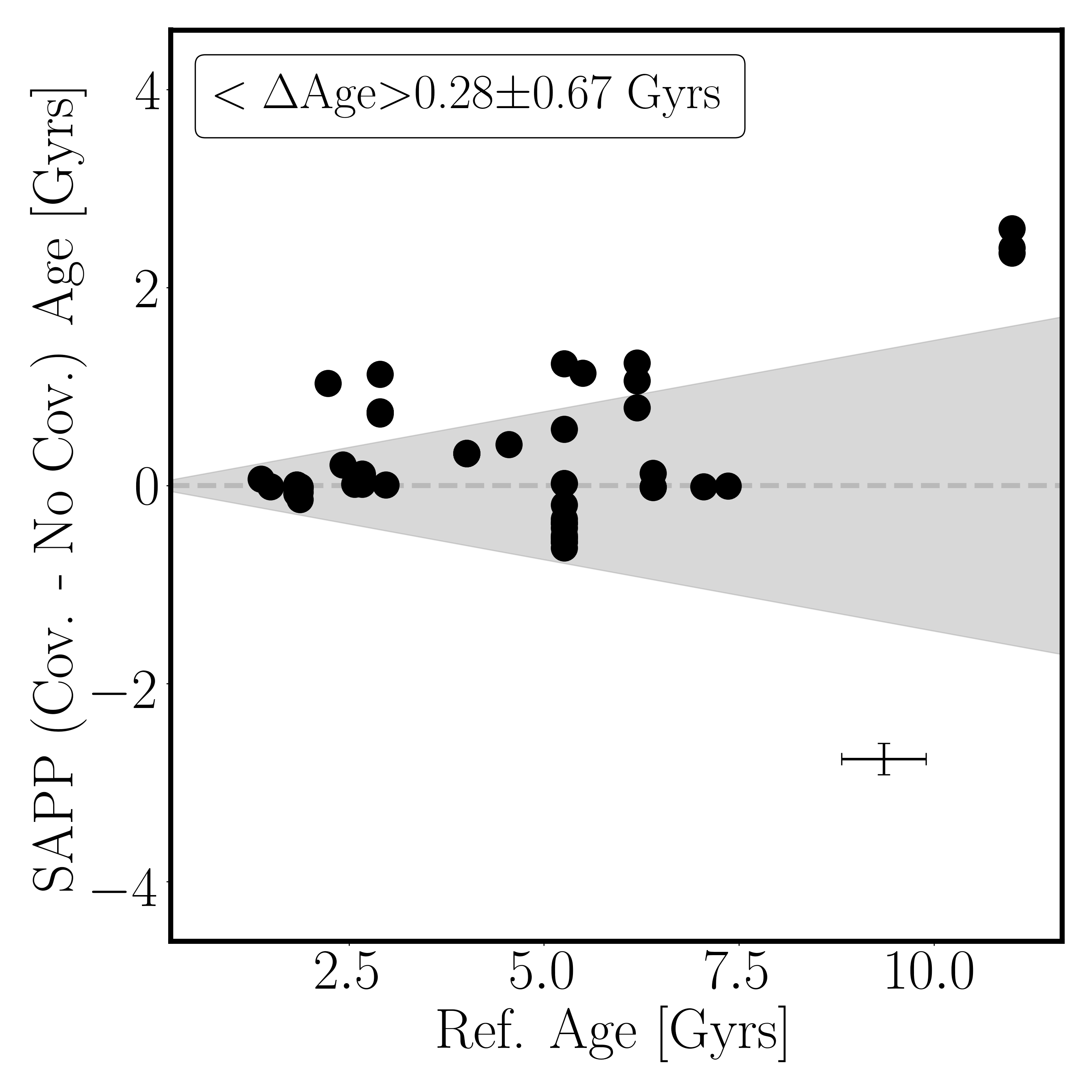} \\
\end{tabular}
}
\caption{The differences between SAPP best-fit parameters from  Bayesian scheme calculated with and without  covariance matrix from spectroscopy  (vertical axes) versus the reference parameters presented in Table \ref{tab:ref_primary} (horizontal axes). The uncertainty represents the 1 $\sigma$ deviation. The gray shaded region represents the desired errors: 1$\%$ for $\teff$, 0.05 dex for $\logg$ and [Fe/H], 15 $\%$ for mass, 10 $\%$ for age, 2 $\%$ for radius.}
\label{fig:covariance}
\end{figure*}

Figure \ref{fig:covariance} explores the difference between using the covariance matrix from spectroscopy and assuming independence between spectroscopic parameters in the Bayesian scheme. For each parameter, the difference is plotted against the reference value of each star with the average uncertainties annotated in the bottom right corner. The grey shaded region represents the desired tolerance of the final results.

\end{document}